\documentclass[pre, twocolumn, preprintnumbers, amsmath, amssymb, nofootinbib, floatfix, superscriptaddress]{revtex4}

\usepackage{graphicx, bm, xcolor, bbold}
\usepackage{enumerate}
\usepackage[normalem]{ulem} 
\usepackage[breaklinks]{hyperref}

\makeatletter
\def\graphicscale{\twocolumn@sw{0.3}{0.4}}
\def\graphicthreescale{\twocolumn@sw{0.3}{0.4}}

\begin{document}

\title{Quantum quenches across continuous and first-order quantum
  transitions\\ in one-dimensional quantum Ising models}

\author{Andrea Pelissetto}
\altaffiliation{Authors are listed in alphabetic order.}
\affiliation{Dipartimento di Fisica dell'Universit\`a di Roma
  La Sapienza and INFN, Sezione di Roma I, I-00185 Roma, Italy}

\author{Davide Rossini}
\altaffiliation{Authors are listed in alphabetic order.} 
\affiliation{Dipartimento di Fisica dell'Universit\`a di Pisa
  and INFN, Largo Pontecorvo 3, I-56127 Pisa, Italy}

\author{Ettore Vicari}
\altaffiliation{Authors are listed in alphabetic order.}
\affiliation{Dipartimento di Fisica dell'Universit\`a di Pisa,
  Largo Pontecorvo 3, I-56127 Pisa, Italy}

\date{\today}

\begin{abstract}
  We investigate the quantum dynamics generated by quantum quenches (QQs)
  of the Hamiltonian parameters in many-body systems, focusing on
  protocols that cross first-order and continuous quantum transitions,
  both in finite-size systems and in the thermodynamic limit.  As a
  paradigmatic example, we consider the quantum Ising chain in the
  presence of homogeneous transverse ($g$) and longitudinal ($h$)
  magnetic fields.  This model exhibits a continuous quantum transition (CQT)
  at $g=g_c$ and $h=0$, and first-order quantum transitions (FOQTs)
  driven by $h$ along the line $h=0$ ($g<g_c$).  In the integrable
  limit $h=0$, the system can be mapped onto a quadratic
  fermionic theory; however, any nonvanishing longitudinal field
  breaks integrability and the spectrum of the resulting Hamiltonian
  is generally expected to enter a chaotic regime.  We analyze QQs in
  which the longitudinal field is suddenly changed from a
  negative value $h_i < 0$ to a positive value $h_f>0$. We focus on
  values of $h_f$ such that the spectrum of the post-QQ Hamiltonian
  $\hat H(g,h_f)$ lies in the chaotic regime, where thermalization
  may emerge at asymptotically long times.  We study the
  out-of-equilibrium dynamics for different values of $g$, finding
  qualitatively distinct behaviors for $g > g_c$
  (where the chain is in the disordered phase), for $g = g_c$
  (QQ across the CQT), and for $g<g_c$ (QQ across the FOQT line).
\end{abstract}

\maketitle

\section{Introduction}
\label{intro}

The evolution of many-body systems in the presence of a time-dependent
Hamiltonian is generally characterized by an out-of-equilibrium
behavior, especially when the parameters cross values at which the
system undergoes a phase
transition~\cite{Fisher-74,Wilson-83,Binder-87,Ma-book,PV-02,SGCS-97,
  Sachdev-book,Pfleiderer-05,RV-21,PV-24}. In such situations,
large-scale modes are unable to equilibrate, even when the parameters
are varied very slowly~\cite{Kibble-80,Zurek-96,RV-21}.  Hysteresis,
spinodal-like and coarsening phenomena, critical aging, and
Kibble-Zurek defect production are typical examples of nonequilibrium
phenomena that are observed at first-order and continuous transitions,
driven by thermal or quantum fluctuations (see, e.g.,
Refs.~\cite{Kibble-80, Binder-87, Bray-94, Zurek-96, CA-99,
  CG-04, PSSV-11, Biroli-16, RV-21, PV-24} and references therein).

In this paper, we investigate the quantum dynamics induced by
instantaneous variations of the Hamiltonian parameters (quench dynamics)
across first-order and continuous quantum
transitions~\cite{SGCS-97, Sachdev-book, Pfleiderer-05, CPV-14, RV-21, PV-24},
considering both finite-size and infinite-size systems. 
We focus on cases in which the post-quench Hamiltonian has a chaotic
spectrum~\cite{GMGW-98, Alhassid-00, Mehta-book, DKPR-16}, where the
quantum system is expected to thermalize for typical initial states
and for sufficiently long times (see, e.g.,
Refs.~\cite{DKPR-16, Izrailev-90, ZBFH-96, BT-77, DH-90, OH-07, RDO-08,
  SR-10, BCH-11, SPR-11, RS-12, SBI-12, SBI-12b, ABGR-13, KH-13, KIH-14,
  MIKU-17, LM-17, PC-22}).  As a paradigmatic model, we consider the
quantum nearest-neighbor Ising chain in the presence of homogeneous
transverse ($g$) and longitudinal ($h$) magnetic fields. The system exhibits a
continuous quantum transition (CQT) at $g=g_c$ and $h=0$ (with $g_c=1$, in units
of the ferromagnetic coupling $J$), and first-order quantum
transitions (FOQTs) driven by the longitudinal field, along the line
$h=0$, for $g<g_c$. Any nonvanishing value of $h$ breaks integrability of
the corresponding Hamiltonian $\hat H(g,h)$, which can be mapped onto
a quadratic fermionic system only for $h=0$.  Consequently, the
spectrum of $\hat H(g,h)$ is generally expected to enter a chaotic
regime, characterized by a Wigner-Dyson distribution of level
spacings~\cite{DKPR-16}.

One of the simplest protocols to probe the out-of-equilibrium dynamics
of many-body systems is the so-called quantum quench (QQ) (see, e.g.,
Refs.~\cite{Greiner-02, Weiss-06, Schmiedmayer-07, Trotzky-12,
  Cheneau-12, Schmiedmayer-12, Guardado-etal-18, Ali-etal-24}).
This is generally implemented by suddenly changing one of the
Hamiltonian parameters, such as the transverse ($g$) or the
longitudinal ($h$) field in quantum Ising chains.  For example, when
only $h$ is varied while $g$ is kept fixed, one starts from the ground
state $|\Phi_0(g,h_i)\rangle$ of the Hamiltonian $\hat H(g,h_i)$
associated with an initial value of the parameter $h_i$, and then
suddenly quenches it to a different value $h_f \neq h_i$, at a
reference time $t=0$.  The subsequent quantum evolution is governed
by the new Hamiltonian $\hat H(g,h_f)$, so that, in units of
$\hbar=1$,
\begin{equation}
  |\Psi(t)\rangle = e^{-i \hat H(g,h_f) t} |\Phi_0(g,h_i)\rangle  \,.
\end{equation}
The post-QQ dynamics can reveal a wide range of interesting
phenomena---see, e.g., Refs.~\cite{RV-21, PSSV-11, Niemeijer-67, BMD-70,
  DMCF-06, RDYO-07, KLA-07, MWNM-07, RDO-08, ZPP-08, Bloch-08, PZ-09,
  Roux-10, IR-11, CEF-12-1, CEF-12-2, CE-13, FCEC-14, NH-15, CTGM-15,
  BDD-15, CC-16, BD-16, IMPZ-16, LGS-16, VM-16, EF-16, CKT-16, NRVH-17,
  Heyl-17, RMD-17, KCTC-17, PRV-18, HPD-18, TIGGG-19, Silva-08, GS-12,
  MS-14, MIKU-17, MR-18, MRD-19, NRV-19-w, NRV-19, STT-19, RV-20-qu, MR-21,
  HMHPRD-21, ZDZ-05, PG-08, SUXF-06, USF-10, CEGS-12, PRV-18-def, RV-20,
  TV-22, DV-23, RV-24, MW-78, Rutkevich-99, LZW-19, PRV-20, DRV-20,
  RV-20-qm, SCD-21, LSKC-21, TS-23, Surace-etal-24, PRV-25, PRV-25b,
  MHV-25, JRCB-25, Moss-25} and references therein (this should
not be regarded as an exhaustive list).  We mention: the post-QQ time
dependence of quantum correlations and
entanglement~\cite{Niemeijer-67, BMD-70, DMCF-06, Bloch-08, PZ-09,
  Roux-10, IR-11, CEF-12-1, CEF-12-2, CE-13, FCEC-14, CC-16, EF-16,
  CKT-16, NRVH-17, STT-19} and the eventual approach to stationary
states that admit a microcanonical or canonical statistical
characterization in the thermodynamic limit, when probed
locally~\cite{RDYO-07, RDO-08, NH-15, IMPZ-16, VM-16, DKPR-16}; the
emergence of preasymptotic regimes~\cite{CTGM-15, LGS-16}, especially
when the QQ crosses a FOQT, potentially leading to (quasi)-metastable
states and spinodal-like (nucleation)
phenomena~\cite{MW-78, Rutkevich-99, SCD-21, Surace-etal-24, PRV-25, PRV-25b};
the appearance of peculiar out-of-equilibrium scaling regimes at both
CQTs and FOQTs in the limit of small variations across quantum
transitions~\cite{PRV-18, PRV-18-def, PRV-20, TV-22, TS-23,
  ZDZ-05, PG-08, SUXF-06, USF-10, CEGS-12, RV-20, DV-23, RV-24};
signatures of quantum transitions imprinted by the post-QQ
dynamics~\cite{BDD-15, RMD-17, HPD-18, TIGGG-19, RV-20-qu, HMHPRD-21};
the statistics of the work associated with the
QQ~\cite{Silva-08, GS-12, MS-14, NRV-19-w}; prethermalization regimes
preceding eventual thermalization~\cite{MS-14, MIKU-17, MRD-19, MR-21};
dynamic singularities, often interpreted as dynamical phase
transitions~\cite{Heyl-17}; the role of dissipation arising from
interactions with an external
environment~\cite{NRV-19, DRV-20, RV-20-qm}; etc.

The post-QQ dynamics at CQTs develops an out-of-equilibrium
finite-size scaling (OFSS) behavior for {\em soft} QQs across the
transition~\cite{PRV-18,RV-21}, i.e., when the initial ($h_i$) and
final ($h_f$) Hamiltonian parameters remain close to the critical
point and are tuned to zero with the system size $L$ as
$h_i\sim h_f \sim L^{-y_h}$, where $y_h=15/8$ is the renormalization-group
dimension of the longitudinal field at criticality. This condition
ensures that the system remains in the critical region throughout the
entire post-QQ evolution.  Analogous considerations apply to FOQTs,
where OFSS behaviors also emerge~\cite{PRV-18,PRV-20,RV-21,PV-24}, the
main difference residing in the fact that the key features of the OFSS
are particularly sensitive to the choice of boundary conditions.

By contrast, for standard (hard) QQs, in which $h$ is kept fixed as
$L$ increases, one does not expect to observe any OFSS controlled by
the universality class of the equilibrium quantum transitions. This is
because the energy injected into the system is extensive, i.e.,
$O(L)$, and therefore much larger then the energy scale $E_c \sim
L^{-z}$ of the low-energy critical modes, where $z=1$ is the dynamic
exponent governing the finite-size gap $\Delta\sim L^{-z}$ at the
critical point.  Moreover, in nonintegrable systems which are expected
to thermalize, local observables are generally smooth functions of the
temperature, and thus are not expected to develop asymptotic long-time
singularities after a QQ.  This raises a key open question: do any
footprints of the universal low-energy features of quantum transitions
persist in the post-QQ unitary dynamics at finite values of $h$, where
the system should enter a chaotic regime and thermalize in the
long-time limit?

In fact, as argued in
Refs.~\cite{RV-20-qu,BDD-15,RMD-17,TIGGG-19,HPD-18,HMHPRD-21},
signatures of quantum transitions may appear in certain models under
standard QQs.  Standard QQs have been thoroughly investigated in
quantum Ising chains in the absence of longitudinal magnetic field,
therefore in the integrable regime, varying $g$ across the CQT, that
separates the quantum paramagnetic and ferromagnetic
phases~\cite{Niemeijer-67,BMD-70,RV-20-qu,CEF-12-1,CEF-12-2,BDD-15,EF-16}.
These QQs typically start from the ground state corresponding to an
initial value $g_i$ and then suddenly change it to $g\ne g_i$. In the
thermodynamic limit, the time-dependent energy density, as well as the
longitudinal and transverse magnetizations, develop a singular
behavior in the long-time limit, when $g\to
g_c$~\cite{BDD-15,RV-20-qu,RV-21}.  Indeed, in a QQ from the ground
state at $g_i>g_c$ to values of $g<g_i$, the $g$-derivative of the
large-time transverse magnetization $M_{\rm tr}$ turns out to be
discontinuous at $g=g_c$~\cite{RV-20-qu,RV-21}:\footnote{For generic
quantum XY chains, involving an additional anisotropy parameter
$\gamma>0$~\cite{Sachdev-book,RV-21}, one finds
\begin{equation}
  \lim_{g\to g_c^+} {\partial M_{\rm tr}(g_i,g)\over \partial g} -
  \lim_{g\to g_c^-} {\partial M_{\rm tr}(g_i,g)\over \partial g} =
  \gamma^{-1}.
  \label{singdermtga}
\end{equation}
This equation corrects a misprint appeared in
Refs.~\cite{RV-20-qu,RV-21}.}
\begin{equation}
  \lim_{g\to g_c^+} {\partial M_{\rm tr}(g_i,g)\over \partial g} -
  \lim_{g\to g_c^-} {\partial M_{\rm tr}(g_i,g)\over \partial g} =
  1,
  \label{singdermt}
\end{equation}
independently of $g_i$.  Nonanalytic behaviors around $g_c$ have also
been observed in the long-time behavior of the longitudinal
magnetization following QQs from an initially ordered
state---corresponding to the ground state at large $h$---when the
post-QQ evolution is governed by the Ising Hamiltonian with $h=0$.
However, these nonanalyticities at $g_c$ seem to be related to the
specific spectral properties of the post-QQ Hamiltonian in the absence
of a longitudinal field, which can be mapped onto a quadratic
fermionic Hamiltonian (see, e.g., Ref.~\cite{Sachdev-book}).

In this paper we rather consider QQs of the longitudinal field $h$,
focusing on regimes where the spectrum of the post-QQ Hamiltonian is
chaotic.  We investigate several features of the resulting
out-of-equilibrium dynamics for various values of the
transverse field $g$, observing different qualitative behaviors
for $g >g_c$ (QQ in the disordered phase), for
$g= g_c$ (QQ across the CQT), and for $g<g_c$ (QQ across the FOQT
line).  To this end, we present a detailed analysis of the spectrum
and of several observables in the diagonal ensemble for lattice sizes
up to $L=20$---these calculations require exact diagonalization of
the Hamiltonian. As for the post-QQ evolution of the system, we are
able to obtain exact results for significantly longer chains, up to
$L=28$, and for quite long times, by using Lanczos methods followed
by a fourth-order Runge-Kutta integration of the post-QQ unitary
dynamics (the time step can be adjusted to achieve accuracies
of the order of machine precision).
Depending on the values of the parameters $g$ and $h$, these times
may reach several hundreds (in units of $\hbar/J$) and be sufficient
to achieve the thermodynamic-limit behavior.
To access such chain sizes and times, it is crucial to exploit
the symmetries of the system, projecting onto
the zero-momentum Hilbert subspace~\cite{PRV-25,Sandvik-10}.
\footnote{There are other numerical methods, such as tensor-network
techniques (e.g., the iTEBD algorithm operating on an infinite
matrix-product-state representation of the system wave function) that
can be used to assess the post-QQ evolution directly in the
thermodynamic limit. The caveat is that reachable times are limited by
entanglement growth [particularly in highly entangling dynamics, where
  simulations can be typically considered reliable up to times
  $t=O(10^1)$ in units of the characteristic energy scale]---see,
e.g., Refs.~\cite{DMCF-06, KLA-07, MWNM-07, BCH-11, FCEC-14, CKT-16,
  KCTC-17, LSKC-21}.  One would thus need to carefully scrutinize the
effects of a finite truncation of the bond links of the various
tensors, rather than controlling finite-size effects.  Our approach is
complementary to these methods and turns out to be particulatly useful
for hard QQs crossing quantum transitions, such as those considered
here. \label{note:TEBD}}

The paper is organized as follows.  In Sec.~\ref{model} we introduce
the quantum Ising-chain Hamiltonian of interest, together with the QQ
protocols considered and all the quantities addressed throughout the
work.  Section~\ref{prelspe} presents an analysis of the spectral
properties of the system as a function of $g$ and $h$, within the
zero-momentum sector.  In Sec.~\ref{softqu}, we focus on soft QQs
across the CQT and FOQTs of the Ising chain.  The remaining
sections~\ref{disphase}, \ref{ququcqt}, and \ref{ququfoqt} contain a
detailed numerical study of the dynamics of the system following hard
QQs of the longitudinal field $h$, respectively within the disordered
phase ($g>g_c$), across the CQT ($g=g_c$), and across FOQTs ($g<g_c$).
In Sec.~\ref{conclu} we summarize the results and draw our
conclusions. Finally, in the Appendix we present some results for a QQ
in the presence of oppositely fixed boundary conditions in the
small-$g$ and small-$h$ regime, in which the one-kink approximation
holds.

\section{Model and quantum quench protocol}
\label{model}

\subsection{The quantum Ising chain}
\label{quisch}

The nearest-neighbor quantum Ising chain in a transverse field $g$ is
a paradigmatic model showing a CQT and {\em magnetic} FOQTs driven by
the longitudinal field $h$. Its Hamiltonian reads
\begin{equation}
  \hat H(g,h) = - \sum_{x=1}^L \left[ J\,\hat\sigma^{(1)}_x
  \hat\sigma^{(1)}_{x+1}
  + g\, \hat\sigma^{(3)}_x + h \,\hat\sigma^{(1)}_x\right] ,
  \label{hedef}
\end{equation}
where $\hat\sigma_x^{(k)}$ are Pauli matrices ($k=1,2,3$) defined on
the $x$th lattice site. We consider chains with $L$ sites and
periodic boundary conditions
($\hat\sigma^{(k)}_x=\hat\sigma^{(k)}_{L+x}$), which preserve
translational invariance (no boundaries are present) and the
${\mathbb Z}_2$ symmetry when $h=0$.  Without loss of generality,
we assume $J = 1$ and $g>0$. We also set $\hbar = k_B = 1$.

At zero temperature and for $g=g_c=1$, $h=0$, the model~\eqref{hedef}
undergoes a CQT belonging to the two-dimensional Ising universality
class (with critical exponents $\nu=1$, $\eta=1/4$, and $z=1$), which
separates a disordered phase ($g > g_c$) from an ordered one
($g<g_c$).  For any $g<g_c$, the longitudinal field drives FOQTs along
the $h=0$ line, with the spontaneous breaking of the ${\mathbb Z}_2$
symmetry in the thermodynamic limit.  For any $g<g_c$ and for
$|h|\to 0$, the magnetized states $|+\rangle$ and $|-\rangle$ along
the longitudinal direction are the ground states of the system
for $h>0$ and $h<0$, respectively, giving rise to a discontinuous
infinite-volume longitudinal magnetization $m_0$:~\cite{Pfeuty-70}
\begin{equation}
  \lim_{h\to 0^\pm} \lim_{L\to\infty} M = \pm \, m_0, \qquad m_0 = (1
  - g^2)^{1/8},
  \label{sigmasingexp}
\end{equation}
where
\begin{equation}
  M = {1\over L} \langle \Phi_0(g,h) | \hat M | \Phi_0(g,h) \rangle, \quad
  \hat M \equiv {1\over L} \sum_x \hat\sigma_x^{(1)},
  \label{lomagn}
\end{equation}
and $|\Phi_0(g,h)\rangle$ is the ground state.

\subsection{Quantum quench protocols}
\label{quprot}

We focus on the out-of-equilibrium quantum dynamics arising from QQs
of the longitudinal-field parameter $h$, from $h_i\le 0$ to $h_f>0$, keeping
the transverse-field parameter $g$ fixed (for the sake of clarity, in
the following we omit the $g$ dependence in the formulas).  We
consider a standard QQ protocol: the system is initially in the ground
state $|\Phi_{0}(h_i)\rangle$. At $t=0$, the longitudinal field is
instantaneously changed to $h_f$ and the quantum evolution is
described by the density matrix
\begin{equation}
  \hat\rho(t) = |\Psi(t) \rangle \langle\Psi(t)|,
  \label{rhot}
\end{equation}
where $|\Psi(t) \rangle$ is the solution of the Schr\"odinger equation
\begin{equation}
i \partial_t |\Psi(t) \rangle = \hat H(h_f) |\Psi(t) \rangle,
\quad |\Psi(t=0) \rangle = |\Phi_0(h_i) \rangle.
\label{scheq}
\end{equation}
The evolving state $|\Psi(t)\rangle$ can be written as a sum over the
eigenstates $|\Phi_n(h)\rangle$ of the post-QQ Hamiltonian $\hat H(h_f)$,
\begin{equation}
  |\Psi(t)\rangle = \sum_n c_n \,e^{-iE_n(h_f)t}\,|\Phi_n(h_f)\rangle,
  \label{formevo}
\end{equation}
where $c_n = \langle \Phi_n(h_f) | \Phi_0(h_i)\rangle$ and $E_n(h_f)$
are the eigenvalues of $\hat H(h_f)$.

To simplify the analysis, we consider a QQ with
\begin{equation}
  h_i = -h < 0, \qquad h_f = h > 0,  \label{hih}
\end{equation}
so that the QQ protocol has only one free parameter $h$. Along the
FOQT line for $g<g_c$, we also consider QQ protocols starting from
$h_i=0^-$, where we specify that $h_i$ approaches zero from below,
because of the discontinuity~\eqref{sigmasingexp} in the thermodynamic
limit.

We mention that other types of QQs from magnetized states
(corresponding to large $h_i$) to $h_f=0$ have been already
investigated~\cite{CEF-12-1,CEF-12-2,RV-20-qu,RV-21}, revealing the
emergence of singular behaviors in the thermodynamic and long-time
limits.  In particular, the large-time asymptotic stationary value of
the longitudinal magnetization is nonanalytic at
$g=g_c$~\cite{RV-20-qu,RV-21}.  However, as already noted in the
introduction, this phenomenon is expected to be related to the
specific integrability properties of the post-QQ Hamiltonian when
$h=0$.

The system evolution for $t>0$ can be monitored by evaluating the
instantaneous longitudinal magnetization
\begin{equation}
  M(t) = {\rm Tr} \big[ \hat\rho(t) \, \hat M \big]
  = {\rm Tr} \big[ \hat\rho(t) \,\hat\sigma_{x}^{(1)} \big] ,
  \label{mxm}
\end{equation}
where we used translational invariance.  We also consider the excess
bond energy,
\begin{subequations}
  \begin{eqnarray}
    K(t) &=& {\rm Tr} \big[ \hat\rho(t) \, \hat K \big]
    - \langle\Phi_0(h)| \hat K | \Phi_0(h) \rangle, \label{boexdef}\\
    \hat K &=& - {1\over L}
    \sum_x \hat\sigma^{(1)}_x \hat\sigma^{(1)}_{x+1}.
    \label{Kdef}
  \end{eqnarray}
  \label{boexdefF}
\end{subequations}

At the CQT and at FOQTs, one may consider soft QQs, in which the
initial ($h_i$) and final ($h_f$) longitudinal fields are sufficiently
small to only affect the low-energy and large-scale modes associated
with the transition~\cite{RV-21,PRV-18}. This generally requires an
appropriate tuning of $h_i$ and $h_f$ to zero in the large-$L$ limit.
By contrast, in standard (hard) QQs, where $h_i$ and $h_f$ are kept
fixed as $L$ increases, noncritical, short-range modes are eventually
activated. A key issue is whether---and to what extent---any
footprints of the phase transition survive in the dynamics following a
hard QQ across the quantum transition point (see, e.g.,
Refs.~\cite{RV-20-qu,BDD-15,RMD-17,TIGGG-19,HPD-18,HMHPRD-21}).

\subsection{Remarks on the post-quench evolution}
\label{remarks}

The time dependence of a generic observable $\hat O$ can be written as
\begin{eqnarray}
O(t) &=& {\rm Tr} \big[ \hat\rho(t)\,\hat O \big] =
  \sum_{m,n} c_m^* c_n e^{i(E_m-E_n)t} O_{mn} \nonumber \\
  &=& \sum_{n} w_n O_{nn} +  \sum_{m\neq n} c_m^* c_n e^{i(E_m-E_n)t} O_{mn},
  \qquad
  \label{oevol}
\end{eqnarray}
where
\begin{equation}
  w_n = |c_n|^2 = | \langle \Phi_n(h) | \Phi_0(h_i)\rangle |^2,
  \quad \sum_n w_n = 1,
  \label{wndef}
  \end{equation}
and $O_{mn} = \langle \Phi_m(h) | \hat O |\Phi_n(h)\rangle$.
Averaging the expectation value $O(t)$ over a sufficiently long time
(assuming no degeneracies of the involved Hamiltonian eigenstates),
the oscillating last term in the second line of Eq.~\eqref{oevol} does
not contribute.  Therefore, one can define a time-independent
{\em diagonal ensemble} density matrix~\cite{RDO-08}
\begin{eqnarray}
  \hat\rho_D = \sum_n w_n \, |\Phi_n(h)\rangle \langle \Phi_n(h)|,
  \label{rhoddef}
\end{eqnarray}
which is expected to provide an effective approximation of the
long-time average of few-body observables (see, e.g.,
Ref.~\cite{DKPR-16}).  Corresponding expectation values $M_D$ and
$K_D$ can then be defined as
\begin{eqnarray}
  M_D & = & {\rm Tr} \big[ \hat\rho_D\,\hat M \big],  \label{mxmD}\\
  K_D & = & {\rm Tr} \big[ \hat\rho_D \,\hat K \big]
  - \langle\Phi_0(h)| \hat K | \Phi_0(h) \rangle.     \label{boexdefD}
\end{eqnarray}

The post-QQ evolution of the state is characterized by the
energy-density distribution
\begin{equation}
  D(e) = \sum_n w_n \delta(e-e_n),\quad e_n = {E_n(h)-E_0(h)\over L},
  \label{pedistr}
\end{equation}
where $E_n= \langle \Phi_n(h) | \hat H(h) | \Phi_n(h) \rangle$.  The
corresponding mean and variance are given by
\begin{align}
\bar e = & \frac{\langle \Phi_0(h_i) | \hat H(h) | \Phi_0(h_i) \rangle 
    - E_0(h)}{L}, \label{meansigmaE}\\
  \sigma_e^2 = & \frac{\langle \Phi_0(h_i) | \hat H(h)^2 | \Phi_0(h_i) \rangle
    - \langle \Phi_0(h_i) | \hat H(h) | \Phi_0(h_i)\rangle^2}{L^2}.
  \nonumber
\end{align}
For a QQ, the normalized energy distribution $D(e)$ is generally
expected to approach a Gaussian, independently of the specific nature
of $\hat H(h_i)$, provided the system is sufficiently large (this is
essentially due to the exponentially growing density of states with
the system size)~\cite{DKPR-16,SBI-12,SBI-12b}.  Moreover, the ratio
$\sigma_e/\bar e$ is expected to decrease as $V^{-1/2}$ with
increasing volume $V$~\cite{DKPR-16}, or equivalently as $L^{-1/2}$
for one-dimensional lattice systems, leading to an effective
microcanonical energy window.

Since $\sigma_e/e\to 0$ in the large-$L$ limit, only states with given
energy density $e$ are relevant. Assuming that all these states have
the same probability, i.e., an effective asymptotic thermalization,
the chain behaves as an effectively microcanonical statistical system
in the large-$L$ and large-time limit. If this is the case, one can
define a corresponding effective inverse temperature $\beta_{\rm th}$,
which can be determined by considering a small subsystem, since its
reduced density matrix should be proportional to the
canonical-distribution density matrix $e^{-\beta_{\rm th} \hat{H}_s}$,
where $\hat H_s$ is the restriction of the Hamiltonian $\hat{H}$ to
the subsystem. Explicitly, we consider a subsystem of size $\ell$,
with sites $x\in [1,\ell]$ with $1\ll \ell\ll L$ (the choice of the
first site is arbitrary, due to translation invariance) and with
Hamiltonian
\begin{equation}
  \hat H_s = - \sum_{x=1}^{\ell-1} \hat\sigma^{(1)}_x
    \hat\sigma^{(1)}_{x+1} - \sum_{x=1}^\ell \Big[ g \hat\sigma^{(3)}_x +
    h \,\hat\sigma^{(1)}_x \Big] ,
  \label{hedefs}
\end{equation}
with open boundary conditions. In practice, the effective temperature
can be determined by imposing that
\begin{equation}
  \frac{\langle \Phi_0(h_i)|\hat H(h)|\Phi_0(h_i)\rangle}{L} = \frac{1}{\ell} \:
       { {\rm Tr} \big[ e^{-\beta_{\rm th}\hat H_s(h)} \hat H_s(h) \big] \over
         {\rm Tr} \big[ e^{-\beta_{\rm th} \hat H_s(h)} \big] }, 
    \label{tthdef}
\end{equation}
namely, that the canonical mean energy per site of the subsystem equals
the energy per site of the full chain in the initial state.

The same relation should hold for
the asymptotic large-time values of all local observables, which can
also be obtained as canonical averages at the effective inverse
temperature $\beta_{\rm th}$.  For instance, the asymptotic long-time
values of $M$ should satisfy
\begin{equation}
  \lim_{t\to \infty} M(t) \approx 
      { {\rm Tr} \big[ e^{-\beta_{\rm th}\hat H_s(h)} \hat M_s \big] \over
      {\rm Tr} \big[ e^{-\beta_{\rm th} \hat H_s(h)} \big] } ,
  \label{asyM}
\end{equation}
where $\hat M_s= \ell^{-1} \sum_{x=1}^\ell \hat \sigma_x^{(1)}$;
analogous relations should hold for all local observables, for
instance for the excess bond energy. Thus, $\beta_{\rm th}$ can also
be obtained be enforcing Eq.~(\ref{asyM}) or the analogous relation
for the excess bond energy. If these equations yield significantly
different values of $\beta_{\rm th}$, thermalization does not hold.
Note that, as we shall see, one may also obtain negative values of
$\beta_{\rm th}$, i.e., asymptotic stationary states locally described
by Gibbs ensembles with negative temperature, giving rise to a
peculiar type of equilibrium states (see Ref.~\cite{BILV-21} for a
through discussion of the relevance and thermodynamic meaning of
negative $\beta_{\rm th}$ to describe statistical systems).

The degree of spectral delocalization of the post-QQ
Hamiltonian $\hat H(h)$ can be quantified through the
diagonal-ensemble
entropy~\cite{SPR-11,RS-12,SR-10,Izrailev-90,ZBFH-96}
\begin{equation}
  S_D = - {\rm Tr}\,[\hat\rho_D \ln \hat\rho_D] = - \sum_n w_n \ln w_n.
  \label{sddef}
\end{equation}
This quantity vanishes if there exists a single element $m$ for which
$w_m=1$, and $w_n=0$ for $m \neq n$, while it reaches its maximal value
$S_D=\ln D_{\cal H}$ when all $w_n$ are equal ($D_{\cal H}$ is the
dimension of the Hilbert subspace spanned by the initial state, taking
into account the symmetries). Further information can be obtained from
the large-$L$ decay of the largest value of $w_n$,
\begin{equation}
  w_{\rm max} = {\rm Max}[w_n],
  \label{W1def}
\end{equation}
or from the average $w_{{\rm max},k}$ of the largest $k$ values of $w_n$.
This tells us whether few eigenstates of the post-QQ Hamiltonian
dominate the quantum evolution.

\section{Analysis of the spectrum}
\label{prelspe}

Before studying the post-QQ out-of-equilibrium dynamics of the quantum
Ising chain, it is instructive to first examine its spectrum for
generic finite values of $g$ and $h$, where the model is
nonintegrable.  We recall that integrability is recovered for $h=0$ at
any $g$, in which case the Hamiltonian can be exactly mapped onto a
quadratic fermionic model via a Jordan-Wigner transformation, as well
as in the trivial limits $g\to 0$, $g\to\infty$, and $h\to\infty$.

\subsection{The zero-momentum sector}
\label{zerosec}

We consider chains with an even number $L$ of sites and periodic
boundary conditions. These boundary conditions preserve translational
invariance, allowing the spectrum to be decomposed into sectors with
different momenta $p = {2\pi k/L}$, where $k=0, \ldots ,L-1$.  Since
the dynamics starts from a zero-momentum ground state, we can
restrict our computations to the zero-momentum sector ${\cal H}_0$.
The dimension ${\cal D}_{0}(L)$ of this sector exhibits a nontrivial
dependence on $L$, whivh can be computed by counting the number of binary
necklaces~\cite{Necklaces-ref1,Necklaces-ref2}, i.e., of distinct
cyclic arrangements of $L$ colored beads, chosen from two
available colors. One obtains
\begin{equation}
  {\cal D}_{0}(L) = \frac{1}{L} \sum_{n \in \{{\rm Div}(L)\}} 2^{L/n} \varphi_E(n),
  \label{p0states}
\end{equation}
where ${\rm Div}(L)$ is the set of integer numbers that divide $L$,
and $\varphi_E(n)$ is Euler totient function.  For example,
${\cal D}_{0}= 6,14,36,108,352,1182,4116$, for $L=4,6,8,10,12,14,16$,
respectively. Note that, in the large-$L$ limit, ${\cal D}_{0}(L)$
approaches the asymptotic behavior
\begin{equation}
  \lim_{L\to\infty} {L \, {\cal D}_{0}(L) \over {\cal D}(L)} = 1,
  \label{asyp0states}  
\end{equation}
where ${\cal D}(L) = 2^L$ is the dimension of the full Hilbert space.

The zero-momentum sector of the quantum Ising Hamiltonian
(\ref{hedef}) is invariant under the ${\mathbb Z}_2$ symmetry
transformation $\hat P_r$ that corresponds to the coordinate
reflection $x \to L+1-x$. As a consequence, the $p=0$ eigenstates of
the Hamiltonian can be divided into two sets, belonging to the
subsectors ${\cal H}_{0+}$ and ${\cal H}_{0-}$, with parity $P_r=\pm 1$
under reflection. The numbers ${\cal D}_{0 \pm}(L)$ of $p=0$ states
with positive and negative parity differ.  For even $L$ one finds:
\begin{equation}
  {\cal D}_{0+}(L) - {\cal D}_{0-}(L) = 3 \times 2^{L/2-1}.
  \label{diffdp0pm}
\end{equation}
Nevertheless, this imbalance becomes asymptotically negligible
in the large-$L$ limit, since
\begin{equation}
  \lim_{L\to\infty} {{\cal D}_{0+}(L) - {\cal D}_{0-}(L)
    \over {\cal D}_{0}} = 0 ,
  \label{asydiffp0}
\end{equation}
where ${\cal D}_{0+} + {\cal D}_{0-} = {\cal D}_{0}$.  The
reflection symmetry of the system implies that, if the initial state
is a $p=0$ state with parity $P_r=\pm 1$, the subsequent time evolution
remains confined within the corresponding subspace ${\cal H}_{0\pm}$.

\subsection{Chaotic quantum regime}
\label{chaoreg}

The energy spectrum of a quantum many-body system can be characterized
by the distribution of the differences between adjacent energy levels
\begin{equation}
  s_n = E_{n+1}-E_n.
  \label{omedef}
\end{equation}
Their average value decreases exponentially with increasing $L$.
Indeed
\begin{equation}
  \langle s \rangle = {\sum_n s_n \over {\cal D}_{0+}-1} = {E_{\rm max}
    - E_{\rm min} \over {\cal D}_{0+}-1} \sim L^2 \, 2^{-(L-1)},
  \label{omegaav}
\end{equation}
where we used~Eqs.~(\ref{asyp0states}) and (\ref{asydiffp0}), and the
fact that $E_{\rm max}- E_{\rm min}\sim L$.  For this reason, to
obtain large-$L$ predictions, one should consider the statistics
$P(\{\bar s_n\})$ for the rescaled quantities $\bar s_n = s_n /
\langle s_n \rangle$, so that spacings are normalized by their average
value $\langle s_n \rangle$ in a local energy window, where the mean
level density is set to unity, a procedure which is usually referred
to as the unfolding of the spectrum~\cite{Mehta-book}.  In most
situations, it is sufficient to normalize the spacings over the global
energy average, $\bar s_n = s_n / \langle s \rangle$.

The shape of the distribution $P(\{\bar s_n\})$ allows one to
distinguish between integrable and chaotic regimes (see, e.g.,
Ref.~\cite{DKPR-16} for an extended discussion).  Indeed, for models
with time-reversal symmetry such as the quantum Ising
chain~\eqref{hedef}, when the system is in a chaotic regime,
$P(\{\bar s_n\})$ is expected to follow the Wigner-Dyson (WD)
level-spacing distribution associated with the Gaussian orthogonal
ensemble (GOE). This distribution can be accurately
approximated~\cite{DH-90,ABGR-13} by the Wigner surmise (exact for the
random two-level problem):
\begin{equation}
  P_{\rm W}(s) = {\pi s \over 2} e^{- \pi s^2 / 4},
\label{wigdis}
\end{equation}
with an average level spacing normalized to one.  Another possibility
(the Berry-Tabor conjecture~\cite{BT-77}) is that the distribution of the
energy-level separations follow a Poissonian distribution
\begin{equation}
  P_{\rm P}(s) = e^{-s}.
  \label{Poissondistr}
  \end{equation}
Note that a Poissonian distribution does not necessarily occur outside
the chaotic regime for $h \not = 0$---see the Appendix for an explicit
example in the small-$g$ and small-$h$ regime---as it generally occurs
when the Hamiltonian can be expressed as a sum of many independent
contributions. In such case, the energy eigenvalues behave as sums of
independent random variables, and consequently exhibit Poisson
statistics~\cite{BT-77}.\footnote{For example, this situation is
realized in the quantum Ising chain with Hamiltonian~\eqref{hedef} and
$h=0$, where, after a Jordan-Wigner transformation, such Hamiltonian
becomes quadratic in the fermionic creation and annihilation
operators~\cite{Sachdev-book}.}

\begin{figure}[!t]
  \includegraphics[width=0.47\textwidth]{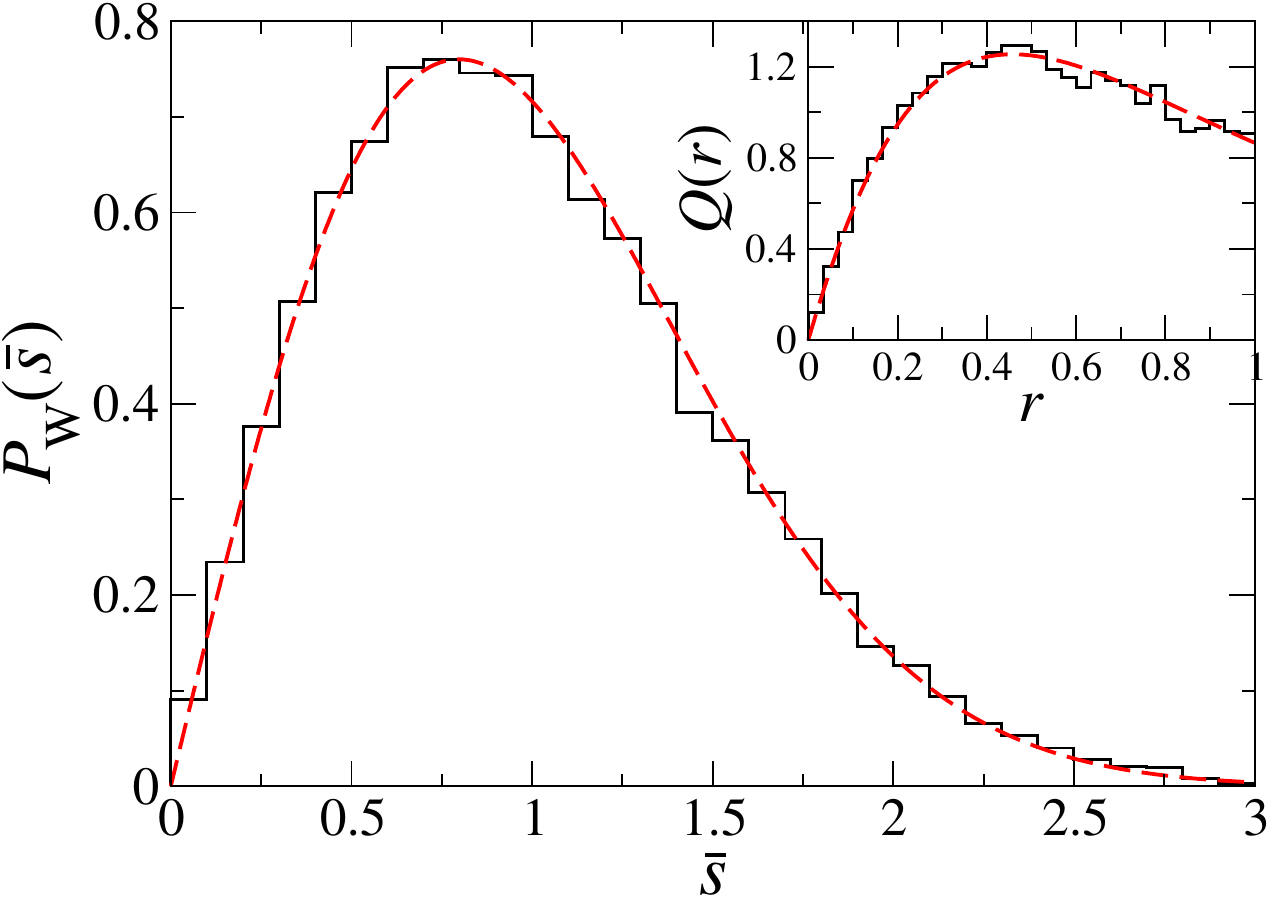}
  \caption{The distribution of the level spacings $s_n$, normalized by
    the average value $\langle s \rangle$, for a quantum Ising chain
    with $L=20$ spins and $g=h=1$. We only consider states in the
    Hilbert subspace ${\cal H}_{0+}$, with dimension $\mathcal{D}_{0+}
    = 27012$.  The dashed red curve shows the expected Wigner surmise
    for a GOE, given by Eq.~\eqref{wigdis}.  The inset displays the
    corresponding distribution of the ratios $r_n$ for the same set of
    parameters, together with the analytic prediction~\eqref{distr_R}
    obtained from the Wigner surmise.}
\label{figexW}
\end{figure}

We emphasize that the WD distribution is expected to be observed in
chaotic regimes only in the absence of symmetries, which may otherwise
induce degeneracies in the
spectrum~\cite{Bohigas-91,BCMS-03,ABGR-13}. Accordingly, for the Ising
chain in Eq.~\eqref{hedef}, the analysis should be restricted to
states within a single parity sector, ${\cal H}_{0+}$ or ${\cal H}_{0-}$.
We performed such an analysis for generic values of $g$
and $h$ sufficiently far from the integrable limits of the model. For
example, Fig.~\ref{figexW} shows that our numerical results obtained
in the subspace ${\cal H}_{0+}$, for $g=h=1$, nicely match the Wigner
surmise~\eqref{wigdis}.  In contrast, computing the level-spacing
distribution over the full ${\cal H}_{0}$ subspace would fail to
reproduce the WD statistics, as it would reflect the underlying
${\mathbb Z}_2$ reflection symmetry which is present in ${\cal H}_{0}$.
This would eventually lead to a variant distribution,
appropriate for systems with this symmetry (see, e.g.,
Refs.~\cite{Bohigas-91,BCMS-03}).

%%%%%%%%%%%%%%%%%%%%%%%%%%%%%%%%%%%%%%%%%%%%%%%%%%%%%%%%%%%%%%%%%%%%%%%%%%%%%%%%%%
%%%%%%%%%%%%%%%%%%%%%%%%%%%%%%%%%%%%%%%%%%%%%%%%%%%%%%%%%%%%%%%%%%%%%%%%%%%%%%%%%%

A useful parameter to quantify the closeness to the chaotic WD regime
can be obtained from the average~\cite{OH-07}
\begin{equation}
  R = \langle r_n \rangle
  \label{Rdef}
  \end{equation}
  of the ratios
\begin{equation}
  r_n={{\rm min} \, [s_n, s_{n-1}]\over {\rm max} \, [s_n, s_{n-1}]}
  = {\rm min}\left[{s_n \over s_{n-1}},{s_{n-1} \over s_{n}}\right].
  \label{sndef}
\end{equation}
Contrary to the bare energy differences $s_n$, the ratios of
consecutive level spacings are independent of the local density of
states, thus unfolding is not necessary in this case. For that reason,
the study of the statistics of the $r_n$ has nowadays been established
as the standard tool to ascertain the Hamiltonian spectral properties
in quantum many-body systems.  The corresponding distributions
obtained from the Wigner surmise~\eqref{wigdis} and the
Poissonian~\eqref{Poissondistr} statistics are respectively given
by~\cite{ABGR-13}
\begin{equation}
  Q_W(r) = \frac{27}{4} \frac{r+r^2}{(1+r+r^2)^{5/2}} , \quad
  Q_P(r) = \frac{2}{(1+r)^2} .
  \label{distr_R}
\end{equation}
The average quantity $R$ takes the value~\cite{ABGR-13,OH-07}
$R_W = 0.5307(1)$ for the level-spacing WD distribution of GOE
(note that the approximate analytical Wigner surmise would give
$R_W = 4 - 2\sqrt{3} \approx 0.5359$), and
$R_P = 2\ln 2 - 1 \approx 0.3863$ for the Poisson distribution.

%%%%%%%%%%%%%%%%%%%%%%%%%%%%%%%%%%%%%%%%%%%%%%%%%%%%%%%%%%%%%%%%%%%%%%%%%%%%%%%%%%
\begin{figure}[!t]
  \includegraphics[width=0.47\textwidth]{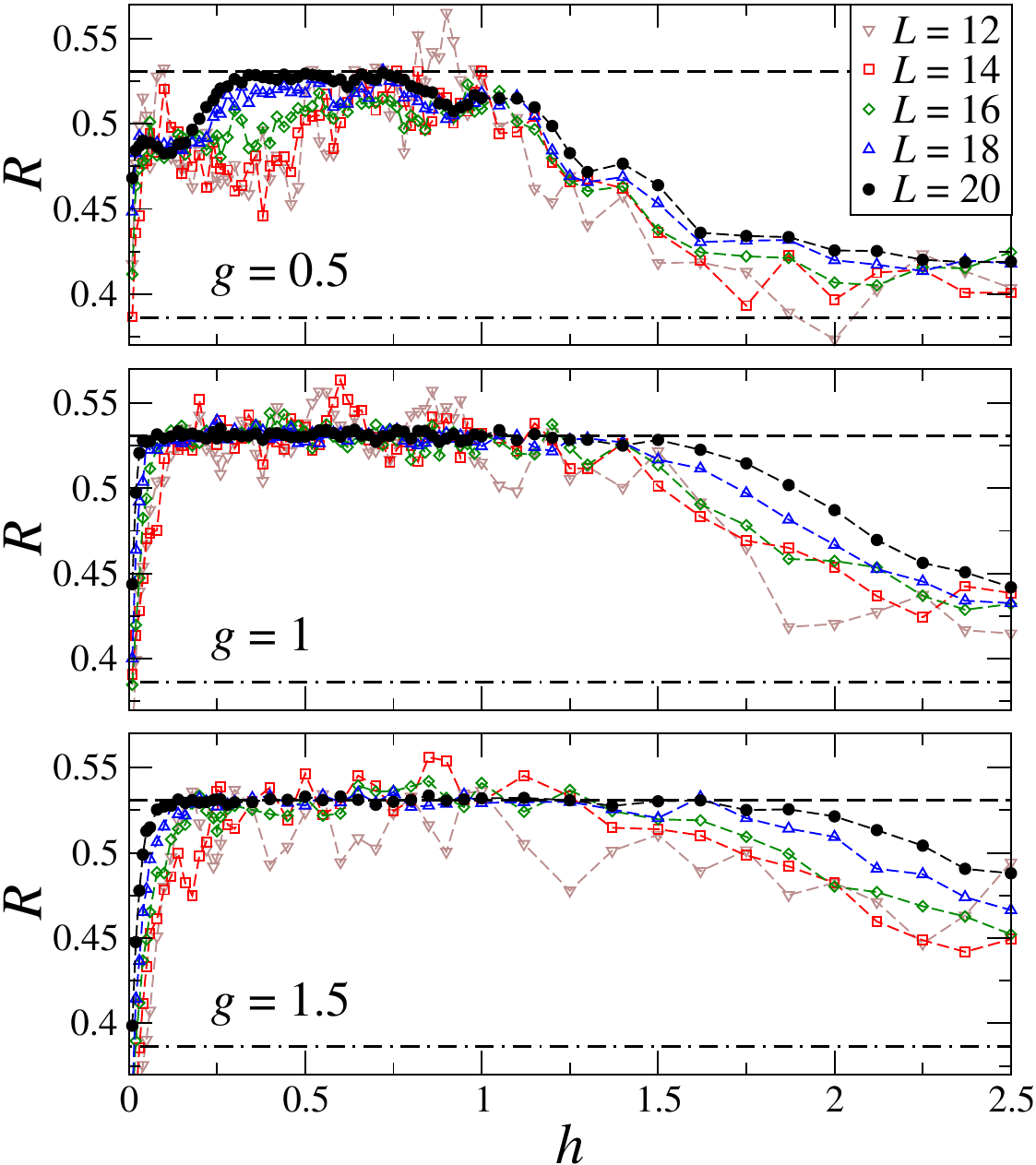}
  \caption{The average value $R$ of the ratios $r_n$,
    cf.~Eqs.~\eqref{Rdef} and~\eqref{sndef}, as a function of the
    longitudinal field $h$, for three fixed values of $g=0.5$ (top),
    $g=1$ (middle), and $g=1.5$ (bottom), for the eigenstates in the
    sector ${\cal H}_{0+}$.  Data are shown for different system
    sizes, as indicated in the legend. Horizontal lines correspond
    to the WD value $R_W=0.5307$ (dashed line) and the Poisson
    value $R_P=0.3863$ (dot-dashed line).}
  \label{figR}
\end{figure}
%%%%%%%%%%%%%%%%%%%%%%%%%%%%%%%%%%%%%%%%%%%%%%%%%%%%%%%%%%%%%%%%%%%%%%%%%%%%%%%%%%

In Fig.~\ref{figR} we show some numerical results for the ratio $R$,
as computed for the Ising chain with $g=0.5$ (top), $1$ (middle), and
$1.5$ (bottom), and several sizes, as a function of the longitudinal
field $h$, for states in the sector ${\cal H}_{0+}$.  The available
data up to $L=20$ provide a good evidence of convergence to the WD
value $R_W$ for large intervals of values of $h$, with oscillations
that tend to attenuate with increasing $L$.  In particular, we observe
an apparent convergence for $0.3\lesssim h \lesssim 1$ at $g=0.5$,
$0.1\lesssim h \lesssim 1.5$ at $g=1$, and $0.1\lesssim h \lesssim 2$
at $g=1.5$.  As expected, deviations are found for small and large
values of $h$, in view of the fact that the system approaches the
integrable regime for $h=0$ and $h\to \infty$.  We also note that,
with increasing $L$, the approach to the WD value $R_W$ is faster for
$g=1$ and $1.5$, rather than for $g=0.5$, which may be explained by
noting that the system tends to become integrable as soon as $g\to 0$.
A reasonable hypothesis that may explain the behavior for small and
large values of $h$ is that the large-$L$ convergence to the chaotic
regime is nonuniform in $h$, so larger and larger sizes are necessary
to observe convergence when approaching the integrability regimes
occurring for $h\to 0$ and $h\to \infty$.

In the following Secs.~\ref{disphase}, \ref{ququcqt},
and~\ref{ququfoqt} discussing hard QQs, we will mostly concentrate on
values of the field $h$ for which the system is chaotic (i.e., $R$ is
close to $R_W$).  In these cases, asymptotic thermalization is
expected to occur, provided that the initial ground state 
is somehow connected with the typical states of the
post-QQ Hamiltonian. Otherwise, different dynamical
regimes may emerge during the post-QQ evolution. When thermalization
takes place, the predictions of the diagonal ensemble should agree
with those of the microcanonical ensemble at the same energy density,
or equivalently, with those of the canonical ensemble for a small
subsystem at the corresponding temperature.

\section{Soft quenches across transitions}
\label{softqu}

We now address the out-of-equilibrium scaling behaviors emerging at
the CQT and at the FOQTs of the quantum Ising model, focusing
on soft quenches, where the longitudinal magnetic field is small
and decreases with increasing chain size.  In this situation,
the Hamiltonian spectrum remains effectively integrable.

\subsection{Across the CQT}
\label{softcqt}

At the CQT ($g=1$), the longitudinal field $h$ drives large-scale
critical modes when it is varied across $h=0$ at
equilibrium~\cite{Sachdev-book,RV-21,CPV-14}.  We apply the
OFSS theory for soft QQ
across CQTs outlined in Refs.~\cite{PRV-18,RV-21}, focusing on
the protocol which starts from the ground state for
$h_i=-h$ and then evolves according to the post-QQ Hamiltonian
$\hat H(h_f=h)$.  We first define the scaling variables
\begin{equation}
  \Phi = h L^{y_h}, \quad \Theta = t L^{-z},\quad
  \Psi = \Phi^{z/y_h} \Theta = h^{z/y_h} t,
  \label{ofsscqt}
\end{equation}
where $y_h=(d+z+2-\eta)/2$ is the renormalization-group dimension of
the longitudinal field at the critical point, and $z$ is the dynamic
exponent controlling the power-law suppression of the gap at the
critical point. For the quantum Ising chain, $y_h = 15/8$ and $z=1$,
since the gap at the critical point behaves as $\Delta(L)\sim 1/L$ (in
the case $h_i\neq -h$, one should add a further dependence on the
ratio $r_h = h_i/h$).  The OFSS regime is defined as the combined
$L\to\infty$, $t\to\infty$, and $h\to 0$ limits, keeping the scaling
variables $\Phi$, $\Theta$, $\Psi$ fixed.  In that case, a generic
observable $A = {\rm Tr} \big[ \hat \rho(t) \, \hat A \big]$ is
expected to behave as
\begin{equation}
  A(t,h,L) \approx L^{-y_a} \tilde{\cal A}(\Phi,\Theta) = h^{y_a/y_h}
  {\cal A}(\Phi,\Psi),
  \label{aofss}
\end{equation}
where $y_a$ is the renormalization-group dimension associated with it,
and ${\cal A}(\Phi,\Psi)= \Phi^{-y_a/y_h} \tilde{\cal A}(\Phi,\Theta)$.
For example, in the case of the magnetization $\hat M$
[cf.~Eq.~\eqref{lomagn}] one has $y_m = d+z-y_h$ (for the
quantum Ising chain, $y_m=1/8$ and $y_m/y_h=1/15$).
The OFSS theory predictions have been confirmed by
numerical results~\cite{PRV-18} and further extended to the
Loschmidt echo, the bipartite entanglement entropy, and the work
fluctuations~\cite{PRV-18,NRV-19-w,RV-21}.

From the above OFSS relations, one can also derive the scaling
behavior in the thermodynamic limit~\cite{RV-21}, i.e., for $L\to
\infty$ while keeping $h$ and $t$ fixed.  This corresponds to taking
$\Phi\to\infty$ in the scaling function ${\cal A}(\Phi,\Psi)$.
Assuming such limit to be regular, one obtains
\begin{equation}
  \lim_{L\to\infty} A(t,h,L) = A_\infty(t,h) \approx h^{y_a/y_h} {\cal
    A}_\infty(\Psi).
  \label{asthlim}
\end{equation}
As anticipated above, this out-of-equilibrium scaling behavior
requires tuning $h$ to keep $\Psi$ fixed in the long-$t$
limit. Therefore, only the low-energy critical spectrum is
substantially involved in the dynamics of such soft QQs.

It is important to note that, when passing from soft to standard QQs,
in which $h$ is kept fixed as $L$ increases, one does not expect to
observe any OFSS controlled by the universality class of the
equilibrium quantum transitions. This is essentially because the
instantaneous change of the Hamiltonian parameters across the CQT
entails an extensive amount of energy exchange, $\Delta E\sim L^d$
(where $d$ is the spatial dimension of the system), which is much
larger then the energy scale $E_c \sim L^{-z}$ of the low-energy
critical modes. Consequently, the post-QQ dynamics generally involves
noncritical higher-energy levels, and the unitary energy-conserving
evolution may not be significantly related to the quantum-critical
features of the low-energy spectrum.

However, as argued in
Refs.~\cite{RV-20-qu,BDD-15,RMD-17,TIGGG-19,HPD-18,HMHPRD-21}, certain
models can still exhibit signatures of the transition.  Hard QQs
have been extensively studied in the integrable quantum Ising and
generalized XY models, under QQs of the transverse field $g$ (keeping
$h=0$) driving the transition that separates the quantum paramagnetic
and ferromagnetic
phases~\cite{Niemeijer-67,BMD-70,RV-20-qu,CEF-12-1,CEF-12-2,BDD-15,EF-16}.
In the thermodynamic limit, the asymptotic long-time limit of the
energy density~\cite{BDD-15}, and the longitudinal and transverse
magnetizations~\cite{RV-20-qu,RV-21}, still develop a singular
dependence on the QQ parameter $g$ around the critical value. However,
these singularities are likely related to the specific spectral
properties of the quantum Ising chain without a longitudinal field
$h$, which can be mapped onto a free-fermion
Hamiltonian~\cite{Sachdev-book}.

\subsection{Across the FOQTs}
\label{softfoqt}

Peculiar scaling behaviors can also be observed at FOQTs (see, e.g.,
Refs.~\cite{RV-21,PV-24,Pfleiderer-05,Rutkevich-08,
  AC-09,YKS-10,JLSZ-10,Coldea-etal-10,Rutkevich-10,
  LMSS-12,CNPV-14,CPV-15,CPV-15b,RV-18,PRV-18-fowb,FB-82,PF-83,
  FP-85,CLB-86,BK-90,LK-91,BK-92,VRSB-93, RTMS-94, NIW-11,
  TB-12,PV-17,PV-17-dyn,SW-18,Fontana-19,PV-25,PRV-25c}).  In
particular, a similar OFSS arises at the transitions driven by the
longitudinal field $h$, for $g<g_c$~\cite{PRV-18,RV-21,PV-24}.  The
main difference, compared to CQTs, lies in the size dependence of the
scaling variables, which may also depend on the choice of boundary
conditions. For the sake of clarity, here we focus on periodic
boundary conditions, which do not favor any of the two magnetized
phases.  The slowest time scale $\tau(L)$ is provided by the inverse
energy difference $\Delta(L)$ of the lowest states at $h=0$, which is
exponentially suppressed in the large-$L$ limit, as~\cite{Pfeuty-70}
\begin{equation}
  \Delta(L)\approx 2\sqrt{1-g^2\over \pi L} \, g^L.
  \label{DeltaLFOQT}
\end{equation}
Following Refs.~\cite{PRV-18,RV-21,PV-24}, we define the scaling
variables
\begin{equation}
  \Phi = {2 m_0 h L \over \Delta(L)},\qquad \Theta = \Delta(L)\,t,
  \label{phithetafoqt}
\end{equation}
where $m_0$ is the spontaneous magnetization defined in
Eq.~(\ref{sigmasingexp}).  Again, the OFSS limit is defined by the
combined limit $L\to\infty$, $t\to\infty$, and $h\to 0$, keeping
$\Phi$ and $\Theta$ fixed. In such limit, the magnetization behaves as
\begin{equation}
  M(t,h,L) \approx m_0 \, {\cal M}(\Phi,\Theta),
  \label{mcheckfoqt}
\end{equation}
${\cal M}(\Phi,\Theta)$ being a scaling function.  The OFSS equations
hold both for $h_i=-h$ and for $h_i=0$ (within OFSS one cannot
distinguish between $h_i = 0$ and $h_i = 0^-$, as in the thermodynamic
limit).

This OFSS limit focuses on a small (exponentially
suppressed) range of values of $h$, i.e., $|h|\sim L^{-1} \Delta(L)$,
where the relevant spectrum is essentially restricted to the two
globally magnetized states.  Indeed, the function ${\cal
  M}(\Theta,\Phi)$ can be analytically computed using an effective
two-level model~\cite{CNPV-14,PRV-18}. For QQs from $h_i=-h$ to $h$
(corresponding to QQs from $-\Phi$ to $\Phi$, in terms of rescaled
quantities) one has
\begin{subequations}
  \begin{equation}
  {\cal M}(\Theta,\Phi) \! = \! - {\Phi\over \sqrt{1 + \Phi^2}}
  + {2\Phi \over (1+\Phi^2)^{3/2}}
  \Bigl[ \! 1 - \cos( \Theta \sqrt{1+\Phi^2} ) \! \Bigr] ,
  \label{m2lscahih}
\end{equation}
while, for QQs from $h_i=0$ to $h$, we obtain
\begin{equation}
  {\cal M}(\Theta,\Phi) =   
  { \Phi \over (1+\Phi^2)} \Bigl[1-\cos(\Theta \sqrt{1+\Phi^2})\Bigl] .
  \label{m2lscahi0}
\end{equation}
\end{subequations}
Numerical results for the Ising chain with PBC~\cite{PRV-18} confirm
the validity of the scaling Ansatz~\eqref{mcheckfoqt} and the above
reported scaling functions obtained by the effective two-level model.
The asymptotic OFSS behavior is approached quite rapidly, as size
corrections turn out to decrease exponentially in $L$.

In the thermodynamic limit $L\to\infty$, while
keeping an appropriate combination of $h$ and $t$ fixed, as the
scaling variable $\Psi$ in Eq.~\eqref{ofsscqt}, the OFSS functions
become trivial, unlike the CQT case discussed in Sec.~\ref{softcqt}.
An analogous situation was recently pointed out in the context of the
out-of-equilibrium dynamics induced by slow Kibble-Zurek protocols
across FOQTs~\cite{PRV-25,PRV-25b}, where the thermodynamic limit of
the corresponding OFSS is trivial and fails to capture the different
mechanisms that are relevant when the system size tends to infinity,
keeping the other parameters fixed.  Higher-energy states, such as
ensembles of noninteracting kink-antikink excitations, are expected to
become relevant in this limit, analogously to the case of Kibble-Zurek
protocols. However, we stress that noninteracting ensembles of
kink-antikink states, such as those considered in
Refs.~\cite{MW-78, Rutkevich-99, Rutkevich-08, Coldea-etal-10,
  Rutkevich-10, KCTC-17, LSKC-21, PRV-25, PRV-25b, MHV-25, JRCB-25},
are not expected to generate a chaotic regime, as occurs for the post-QQ
Hamiltonian with $h>0$.  Therefore, simplified models restricted to
kink-antikink states cannot describe the whole post-QQ evolution toward
a long-time stationary state, though they may be relevant for sufficiently
small longitudinal fields
($h \ll g < g_c$)~\cite{KCTC-17, LSKC-21, PRV-25, PRV-25b, MHV-25, JRCB-25}.

For periodic boundary conditions the OFSS regime is rather trivial, as
only two states are relevant. More complex OFSS behaviors arise with
different boundary conditions. For instance, with oppositely fixed
boundary conditions, kink states become the relevant low-energy excitations.
In this case, OFSS behaviors still emerge~\cite{PRV-20,RV-21}, but their size
dependence is characterized by a power-law scaling $h \sim L^{-3}$,
reflecting the power-law suppression $\Delta\sim L^{-2}$ of the gap at
the transition point, unlike the exponential suppression for periodic
boundary conditions.  A further finite-size regime for which
$h\sim L^{-1}$ is discussed in the Appendix.

\section{Quantum quenches in the disordered phase}
\label{disphase}

We now turn to hard QQs. To avoid complications related to the
presence of phase transitions, we first focus on the post-QQ dynamics
in the disordered phase ($g>1$).  Specifically, we present results for
$g=1.5$, which is far from the CQT and the FOQT line. We consider a QQ
protocol which starts from the ground state for $h_i=-h$, and then
evolves with the Hamiltonian parameter $h_f=+h$ for $t>0$ (the
alternative protocol starting from $h_i=0$ is not particularly
interesting for $g>g_c$ and is therefore not considered).  The time
evolution of observables like $M$ and $K$ is numerically computed up
to $L=28$, whereas the diagonal-ensemble observables $M_D$ and $K_D$
are only obtained up to $L=20$, since their determination requires the
computation of the full spectrum of $\hat H(h)$.
We primarily focus on chains with intermediate values of $h$ (e.g.,
$h\approx 1$), since their spectrum is in the chaotic regime---for
$g=1.5$ and $L=20$ a chaotic spectrum is observed for values of $h$ in
the interval $0.1 \lesssim h \lesssim 2$ (see the lower panel in
Fig.~\ref{figR}).

\subsection{The diagonal ensemble}
\label{diagensdis}

%%%%%%%%%%%%%%%%%%%%%%%%%%%%%%%%%%%%%%%%%%%%%%%%%%%%%%%%%%%%%%%%%%%%%%%%%%%%%%%%%%
\begin{figure}[!t]
  \includegraphics[width=0.47\textwidth]{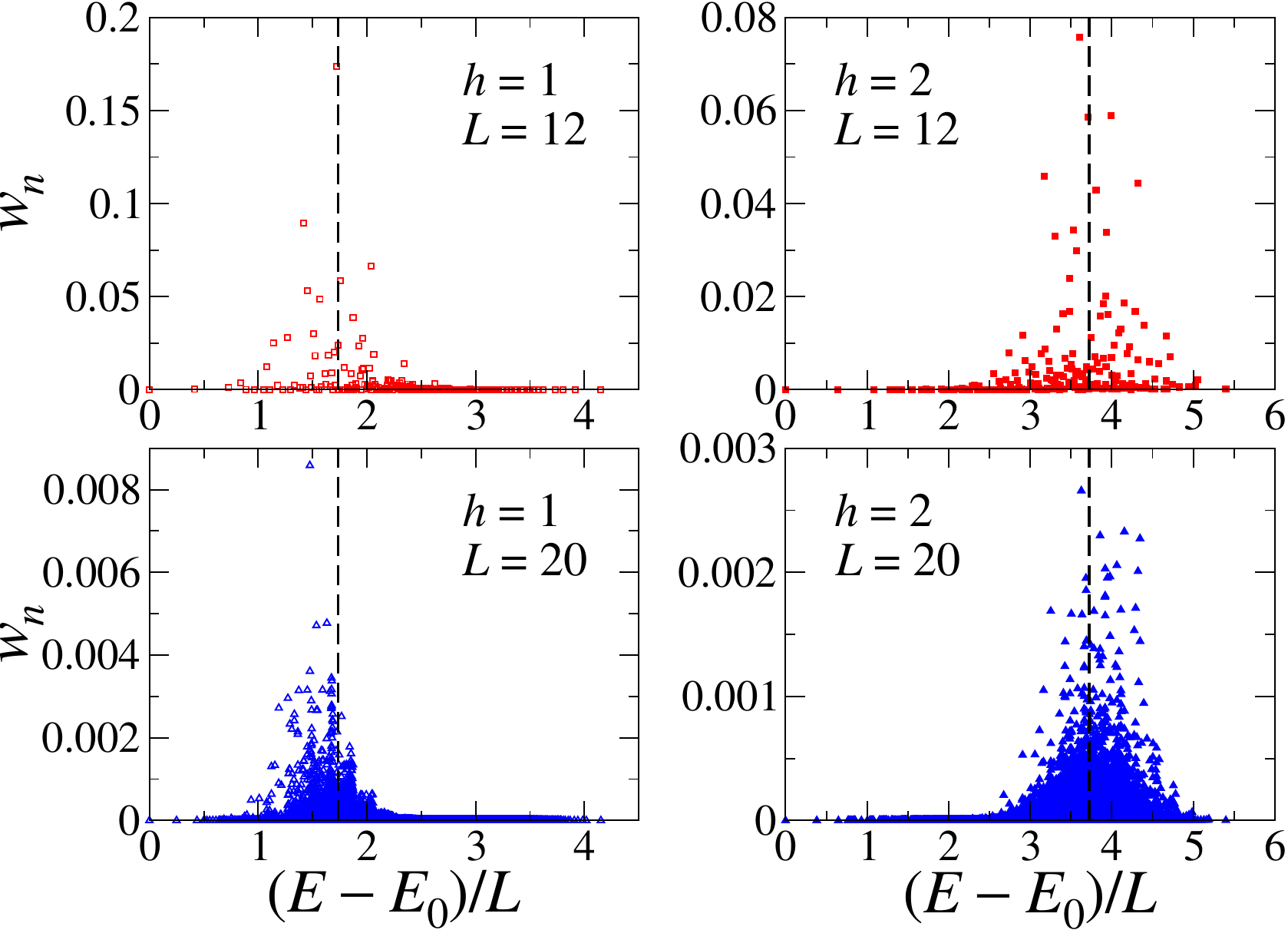}
    \caption{The overlap $w_n=|\langle \Phi_n(h)|\Phi_0(-h)\rangle|^2$
      between the ground state $\Phi_0(-h)\rangle$ of the Hamiltonian
      $\hat H(-h)$ and the eigenstates $|\Phi_n(h)\rangle$ with
      eigenvalues $E_n$ of the post-QQ Hamiltonian $\hat H(h)$ vs the
      energy density $e_n=(E_n-E_0)/L$.  Data are shown for $g=1.5$ and set $h=1$
      (left panels) or $h=2$ (right panels), and for system sizes $L=12$ (upper
      panels) and $L=20$ (lower panels). The vertical lines indicate
      the average energy density $e = \sum_n w_n e_n \approx 1.735047$ (left)
      and $e \approx 3.72211379$ (right).}
  \label{overlapg1p5h12}
\end{figure}
%%%%%%%%%%%%%%%%%%%%%%%%%%%%%%%%%%%%%%%%%%%%%%%%%%%%%%%%%%%%%%%%%%%%%%%%%%%%%%%%%%

To begin with, in Fig.~\ref{overlapg1p5h12} we plot the overlaps
$w_n=|\langle \Phi_n(h)|\Psi_0(-h)\rangle|^2$ of the initial ground
state $|\Phi_0(-h)\rangle$ with the eigenstates $|\Phi_n(h)\rangle$ in
the relevant subsector ${\cal H}_{0+}$ of the Hilbert space of the
post-QQ Hamiltonian $\hat H(h)$ for $h=1$ (left) and $h=2$ (right),
for two different sizes: $L=12$ (top) and $L=20$ (bottom).  Our data
show that the distribution of $w_n$ moves towards larger energies when
increasing $h$ (in Fig.~\ref{overlapg1p5h12} the corresponding mean
energy density is indicated by a vertical dashed line), as expected.
Moreover, as $L$ increases, the distribution becomes larger:
an increasing number of energy levels become relevant---and less
peaked---on average the $w_n$ values decrease.

%%%%%%%%%%%%%%%%%%%%%%%%%%%%%%%%%%%%%%%%%%%%%%%%%%%%%%%%%%%%%%%%%%%%%%%%%%%%%%%%%%
\begin{figure}[!t]
  \includegraphics[width=0.47\textwidth]{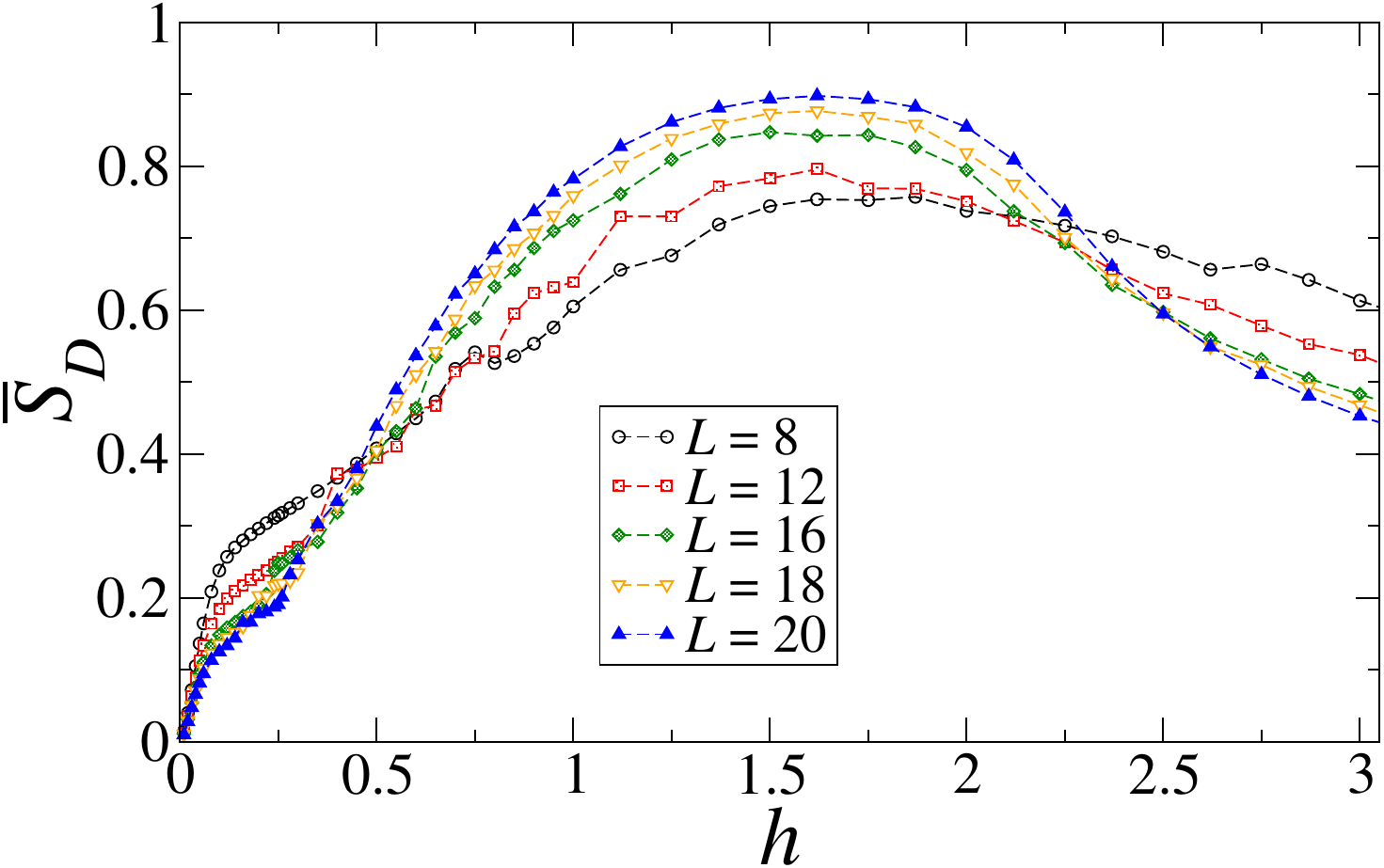}
  \caption{The ratio $\bar S_D(L)$ between the diagonal-ensemble
    entropy $S_D(L)$ and its maximum value $\ln {\cal D}_{0+} \sim L$,
    as a function of the longitudinal field $h$, for different system sizes
    $L$.  Here we set $g=1.5$.}
  \label{dentrog1p5}
\end{figure}
%%%%%%%%%%%%%%%%%%%%%%%%%%%%%%%%%%%%%%%%%%%%%%%%%%%%%%%%%%%%%%%%%%%%%%%%%%%%%%%%%%

To better quantify the delocalization of the values of $w_n$, we
consider the diagonal-ensemble entropy defined in Eq.~\eqref{sddef}.
In particular, we study the ratio
\begin{equation}
  {\bar S}_D(L) \equiv {S_D(L)\over \ln {\cal D}_{0+}(L)},
  \qquad (0\le {\bar S}_D \le 1),
  \label{sdresc}
\end{equation}
where we recall that $\ln {\cal D}_{0+} \sim L$ in the large-$L$
limit.  Note that a value ${\bar S}_D\approx 1$ implies that the
initial state is an approximately homogeneous superposition of the
post-QQ Hamiltonian eigenstates in the Hilbert subspace ${\cal
  H}_{0+}$, while ${\bar S}_D\approx 0$ is obtained when only a finite
number of states is relevant for large values of $L$.
Figure~\ref{dentrog1p5} shows results for several values of $h$
and chain sizes.  The data up to $L=20$ appear to approach the
expected value ${\bar S}_D= 1$ as $L$ increases, at least for
$1 \lesssim h \lesssim 2$, although it is not straightforward to identify
the functional dependence with the size.  In contrast, for
$h \approx 0.5$, still belonging to the chaotic region, the values of
${\bar S}_D$ are significantly smaller.  Nevertheless, convergence to
${\bar S}_D\to 1$ might still be possible for larger $L$.  Indeed, the
approach to the large-$L$ limit should not be uniform as $h$
decreases, potentially giving rise to nontrivial crossover effects for
small $h$. Finally, note that the dependence of ${\bar S}_D$ on $L$ is
not monotonic with increasing $h$, likely because the system moves
toward the integrability region.

%%%%%%%%%%%%%%%%%%%%%%%%%%%%%%%%%%%%%%%%%%%%%%%%%%%%%%%%%%%%%%%%%%%%%%%%%%%%%%%%%%
\begin{figure}[!t]
    \includegraphics[width=0.45\textwidth]{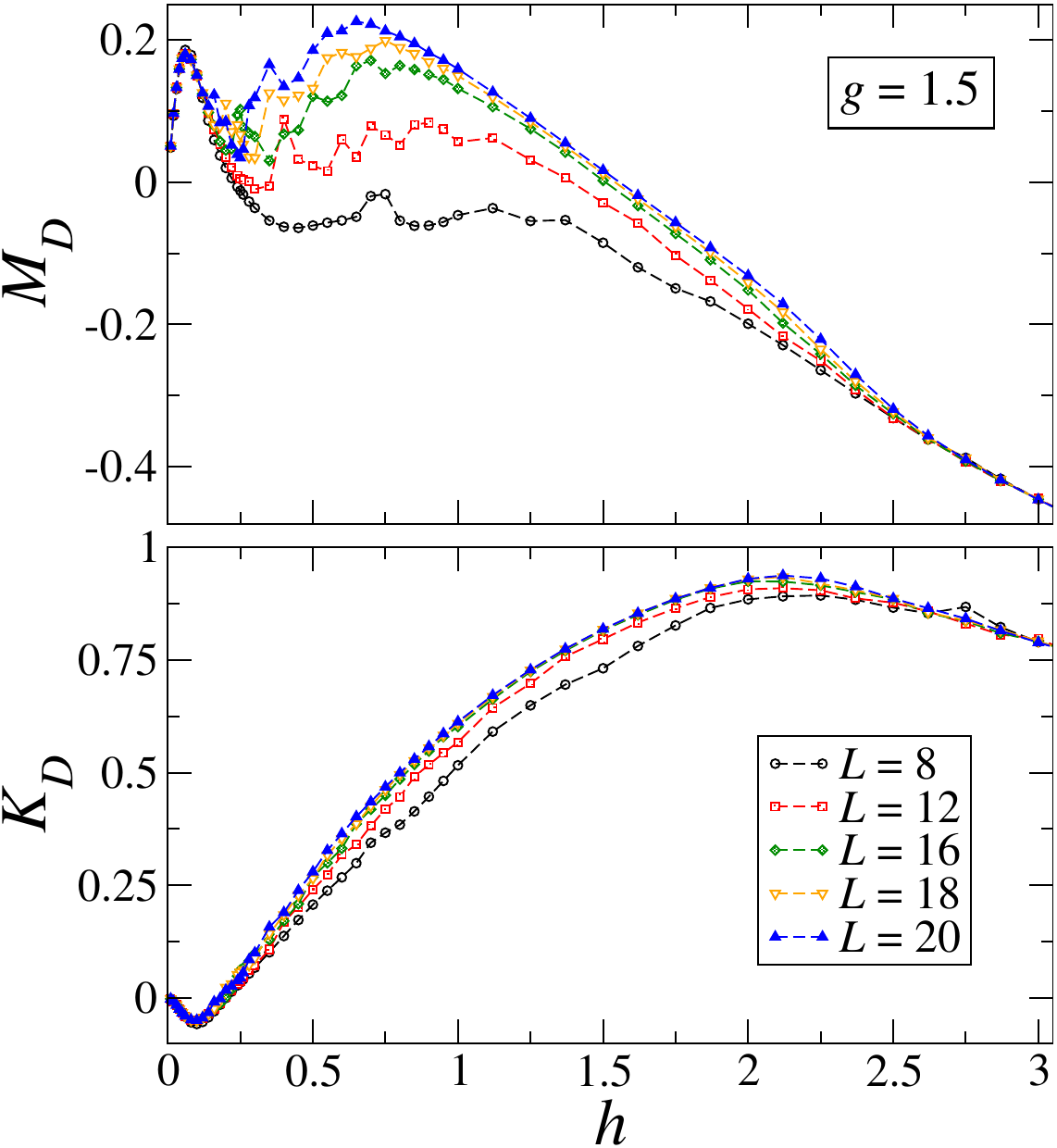}
    \caption{The magnetization $M_D$ (top) and excess-bond energy
      $K_D$ (bottom) in the diagonal ensemble for $g=1.5$ (disordered
      phase), for a QQ from $-h$ to $h$, and for different system sizes.}
  \label{edg1p5}
\end{figure}
%%%%%%%%%%%%%%%%%%%%%%%%%%%%%%%%%%%%%%%%%%%%%%%%%%%%%%%%%%%%%%%%%%%%%%%%%%%%%%%%%%

As already discussed, the diagonal ensemble provides predictions for
the asymptotic behavior of the magnetization and of the excess bond
energy, which should converge to their diagonal-ensemble values $M_D$
and $K_D$, defined in Eqs.~(\ref{mxmD}) and (\ref{boexdefD}).  Data
for $M_D$ and $K_D$ vs $h$ are reported in Fig.~\ref{edg1p5} (top and
bottom panels, respectively), for different system sizes.  They appear
to converge with increasing $L$, thus providing information on their
thermodynamic limit.  The convergence seems to be much faster for
$K_D$, while size effects are larger for $M_D$, in particular around
$h=0.5$.
It is however worth pointing out that the behavior of $M_D$ and $K_D$
in not monotonic with increasing $h$. In particular, the magnetization
decreases for $h \gtrsim 1$. This is related to the properties of the
corresponding overlap $w_n$ between the initial state and the
eigenstates of the post-QQ Hamiltonian, as their distributions move
toward higher energies when increasing $h$, thus exhibiting an
inversion of population (see Fig.~\ref{overlapg1p5h12}).

\subsection{The post-quench evolution}
\label{postqqdis}

%%%%%%%%%%%%%%%%%%%%%%%%%%%%%%%%%%%%%%%%%%%%%%%%%%%%%%%%%%%%%%%%%%%%%%%%%%%%%%%%%%
\begin{figure}[!t]
    \includegraphics[width=0.49\textwidth]{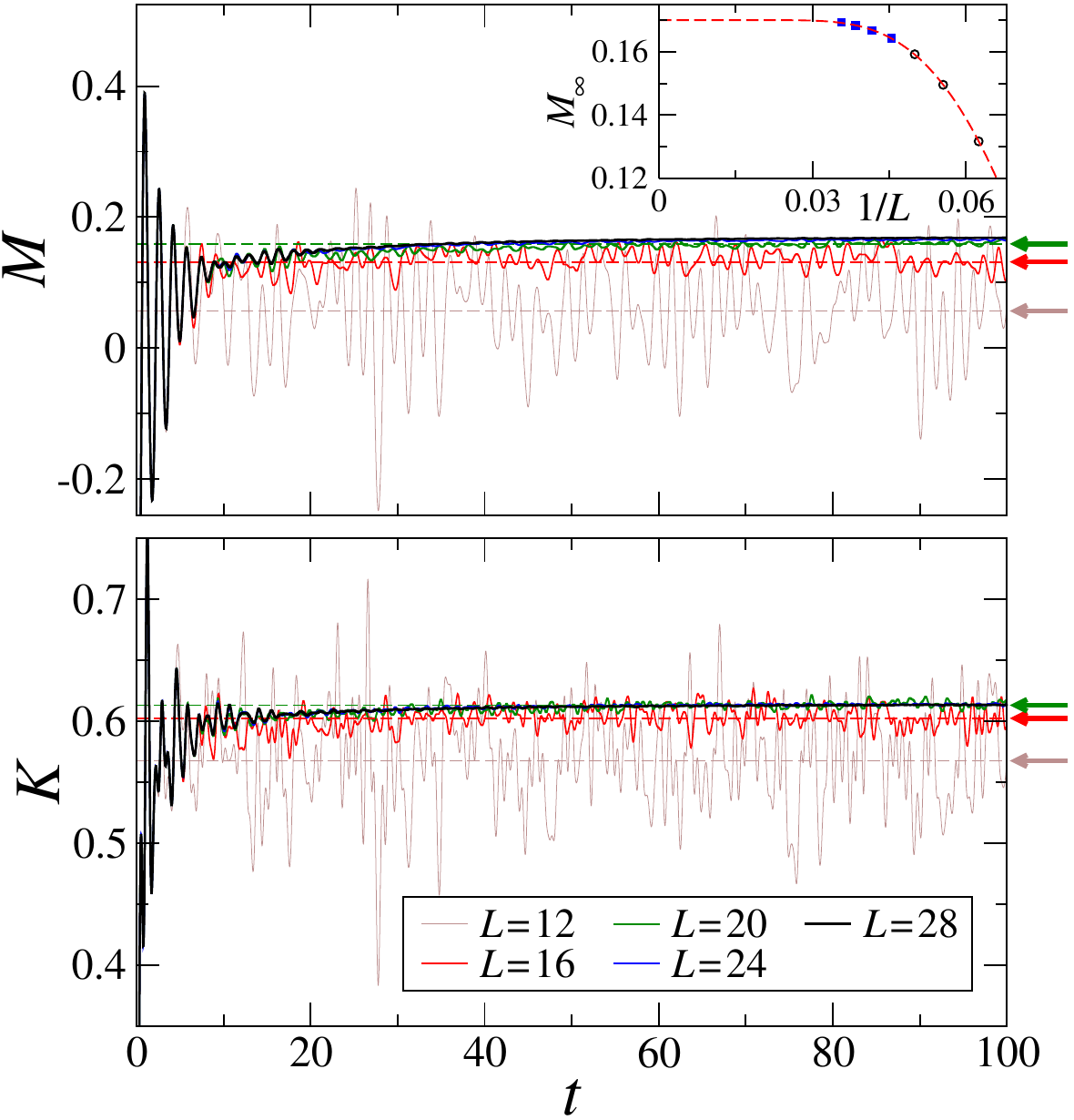}
    \caption{Time behavior after a QQ from $h_i=-1$ to $h=1$ for the
      magnetization (top) and the excess bond energy (bottom), at
      fixed $g=1.5$, and various sizes up to $L=28$ (see legend).
      Data up to $L=20$ are compared against the asymptotic prediction
      of the diagonal ensemble (dashed horizontal lines and arrows
      with the corresponding color code).  The inset shows the
      asymptotic long-time average $M_\infty(L)$ of the magnetization
      versus $1/L$, obtained either in the diagonal-ensemble (empty
      black circles) or by averaging $M(t)$ in the interval $t \in
      [150, 200]$ (filled blue squares); the dashed red line is a fit
      of the numerical data for $L \geq 16$, to $M_\infty(L) =
      M_\infty + a e^{-b L}$, which gives $M_\infty \approx 0.170$.}
  \label{quevg1p5}
\end{figure}
%%%%%%%%%%%%%%%%%%%%%%%%%%%%%%%%%%%%%%%%%%%%%%%%%%%%%%%%%%%%%%%%%%%%%%%%%%%%%%%%%%

We now discuss the post-QQ evolution of the magnetization $M(t)$ (top)
and the excess bond energy $K(t)$ (bottom). Specifically, in
Fig.~\ref{quevg1p5} we show results for $h=1$, and several sizes
$L\le 28$, up to relatively long times $t = O(10^2)$.
Convergence has been ensured by setting a Runge-Kutta time step of
$dt = 2.5 \times 10^{-3}$ and carefully verifying the stability of the
results. With increasing $L$, the curves appear to converge to a
large-$L$ limit, which we interpret as the post-QQ evolution in the
thermodynamic limit (see also footnote~\ref{note:TEBD}).
Moreover, the approach to the long-time
stationary values occurs with fluctuations that decrease with
increasing $L$.  More quantitatively, the inset highlights the
dependence on $L$ of the asymptotic long-time average $M_\infty(L)$ of
the magnetization, clearly converging to a finite value. To
ensure reliable estimates of $M_\infty(L)$ up to $L=28$, we performed
time averages over the time interval $t \in [150, 200]$ (data not
shown). The data shown in the inset of Fig.~\ref{quevg1p5} suggest an
exponential approach to the large-$L$ limit, as $M_\infty(L) =
M_\infty + a e^{-b L}$ (dashed red line is a fit to the numerical data
with this formula).

We also compare the time dependence of $M(t)$ and $K(t)$ with the
corresponding diagonal-ensemble estimates $M_D$ and $K_D$ [they are
  defined in Eqs.~\eqref{mxmD} and~\eqref{boexdefD}], which are
expected to provide the asymptotic large-time value (averaged over a
sufficiently long time interval).  Our data are definitely compatible
with these predictions (dashed lines in Fig.~\ref{quevg1p5}), up to
lattice sizes $L=20$: For sufficiently long times, $M_D$ and $K_D$
provide an excellent approximation of the time average of $M(t)$ and
$K(t)$ for any $L$. Moreover, for the largest available values of $L$,
the need of averaging over time in order to match the diagonal-ensemble
predictions becomes less important, as fluctuations decrease.

\subsection{Thermalization}
\label{sec:termal_disord}

The asymptotic stationary states, developed by the post-QQ unitary
dynamics in the long-time and thermodynamic large-$L$ limit
exhibit several noteworthy properties. According to the thermalization
scenario, the system should approach the statistical
properties of a microcanonical ensemble at fixed energy density
$e = \sum_n w_n e_n$. Equivalently, any sufficiently large subsystem
should approach the canonical statistical behavior at a temperature
corresponding to the given energy density, as discussed at the end of
Sec.~\ref{remarks}.

\begin{figure}[!t]
  \includegraphics[width=0.47\textwidth]{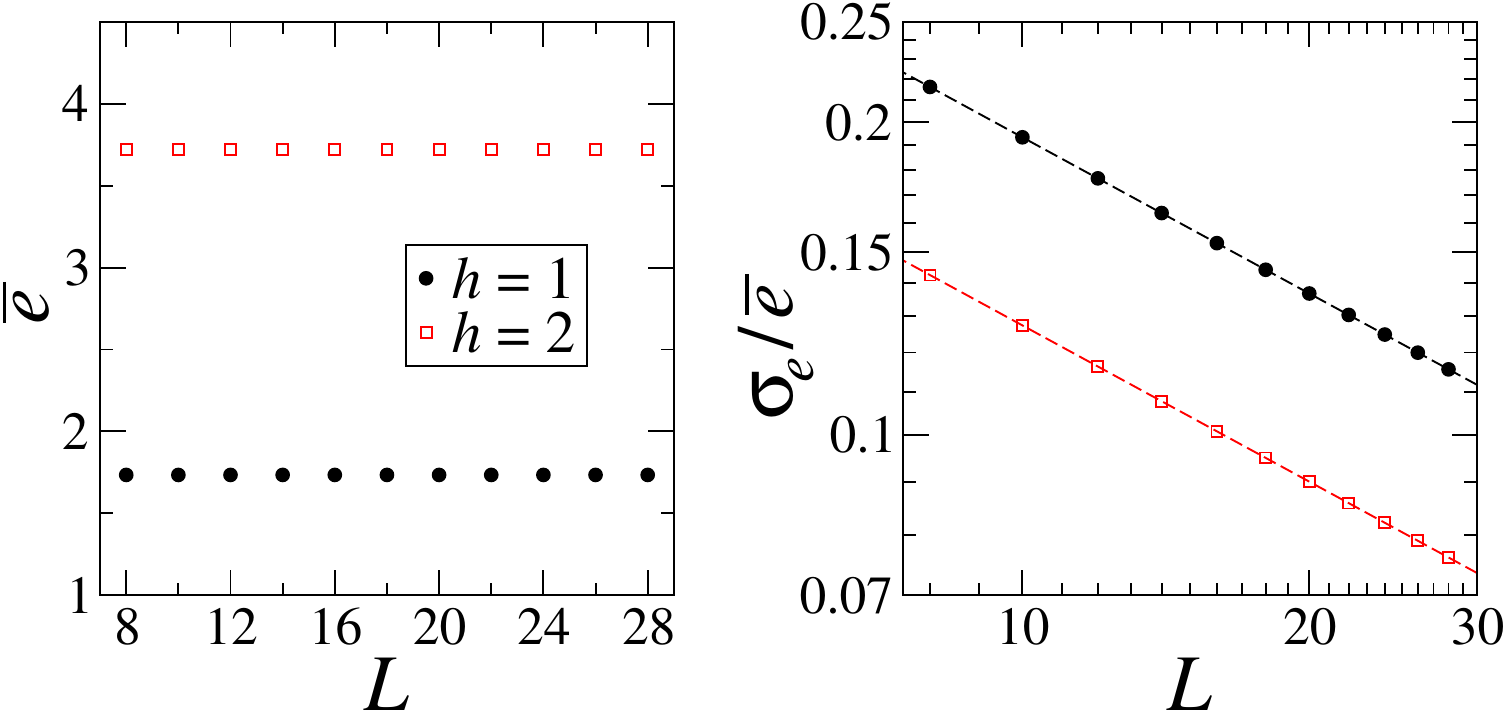}
  \caption{The post-QQ mean energy per site $\bar{e}$ (left) and the
    normalized standard deviation $\sigma_e/\bar{e}$ (right) as a
    function of $L$, for two values of $h$ and for fixed $g=1.5$. In
    the left panel a linear scale is used, while in the right one data
    are plotted using a logarithmic scale on both axes. }
  \label{energyvarfig}
\end{figure}

Here we present results of some of our simulations for fixed
$g=1.5$, considering longitudinal fields in the chaotic window $0.1
\lesssim h \lesssim 2$ identified for $L \approx 20$ (see the lower
panel of Fig.~\ref{figR}).  Figure~\ref{energyvarfig} (left) displays
the energy-density average $e$ defined in Eq.~\eqref{meansigmaE} as a
function of the chain length $L$, for two values of $h$. As is clearly
visible, $e$ varies very little with increasing $L$. We obtain
$\bar{e}=1.73504739$ for $h=1$ and $L \ge 12$ and $\bar{e}=3.72211379$
for $h=2$ and $L\ge 8$.  On the other hand, the ratio of the standard
deviation $\sigma_e$ defined in Eq.~\eqref{meansigmaE} and of the
average $\bar{e}$ decreases as $1/\sqrt{L}$ as expected~\cite{DKPR-16}
[see Fig.~\ref{energyvarfig} (right)].  This confirms that, in the
thermodynamic limit, the system approaches a microcanonical window.

%%%%%%%%%%%%%%%%%%%%%%%%%%%%%%%%%%%%%%%%%%%%%%%%%%%%%%%%%%%%%%%%%%%%%%%%%%%%%%%%%%
%%%%%%%%%%%%%%%%%%%%%%%%%%%%%%%%%%%%%%%%%%%%%%%%%%%%%%%%%%%%%%%%%%%%%%%%%%%%%%%%%%

\begin{figure}[!t]
  \includegraphics[width=0.47\textwidth]{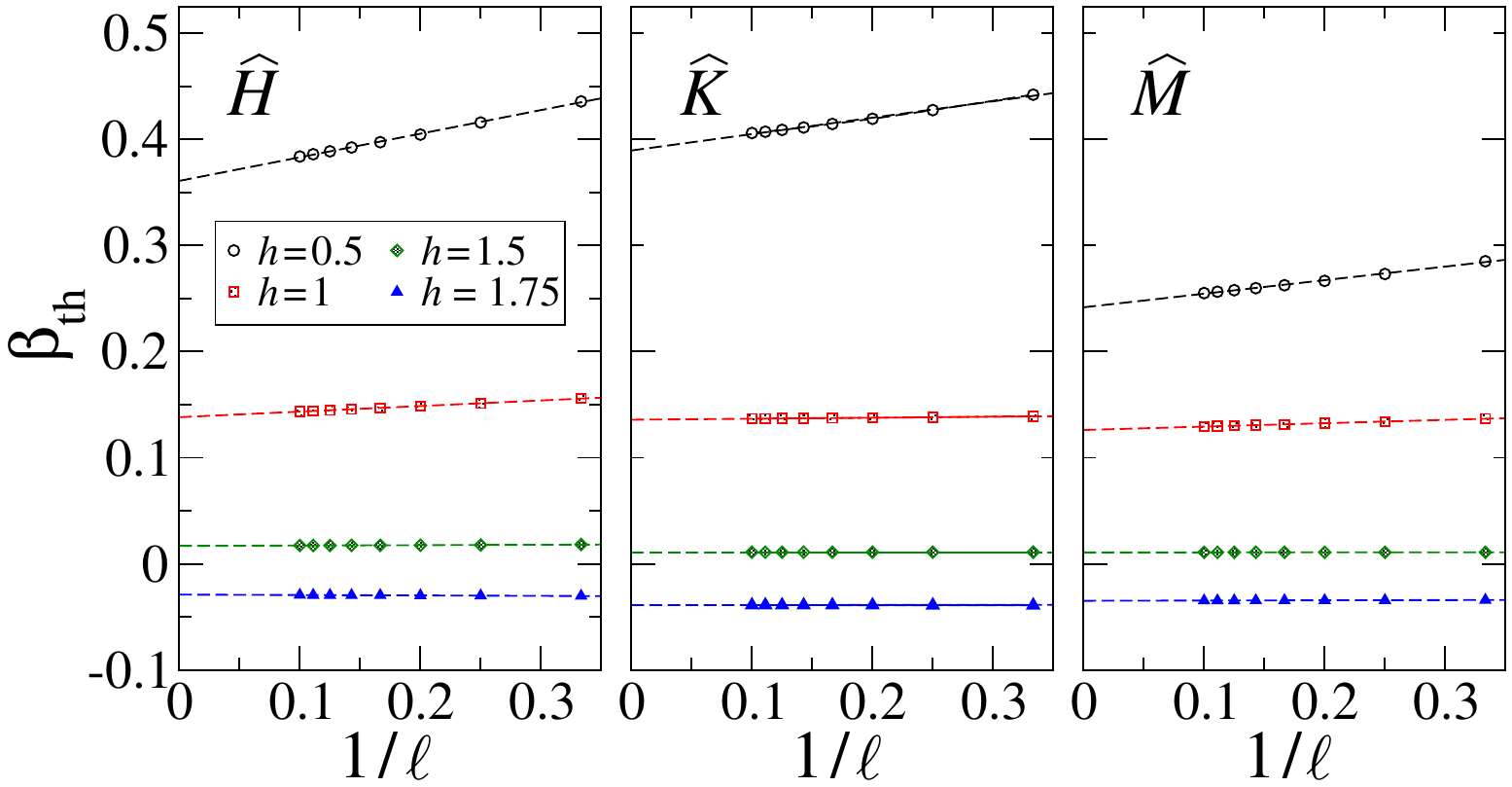}
  \caption{The estimated effective inverse temperature $\beta_{\rm eff}$
    assuming thermalization, using the energy,
    Eq.~\eqref{tthdef}, (left), $K_D$ (middle), and $M_D$ (right).
    Data are plotted against the inverse of the subsystem size $\ell$,
    for $\ell \in [3,10]$ and for a fixed system size $L=20$.  Here we
    fix $g=1.5$, while the different sets are for various values of
    the longitudinal field $h$ in the chaotic window (see legend).
    Straight dashed lines are linear fits in $1/\ell$ of the numerical
    data (symbols) to Eq.~\eqref{eq:fit_beta}. }
  \label{energyvarfig2}
\end{figure}

We have also computed the effective inverse temperature $\beta_{\rm th}$
for subsystems of size $\ell \in [3,10]$, using
Eq.~\eqref{tthdef}. Since the energy density is very stable, also
$\beta_{\rm th}$ turns out to be essentially independent of $L$, with
only a small residual dependence on $\ell$, which can be parametrized
as
\begin{equation}
  \beta_{\rm th}(\ell) = \beta_{\rm th} + b/\ell ,
  \label{eq:fit_beta}
\end{equation}
as is visible from Fig.~\ref{energyvarfig2} (left).  In fact, we
recall that typical corrections of systems with boundaries are
$O(\ell^{-1})$ (see, e.g., Ref.~\cite{CPV-15b}). By fitting our
numerical data for the largest available system size, $L=20$, we
obtain the following estimates in the large-$\ell$ limit:
$\beta_{\rm th} \approx 0.364, \, 0.139, \, 0.017, \, -0.029$,
respectively for $h=0.5, \, 1, \, 1.5, \, 1.75$ (cf.~dashed lines
in left panel of Fig.~\ref{energyvarfig2}).

%%%%%%%%%%%%%%%%%%%%%%%%%%%%%%%%%%%%%%%%%%%%%%%%%%%%%%%%%%%%%%%%%%%%%%%%%%%%%%%%%%
%%%%%%%%%%%%%%%%%%%%%%%%%%%%%%%%%%%%%%%%%%%%%%%%%%%%%%%%%%%%%%%%%%%%%%%%%%%%%%%%%%

Substantially consistent results are obtained by using the
diagonal-ensemble values $K_D$ and $M_D$ for the excess bond energy
and the magnetization (central and right panels of
Fig.~\ref{energyvarfig2}, respectively).  However, these results
appear to be less stable with increasing $L$, in particular at
$h=0.5$, presumably because the estimates $K_D$ and $M_D$ obtained for
chains of size $L\le 20$ do not show a complete convergence with
respect to $L$ (see Fig.~\ref{edg1p5}). Namely, using the estimates of
$K_D$ for $L=20$, we obtain $\beta_{\rm th}\approx 0.394, \, 0.136, \,
0.011, \, -0.039$, for $h = 0.5, \, 1, \, 1.5, \, 1.75$ (we perform a
linear extrapolation of the finite-$\ell$ data using
Eq.~\eqref{eq:fit_beta}, as in the energy-density analysis). An
analogous computation using $M_D$ gives $\beta_{\rm th} \approx 0.244,
\, 0.127, \, 0.011, \, -0.035$, for the same values of $h$. An
apparent larger discrepancy arises for $M_D$ at $h=0.5$, but this may
be attributed to the pronounced finite-size corrections of $M_D$ for
this value of $h$ (see Fig.~\ref{edg1p5}, top panel). Nevertheless, we
emphasize that the signs of $\beta_{\rm th}$ inferred from the various
observables are consistent: for $h \lesssim 1.5$, we always find
positive effective temperatures ($\beta_{\rm th} >0$), while the data
for $h \gtrsim 1.75$ are compatible with an effective negative
temperature ($\beta_{\rm th} <0$), indicating an inverted spectral
population. The physical meaning of negative temperatures and the
nature of the corresponding asymptotic stationary states has been
discussed in Ref.~\cite{BILV-21}.

The comparison of the effective inverse temperatures obtained from
different local observables provides a robust evidence of
thermalization, at least in the interval $0.5\lesssim h \lesssim 2$,
for system sizes $L\approx 20$. This interval is slightly
smaller than the range $0.1\lesssim h \lesssim 2$, in which the
spectrum of $\hat{H}(g=1.5,h)$ is clearly chaotic for $L\approx 20$,
see the lower panel in Fig.~\ref{figR}.  Indeed, for $h=0.25$, fixing
$L=20$ and $\ell=10$, the same analysis gives inconsistent results for
the effective inverse temperature, namely
$\beta_{\rm th} \approx 0.687, \, 1.773, \, 0.113$,
determined from the energy density, $K_D$, and $M_D$, respectively.
We attribute this behavior to the limited system sizes accessible in our study.

\section{Quantum quenches across the CQT}
\label{ququcqt}

In this section, we study the QQ dynamics driven by a longitudinal
field $h$ at $g=1$ (i.e., at the continuous Ising transition),
focusing on QQs from $h_i=-h$ to $h_f=h$. We have not considered QQs
starting from the ground state at $h_i=0$, as this
would introduce strong finite-size effects, due to the divergence
of the correlation length at the critical point ($g=1$, $h=0$).
As we shall see, the out-of-equilibrium post-QQ dynamical
behavior of the chain is qualitatively similar to that observed in the
disordered phase for $g>1$, discussed in Sec.~\ref{disphase}.

\subsection{The diagonal ensemble}
\label{diagenscqt}

%%%%%%%%%%%%%%%%%%%%%%%%%%%%%%%%%%%%%%%%%%%%%%%%%%%%%%%%%%%%%%%%%%%%%%%%%%%%%%%%%%
\begin{figure}[!t]
  \includegraphics[width=0.47\textwidth]{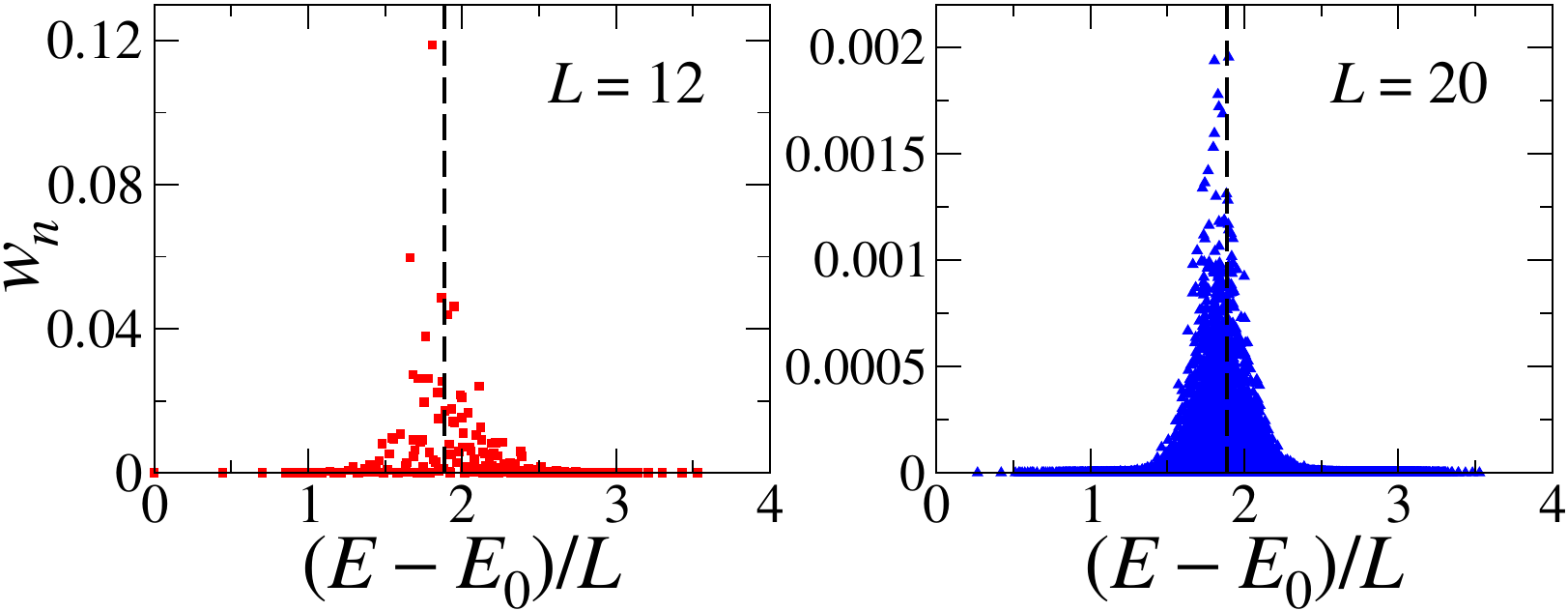}
  \caption{The overlap among the initial ground state of the
    Hamiltonian $\hat H(-h)$ and the eigenstates of the post-QQ
    Hamiltonian $\hat H(h)$ vs the energy density $e_n=(E_n-E_0)/L$,
    for $g=1$ and $h=1$.  The left and right panel report data for
    $L=12$ and $L=20$, respectively. The vertical lines correspond to
    the energy-density average $\bar{e} \approx 1.8849419$.}
  \label{overlapg1p0h1}
\end{figure}
%%%%%%%%%%%%%%%%%%%%%%%%%%%%%%%%%%%%%%%%%%%%%%%%%%%%%%%%%%%%%%%%%%%%%%%%%%%%%%%%%%

Figure~\ref{overlapg1p0h1} shows the distribution of the overlap
$w_n=|\langle \Phi_n(h)|\Phi_0(h_i)\rangle|^2$ for $h=-h_i=1$, in the
${\cal H}_{0+}$ subsector of the Hilbert space, for $L=12$ (left) and
$L=20$ (right). As observed in Fig.~\ref{overlapg1p5h12} for $g=1.5$,
the distribution approaches a Gaussian shape as $L$
increases. Moreover, in the same limit, the distribution becomes wider
and its maximum decreases.  In contrast, the
energy-density average is essentially independent of $L$.

%%%%%%%%%%%%%%%%%%%%%%%%%%%%%%%%%%%%%%%%%%%%%%%%%%%%%%%%%%%%%%%%%%%%%%%%%%%%%%%%%%
\begin{figure}[!t]
    \includegraphics[width=0.47\textwidth]{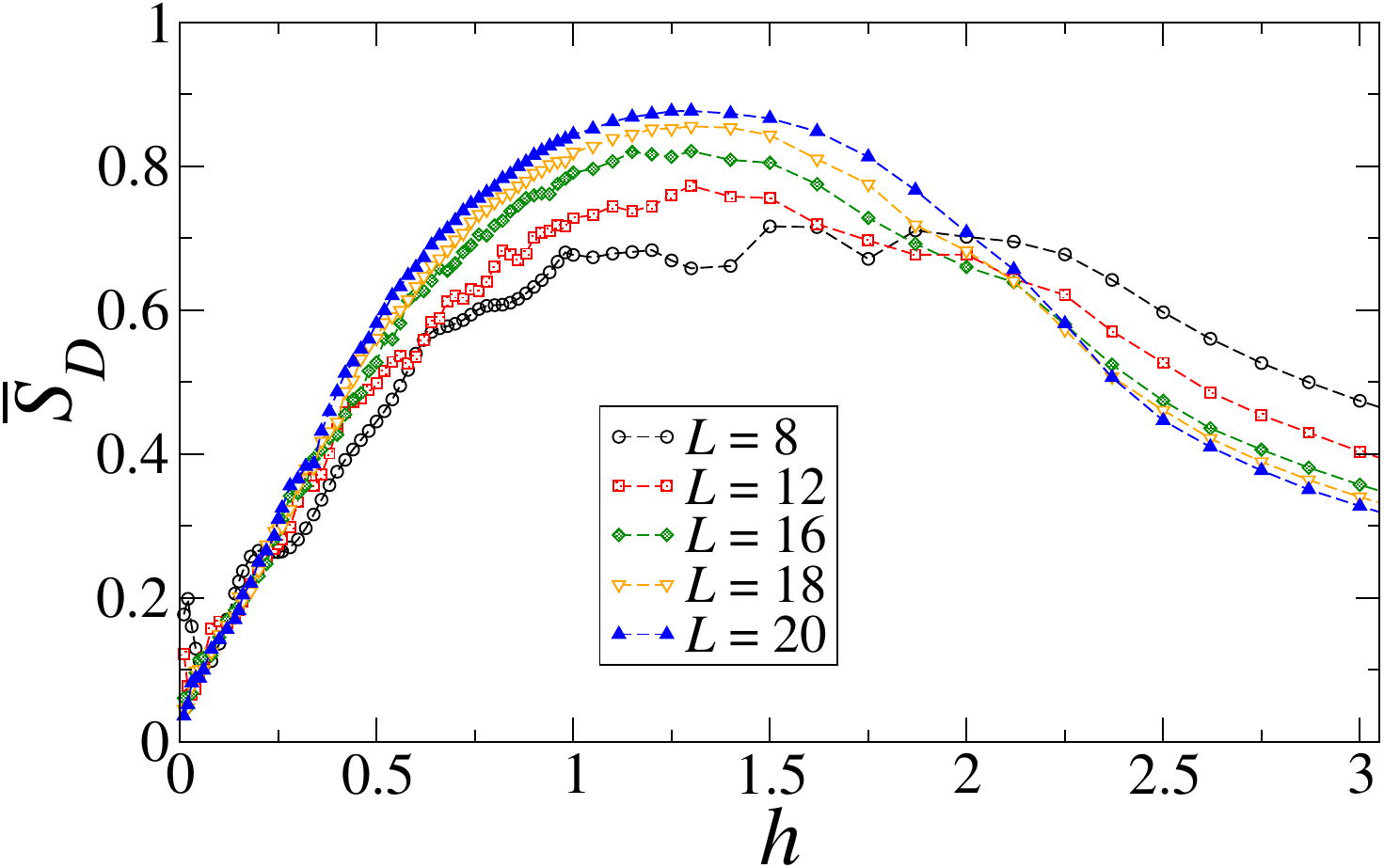}
    \caption{The rescaled diagonal-ensemble entropy ${\bar S}_D(L)$,
      normalized by the logarithm of the dimension of the relevant
      Hilbert subspace $\mathcal{D}_{0+}$, as a function of $h$ and
      for fixed $g=1$ and different sizes (see legend).}
    \label{dentrog1}
\end{figure}
%%%%%%%%%%%%%%%%%%%%%%%%%%%%%%%%%%%%%%%%%%%%%%%%%%%%%%%%%%%%%%%%%%%%%%%%%%%%%%%%%%

To quantify the delocalization of the $w_n$ values, we again consider
the diagonal-ensemble entropy defined in Eq.~\eqref{sddef}.
Fig.~\ref{dentrog1} reports the ratio ${\bar S}_D$, as defined in
Eq.~\eqref{sdresc}, as a function of $h$, for $g=1$ and for several
chain sizes up to $L=20$.  The behavior is qualitatively similar to
the one observed for $g=1.5$, with a nonmonotonic dependence on
$h$. Also in this case, ${\bar S}_D$ is expected to converge to one in
the large-$L$ limit, provided that $h$ lies in an extended
intermediate region where integrability is clearly broken. Comparing
Fig.~\ref{dentrog1} with Fig.~\ref{dentrog1p5}, notice, however, a
slight shift of this window toward smaller values of $h$ (being
roughly $0.5 \lesssim h < 2$, for $g=1$).  This likely indicates that,
for $g=1$, deviations from a fully chaotic behavior begin to emerge
earlier, as $h$ increases, compared with the $g=1.5$ case (see the
central and lower panels of Fig.~\ref{figR}).

\begin{figure}[!t]
    \includegraphics[width=0.45\textwidth]{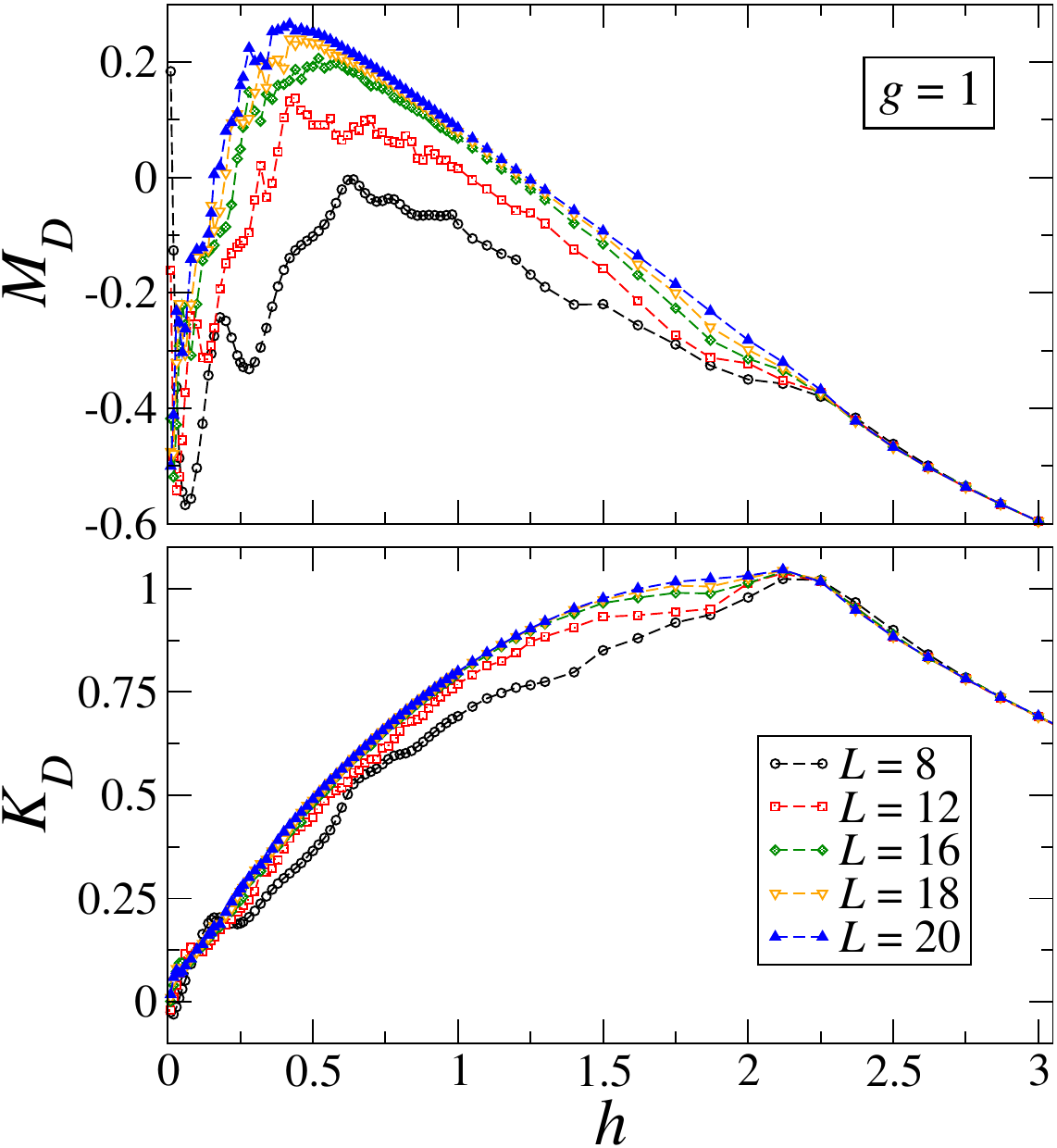}
    \caption{The diagonal-ensemble magnetization $M_D$ (top) and 
     excess bond energy $K_D$ (bottom) for $g=1$, $h=-h_i=1$, and 
      sizes $L\le 20$.}
    \label{edg1p0}
\end{figure}

Results for the diagonal-ensemble magnetization $M_D$ and excess bond
energy $K_D$ are shown in Fig.~\ref{edg1p0} for $L\leq 20$, as a
function of the longitudinal field $h$.  The behavior is qualitatively
similar to that reported in Fig.~\ref{edg1p5} for $g=1.5$, with a
nonmonotonic dependence on $h$, which suggests an inversion of the
population across different energy levels (see also
Fig.~\ref{overlapg1p0h1}).  A reasonable convergence with increasing
$L$ is visible for all values of $h$, although finite-size corrections
of varying magnitude are present.

%%%%%%%%%%%%%%%%%%%%%%%%%%%%%%%%%%%%%%%%%%%%%%%%%%%%%%%%%%%%%%%%%%%%%%%%%%%%%%%%%%
%%%%%%%%%%%%%%%%%%%%%%%%%%%%%%%%%%%%%%%%%%%%%%%%%%%%%%%%%%%%%%%%%%%%%%%%%%%%%%%%%%

\subsection{The post-quench evolution}
\label{postqqcqt}

\begin{figure}[!t]
    \includegraphics[width=0.47\textwidth]{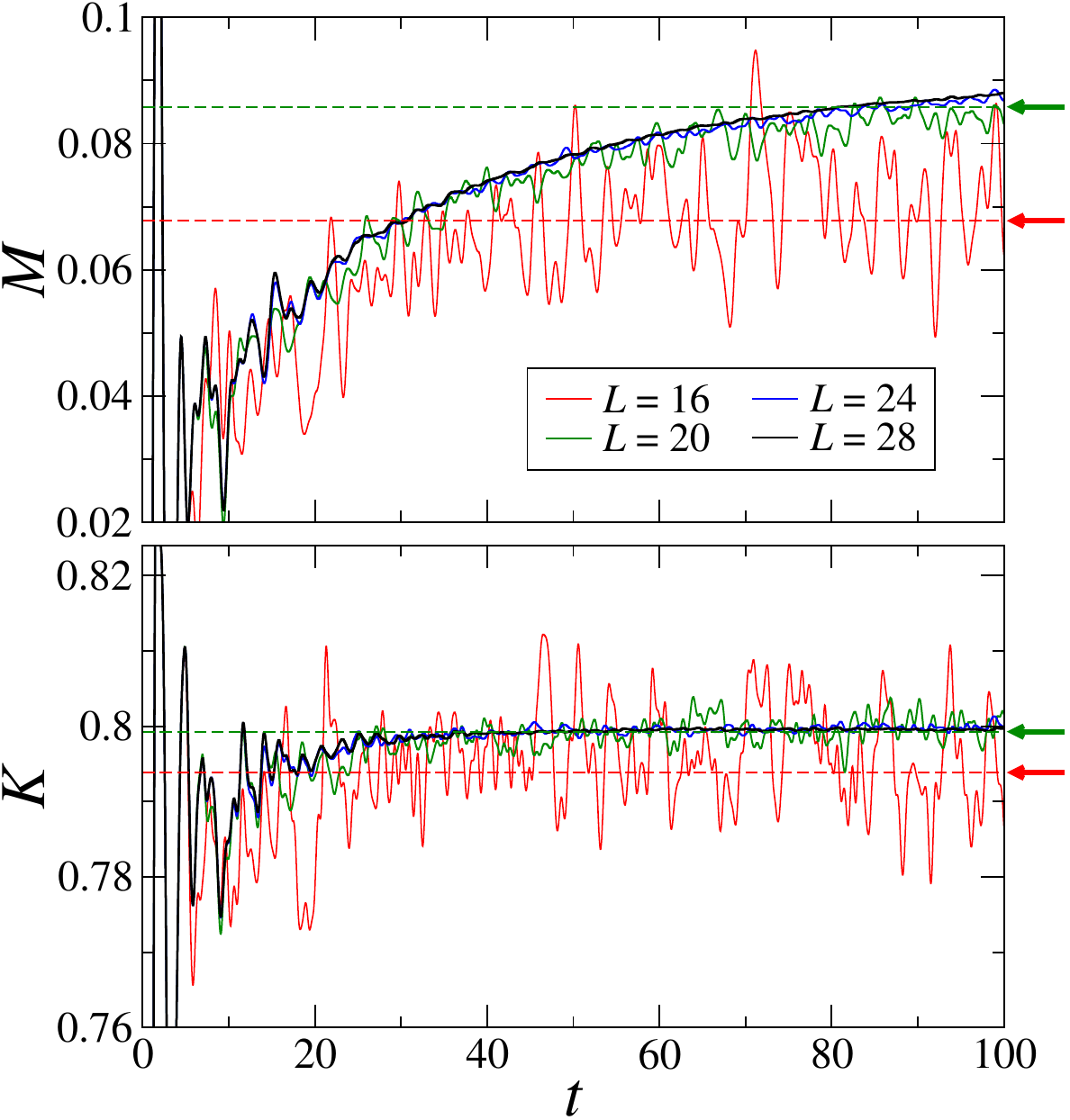}
    \caption{Post-QQ time evolution of the magnetization (top) and the
      excess bond energy (bottom), for $g=1$, $h=-h_i=1$, different
      sizes (see legend).  The dashed horizontal curves and arrows
      indicate the corresponding diagonal-ensemble predictions for $L
      \leq 20$.}
  \label{quevg1p0}
\end{figure}

Turning to the analysis of the post-QQ evolution, Fig.~\ref{quevg1p0}
shows the time dependence of the magnetization (top) and of the excess
bond energy (bottom) for $h=1$.  As $L$ increases, the curves converge
to what we consider the infinite-size limit, as already observed in
the disordered phase, see Fig.~\ref{quevg1p5}.  To better display this
convergence, we use a small $y$-axis interval in both panels (thus,
some large oscillations occurring for small values of $t$ are not
properly displayed) and report results only up to $t=100$. It is,
however, worth noting that it is not straightforward to predict the
average long-time behavior.  Although we expect it to be correctly
predicted by the diagonal ensemble, our calculations in this ensemble
are limited to $L \leq 20$.  In fact, for the larger system sizes the
dynamics is still dominated by a transient behavior (see, for
instance, the blue and black curves for the magnetization for $L=24$
and $L=28$, respectively).  The duration of this transient depends
both on the monitored observable and on the Hamiltonian parameters $g$
and $h$. Moreover, the onset of the asymptotic long-time regime is
pushed to progressively later times as $L$ increases.  Nonetheless, we
verified that convergence with $L$ is still clearly visible up to $t
\approx 500$.

%%%%%%%%%%%%%%%%%%%%%%%%%%%%%%%%%%%%%%%%%%%%%%%%%%%%%%%%%%%%%%%%%%%%%%%%%%%%%%%%%%
%%%%%%%%%%%%%%%%%%%%%%%%%%%%%%%%%%%%%%%%%%%%%%%%%%%%%%%%%%%%%%%%%%%%%%%%%%%%%%%%%%

\subsection{Thermalization}
\label{sec:termal_cqt}

We have also repeated the analysis reported in
Sec.~\ref{sec:termal_disord} for QQs in the disordered phase, studying
the asymptotic thermalization for QQs at the critical point.
Specifically, we have computed the effective inverse temperature
$\beta_{\rm th}$ using energy, Eq.~\eqref{tthdef}, and the
diagonal-ensemble excess energy $K_D$ and magnetization $M_D$.  Across
the entire chaotic window $0.1 \lesssim h \lesssim 1.5$ for systems of
size $L \approx 20$ (cf.~Fig.~\ref{figR}, middle), the results are
qualitatively similar to those obtained for $g=1.5$.  In particular,
once $L$ is fixed, the residual dependence on the subsystem size
$\ell$ is still accurately captured by Eq.~\eqref{eq:fit_beta}.

As an illustration, for a system of size $L=20$, we obtained the
following large-$\ell$ estimates of the effective inverse temperature
(data not shown): for $h=0.5$, we found $\beta_{\rm th} \approx 0.331,
\, 0.334, \, 0.299$ using, respectively, the energy, the excess
energy, and the magnetization; for $h=1$, we found $\beta_{\rm th}
\approx 0.086, \, 0.085, \, 0.074$.  They may be considered in
substantial agreement. We also note the better agreement among the
three estimates for $h=0.5$, compared with the case at fixed
$g=1.5$. This is likely because for $g=1$ the onset of the chaotic
regime at relatively small values of $h$ occurs more rapidly as $L$
increases. Therefore, the analysis of the post-QQ dynamics across the
CQT for systems up to $L=20$ provides strong evidence of the eventual
thermalization in the large-time limit, in the region $0.1\lesssim
h\lesssim 1.5$ where the spectrum of $\hat H(h)$ for $g=g_c$ develops
a chaotic regime (see Fig.~\ref{figR}).

\subsection{Comparison with QQs in the disordered phase}
\label{compdis}

We conclude this section by comparing the $g=1$ results with those
obtained far from the critical point in the disordered phase (see
Sec.~\ref{disphase}, for the data at $g=1.5$).  The post-QQ evolutions
of the magnetization and the excess bond energy, as well as their
asymptotic values obtained from the diagonal ensemble, differ only
quantitatively. We do not observe any qualitative difference in the
post-QQ behavior, when keeping $h$ fixed in the large-$L$ limit,
particularly when the spectrum of the post-QQ Hamiltonian lies within
its chaotic regime.  This confirms that hard QQs, which involve an
extended amount of energy exchange, do not effectively probe the
low-energy large-scale modes responsible for the critical behavior
unlike soft QQs, as thoroughly discussed in Sec.~\ref{softcqt}.

\section{Quantum quenches across FOQTs}
\label{ququfoqt}

The out-of-equilibrium quantum evolution induced by QQs across
the FOQT line presents novel and unexpected features that warrant
a detailed investigation.  We primarily
consider the paradigmatic case in which the transverse field is kept
fixed at $g=0.5$; no significant qualitative differences are expected
as long as $g<1$. Most of the results presented below refer to the
quench protocol from $h_i=-h$ to $h_f=h$; we also examine an
alternative protocol that starts from $h_i=0^-$, finding no
substantial differences.  As discussed in Sec.~\ref{chaoreg}, 
sizes up to $L=20$ are sufficient to show that, for $g=0.5$,
the spectrum lies in the chaotic regime at least in the interval
$0.3\lesssim h \lesssim 1$ (see the upper panel of Fig.~\ref{figR}).
As we shall show, the resulting dynamics is more complex than that
observed for QQs in the disordered phase or across the CQT.
Finally, in Sec.~\ref{sec:gvar} we discuss additional issues that
arise in the diagonal-ensemble description following a QQ of the
longitudinal field, when the transverse field is varied.

\subsection{The diagonal ensemble}
\label{secIII.A}

%%%%%%%%%%%%%%%%%%%%%%%%%%%%%%%%%%%%%%%%%%%%%%%%%%%%%%%%%%%%%%%%%%%%%%%%%%%%%%%%%%
\begin{figure}[!t]
  \includegraphics[width=0.47\textwidth]{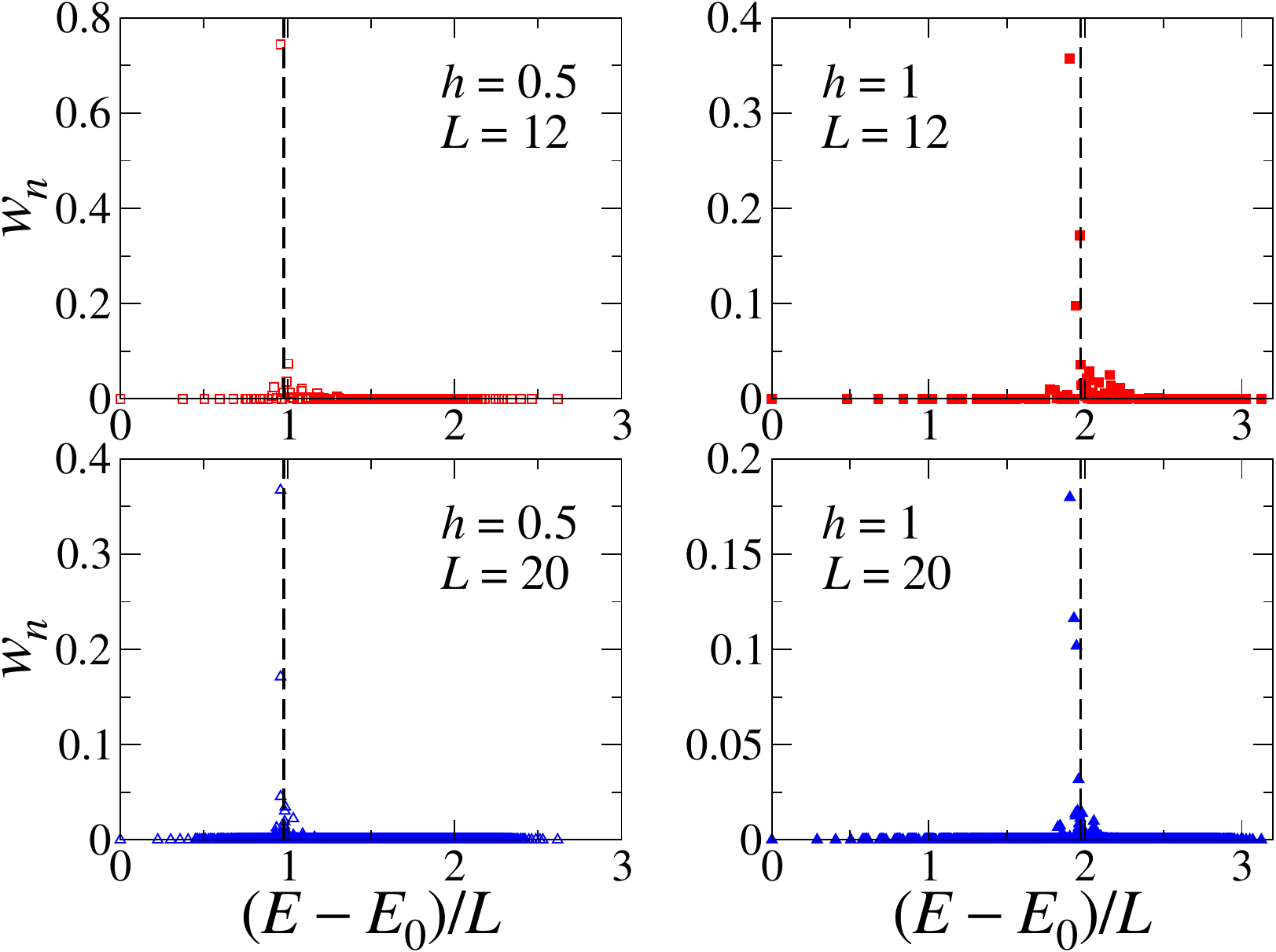}
  \caption{The overlap between the ground state of the Hamiltonian
    $\hat H(-h)$ and the eigenstates of the post-QQ Hamiltonian $\hat
    H(h)$ vs the energy density $e_n=(E_n-E_0)/L$, at fixed $g=0.5$
    and for $h=0.5$ (left panels) or $h=1$ (right panels). In the
    upper panels we show results for $L=12$, while in the lower panels
    we consider $L=20$.  Vertical lines correspond to the
    energy-density averages $e \approx 0.97935239$ (left) and
    $2.9693129$ (right).}
  \label{overlapg0p5}
\end{figure}
%%%%%%%%%%%%%%%%%%%%%%%%%%%%%%%%%%%%%%%%%%%%%%%%%%%%%%%%%%%%%%%%%%%%%%%%%%%%%%%%%%

A first notable difference that emerges when crossing a FOQT, compared
with the previously analyzed QQs, concerns the distribution of the
overlaps $w_n$ between the initial ground state and the eigenstates of
the post-quench Hamiltonian $\hat H(h)$.  Figure~\ref{overlapg0p5}
displays the overlaps $w_n$, for $h=0.5$ (left) and $h=1$ (right),
computed in the subsector ${\cal H}_{0+}$ of the Hilbert space, for
$L=12$ (top) and $L=20$ (bottom). At first glance, it is evident that
the distributions have a sharp peak, with only few eigenvalues $w_n$
been significantly different from zero (fewer than ten,
approximately).  This behavior stands in stark contrast with what
observed for $g<1$ and $g=1$, reported in Figs.~\ref{overlapg1p5h12}
and~\ref{overlapg1p0h1}, respectively.  Indeed, for $L=20$ the
$y$-axis values are here roughly two orders of magnitude larger than
those in Figs.~\ref{overlapg1p5h12} and~\ref{overlapg1p0h1} (see the
blue histograms).  We also observe that the peak shifts toward higher
energies as $h$ increases, as observed in the previous cases. 
Notably, the presence of only a small number of relevant eigenstates
of the post-QQ Hamiltonian---resulting in an overlap distribution
with a pronounced maximum---is not a consequence of spectral chaos.
Indeed, the same behavior is observed (see the Appendix) in the OFSS
regime of a chain with oppositely fixed boundary conditions for small
$g$ and $h$, where the system is manifestly integrable. This feature
is therefore only related with the presence of the FOQT for $h=0$.

%%%%%%%%%%%%%%%%%%%%%%%%%%%%%%%%%%%%%%%%%%%%%%%%%%%%%%%%%%%%%%%%%%%%%%%%%%%%%%%%%%
\begin{figure}[!t]
  \includegraphics[width=0.47\textwidth]{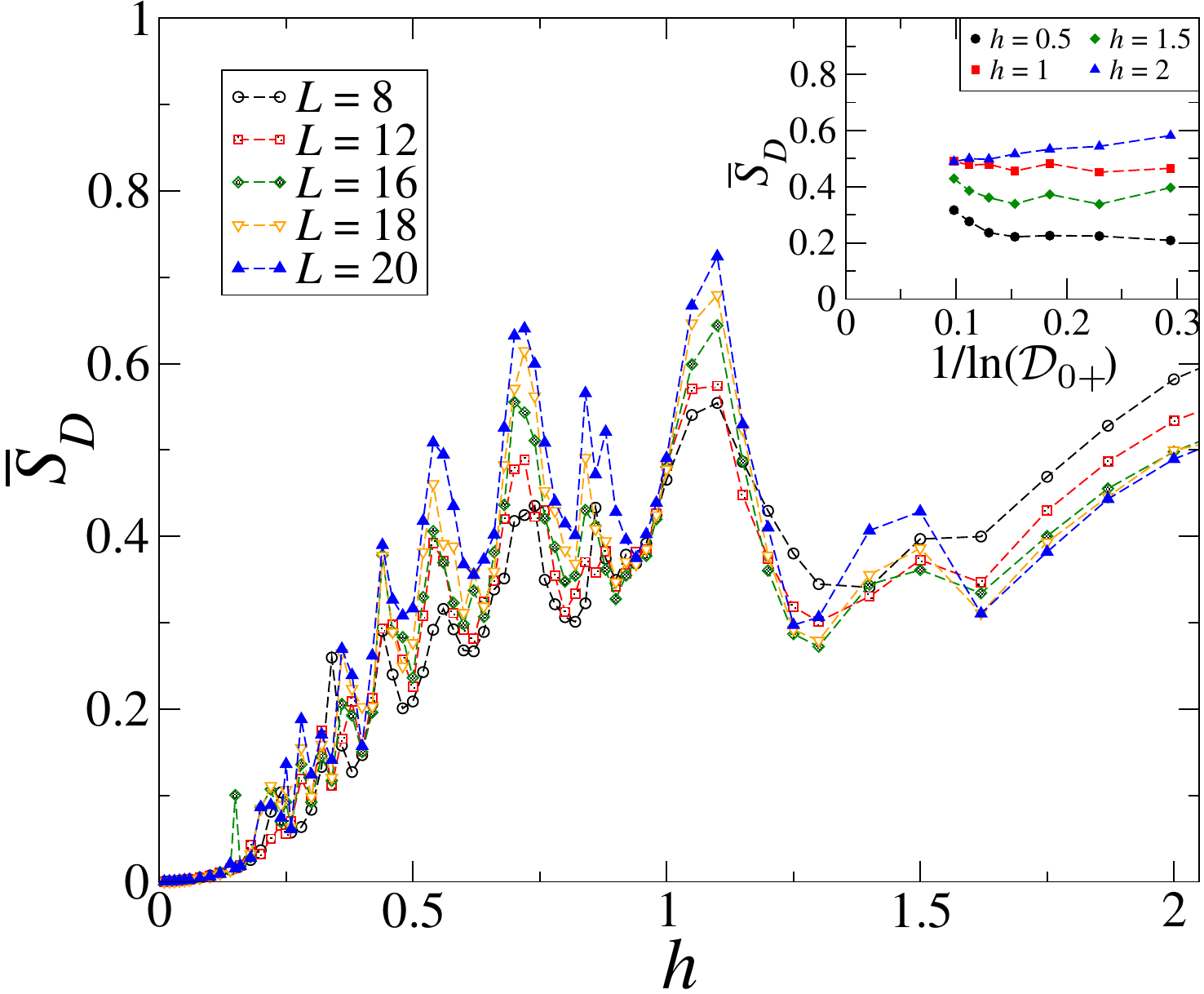}
  \caption{The rescaled diagonal-ensemble entropy ${\bar S}_D(L)$,
    normalized by the logarithm of the dimension of the relevant
    Hilbert subspace $\mathcal{D}_{0+}$, as a function of $h$, for
    different sizes $L$.  Here we fix $g=0.5$.  The inset presents the
    same data plotted against $1/\ln {\cal D}_{0+}$, for selected values
    of $h$, highlighting that numerical results for $L\leq 20$ remain
    far from the hypothetical thermodynamic-limit value of one,
    which would appear for $1/\ln {\cal D}_{0+} \to 0$.}
  \label{dentrog0p5}
\end{figure}
%%%%%%%%%%%%%%%%%%%%%%%%%%%%%%%%%%%%%%%%%%%%%%%%%%%%%%%%%%%%%%%%%%%%%%%%%%%%%%%%%%

To quantify the delocalization of the $w_n$ values, it is again
useful to examine the diagonal-ensemble entropy defined in
Eq.~\eqref{sddef}, and in particular the ratio ${\bar S}_D$ in
Eq.~\eqref{sdresc}. Data for $g=0.5$ are shown in
Fig.~\ref{dentrog0p5}, for various values of $h$ and chain sizes up to
$L=20$. The resulting behavior is radically different from that
observed for $g=1.5$ (Fig.~\ref{dentrog1p5}) and $g=1$
(Fig.~\ref{dentrog1}).  In fact, after an initial region for $h
\lesssim 0.2$, where ${\bar S}_D$ appears to be nearly independent of
$L$, wild oscillations as a function of $h$ set in.  These irregular
features disappear only for large values of $h$ ($h \gtrsim 2$, 
not shown). Moreover, the normalized diagonal-ensemble entropy
typically remains far from one, even at $L=20$ (see, in particular,
the inset, showing results for four representative values of $h$),
consistently with the very peaked distribution of the overlaps $w_n$
discussed above.  Additional comments on the behavior of the overlaps
$w_n$ as a function of $g$ are presented at the end of this section
(see Sec.~\ref{sec:gvar}).

%%%%%%%%%%%%%%%%%%%%%%%%%%%%%%%%%%%%%%%%%%%%%%%%%%%%%%%%%%%%%%%%%%%%%%%%%%%%%%%%%%
\begin{figure}[!t]
  \includegraphics[width=0.47\textwidth]{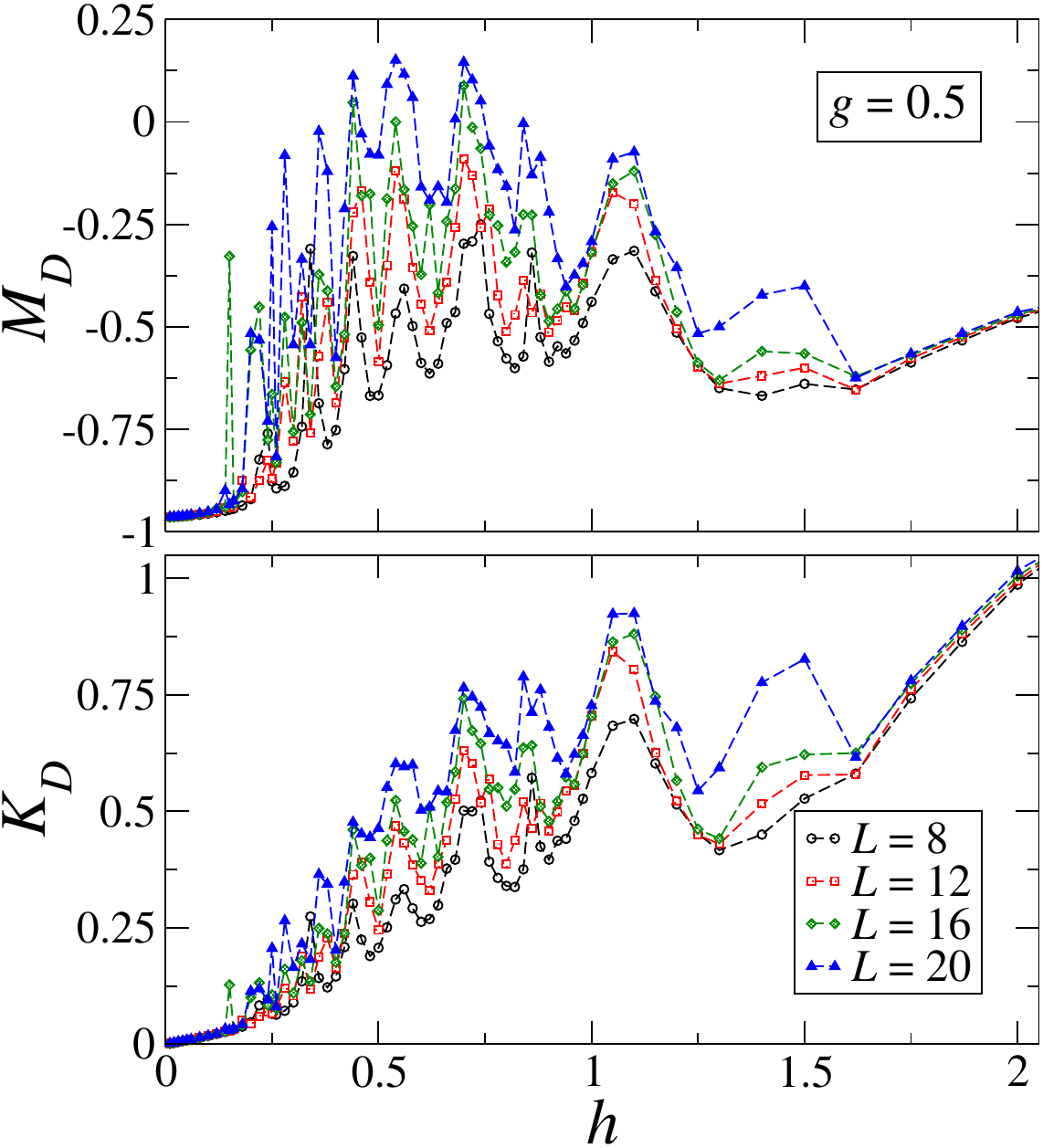}
  \caption{The magnetization (top) and excess bond energy (bottom)
    across the FOQT for $g=0.5$, obtained from the diagonal ensemble
    for QQs from $h_i=-h<0$ to $h_f=h>0$ and for different system sizes
    (see legend).}
  \label{figedg0p5}
\end{figure}
%%%%%%%%%%%%%%%%%%%%%%%%%%%%%%%%%%%%%%%%%%%%%%%%%%%%%%%%%%%%%%%%%%%%%%%%%%%%%%%%%%

A clear evidence of the presence of these singularities is visible in
Fig.~\ref{figedg0p5}, where we report the magnetization (top) and the
excess bond energy (bottom) as a function of $h$, obtained from the
diagonal ensemble corresponding to the QQ from $-h$ to $h$. Unlike
the cases for $g \ge 1$, we note that there is a large interval of
values of $h$ (i.e., $0.3\lesssim h \lesssim 1.5$), where the
diagonal-ensemble values $M_D$ and $K_D$ oscillate wildly as a
function of $h$. This implies that small changes of $h$ give rise to
large changes of the asymptotic long-time behavior.  A similar high
sensitivity occurs also as a function of $L$, for some specific values
of $h$ in such interval (as is the case for $h=0.5$ highlighted in
Fig.~\ref{diagquevg0p5} and discussed below).  This behavior appears
to be qualitatively different from that observed across the CQT and in
the disordered phase.

%%%%%%%%%%%%%%%%%%%%%%%%%%%%%%%%%%%%%%%%%%%%%%%%%%%%%%%%%%%%%%%%%%%%%%%%%%%%%%%%%%
\begin{figure}[!t]
  \includegraphics[width=0.47\textwidth]{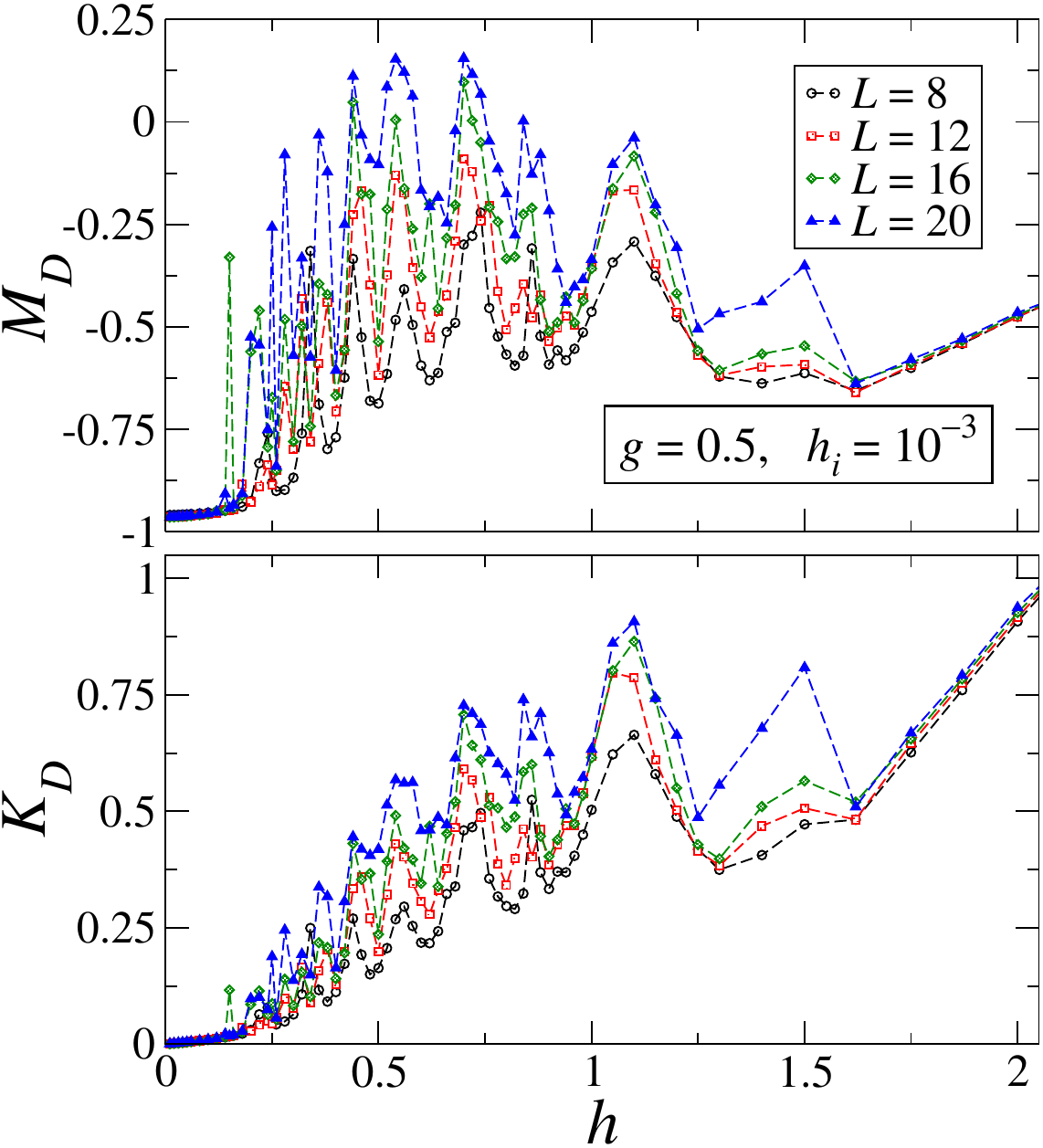}
  \caption{The magnetization (top) and the excess bond energy (bottom)
    for $g=0.5$, obtained from the diagonal ensemble for QQs from
    $h_i=-10^{-3}$ to $h_f>0$ and for different system sizes (see legend).}
  \label{figedg0p5_hi0}
\end{figure}
%%%%%%%%%%%%%%%%%%%%%%%%%%%%%%%%%%%%%%%%%%%%%%%%%%%%%%%%%%%%%%%%%%%%%%%%%%%%%%%%%%

It is likewise instructive to assess the robustness of the above
features by considering an alternative QQ protocol, in which the
system is initialized in the ground state of the quantum Ising
Hamiltonian with an infinitesimal longitudinal field $h_i=0^-$ (we
typically take $h_i=-10^{-3}$, which is sufficient for $L\lesssim 20$).
For $g<g_c$, this weak field explicitly breaks the $\mathbb{Z}_2$
symmetry of the model, thereby selecting a ground state with negative
magnetization ($M<0$).  The system then evolves unitarily
with a finite $h>0$, while keeping $g$ fixed. Some
numerical results for the diagonal-ensemble magnetization and excess
bond energy at $g=0.5$ are shown in Fig.~\ref{figedg0p5_hi0} as a
function of $h$.  Although we observe small quantitative differences
with the QQ protocol from $h_i=-h$ to $h$, especially for the largest
displayed values of $h$, the behavior of $M_D$ and $K_D$ is
qualitatively the same. In particular, both observables exhibit a
pronounced sensitivity to $h$ and $L$, which we identify as a hallmark
of QQs crossing FOQTs.  The independence of the results on the
starting value of the longitudinal field, as long as it selects a
negatively magnetized ground state---that is, for any $h_i < 0$---can
be traced back to the fact that the corresponding ground states differ
only over length scales of the order of the correlation length $\xi$,
which is finite for $g<g_c$ (for $g=0.5$ it is of order one in units
of the lattice spacing).

\subsection{The post-quench evolution}

\begin{figure}[!t]
  \includegraphics[width=0.47\textwidth]{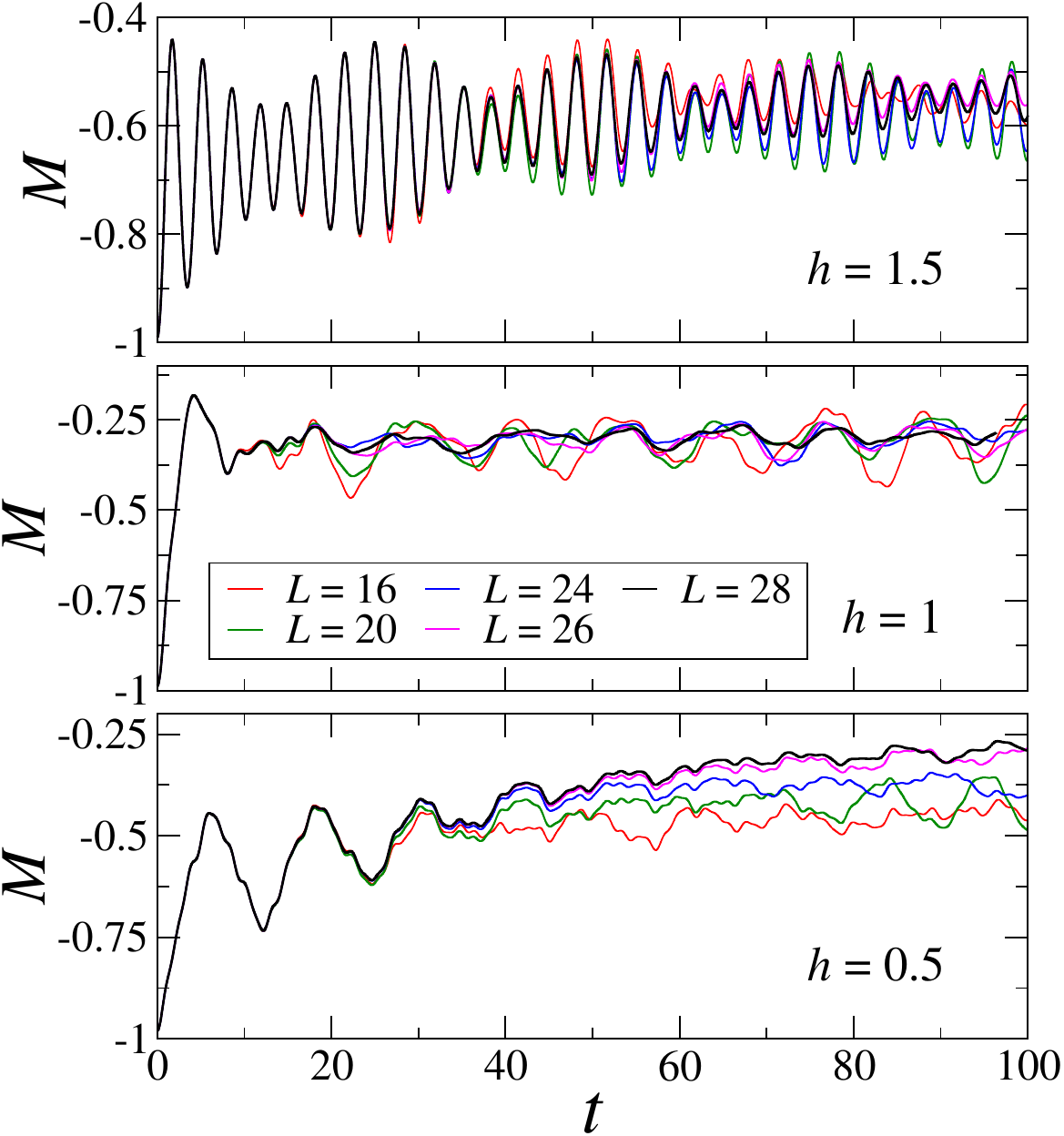}
  \caption{Time dependence of the magnetization after a QQ from
    $h_i=-h$ to $h_f=h$, at fixed $g=0.5$, and for $h=1.5$ (top),
    $h=1$ (middle), and $h=0.5$ (bottom).  The various curves are for
    different lattice sizes up to $L=28$ (see legend).}
  \label{quevg0p5}
\end{figure}

The post-quench evolution of the magnetization $M(t)$ and of the
excess bond energy $K(t)$, for $g=0.5$, is reported in
Fig.~\ref{quevg0p5} for system sizes up to $L=28$ and for three
distinct values of $h$ (the QQ is from $h_i=-h$ to $h_f=+h$).  For
$h=1.5$ (top), the dynamics is characterized by large oscillations
around stable values, indicating that a few energy levels dominate the
post-QQ evolution, even for the largest sizes considered (the data
appear to converge in the large-$L$ limit, at least up to $t\approx 100$).
This can be easily explained by recalling that the value
$h=1.5$ is outside the chaotic region of the spectrum of the post-QQ
Hamiltonian for $L=20$ (see Fig.~\ref{figR}).  For $h=1$ (middle), the
oscillations become less regular, although their amplitude still
decreases as $L$ increases.  Finally, for $h=0.5$ (bottom), the
convergence to the asymptotic stationary state appears significantly
delayed, compared to the other values of $h$.  In fact, the curves for
various system sizes at $t \gtrsim 50$ show no signs of
convergence, indicating that sizes larger than $L=28$ might be
required to clarify the large-$L$ and long-$t$ behavior.

%%%%%%%%%%%%%%%%%%%%%%%%%%%%%%%%%%%%%%%%%%%%%%%%%%%%%%%%%%%%%%%%%%%%%%%%%%%%%%%%%%
%%%%%%%%%%%%%%%%%%%%%%%%%%%%%%%%%%%%%%%%%%%%%%%%%%%%%%%%%%%%%%%%%%%%%%%%%%%%%%%%%%

\begin{figure}[!t]
  \includegraphics[width=0.47\textwidth]{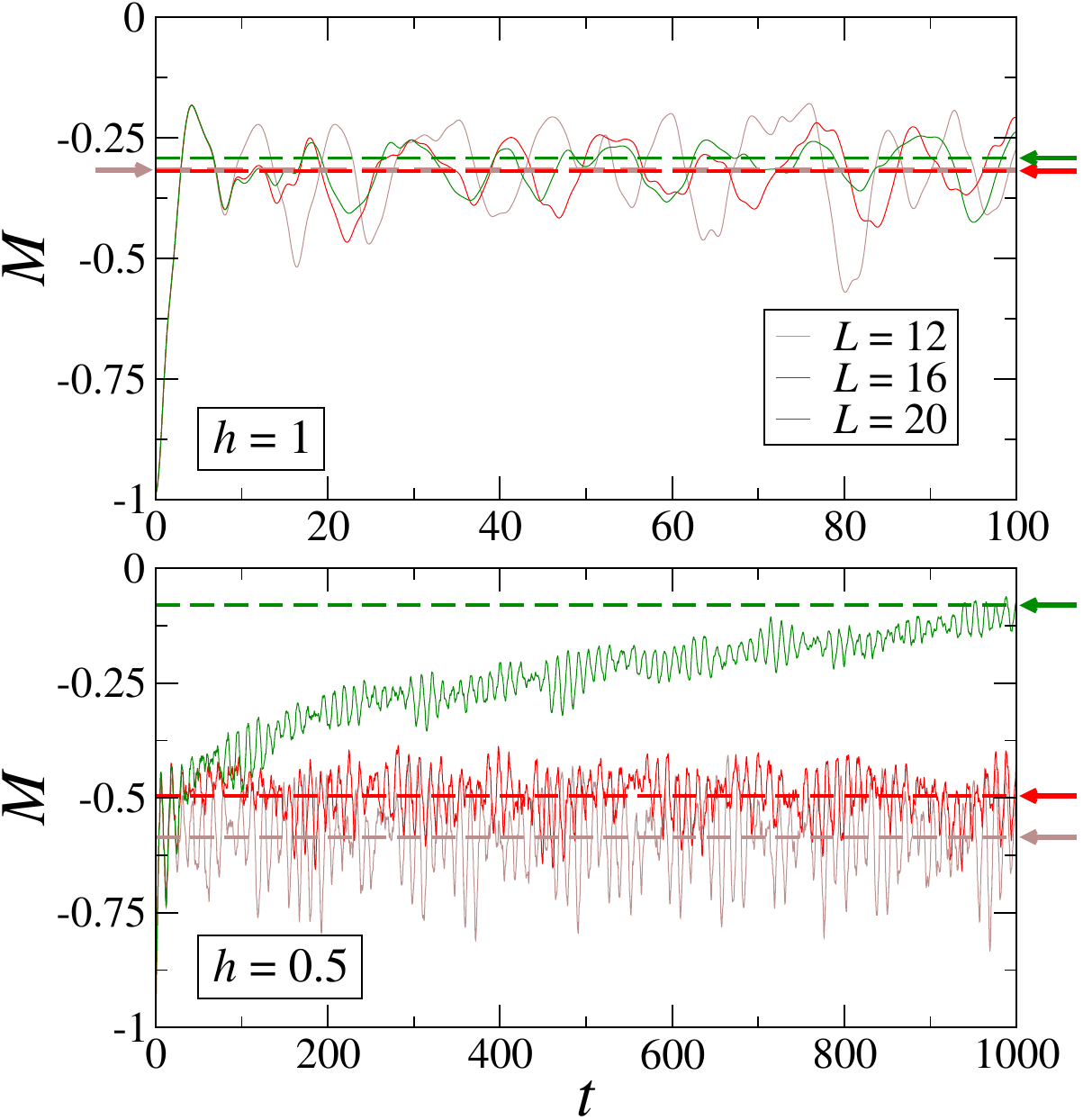}
  \caption{Time dependence of the magnetization, compared
    with the diagonal-ensemble predictions (arrows and horizontal
    dashed lines with the corresponding color code), up to $L = 20$.
    We fix $g \!=\! 0.5$, while $h \!=\! 1$ (top) or $h \!=\! 0.5$ (bottom).}
  \label{diagquevg0p5}
\end{figure}

The above scenario is confirmed by comparing the post-QQ evolutions
with the diagonal-ensemble predictions, shown in
Fig.~\ref{diagquevg0p5}.  In particular, for $h=1$ (top), the curves
on the scale of the figure are already close to convergence, both in
time and with $L$, so that the resulting picture is analogous to those
previously discussed for $g \geq 1$.  In contrast, for $h=0.5$
(bottom), while the curves for $L=12$ and $L=16$ appear to oscillate
around the asymptotic long-time behavior predicted by the diagonal
ensemble already for $L\approx 10$ (with fluctuations that decrease as
$L$ increases), this is not the case for $L=20$.  For that size, the
transient time is significantly longer and appears to reach scales of
order $t \sim 10^3$. Moreover, the asymptotic value exhibits a sudden
jump as a function of $L$.  These features are directly related to the
spikes that may irregularly appear in Figs.~\ref{figedg0p5}
and~\ref{figedg0p5_hi0}, at seemingly random values of $h$ in the
interval $0.3 \lesssim h \lesssim 1.5$, and that also depend on the
system size.

%%%%%%%%%%%%%%%%%%%%%%%%%%%%%%%%%%%%%%%%%%%%%%%%%%%%%%%%%%%%%%%%%%%%%%%%%%%%%%%%%%
%%%%%%%%%%%%%%%%%%%%%%%%%%%%%%%%%%%%%%%%%%%%%%%%%%%%%%%%%%%%%%%%%%%%%%%%%%%%%%%%%%

\subsection{Thermalization issues}

The apparent irregular behavior of $M_D$ and $K_D$ also reflects in
the corresponding estimates of the effective inverse temperature, even
in regimes where asymptotic thermalization would normally be expected.
Recall that, for $L \approx 20$ and $g = 0.5$, a quantum-chaotic
dynamics is likely to occur for $0.3 \lesssim h \lesssim 1$ (see
Fig.~\ref{figR}, top).  Within this interval, we observe
significant discrepancies among the estimated effective inverse
temperatures $\beta_{\rm eff}$ obtained from different observables.
As an illustrative example, for $g = h = 0.5$ and $L = 20$, we
obtained the following large-$\ell$ estimates (data not shown):
$\beta_{\rm th} \approx 0.318, \, 0.437, \, -0.277$ from the energy,
the excess bond energy, and the magnetization, respectively.  Similar
inconsistencies occur at larger longitudinal fields; for instance, at
$h=0.8$, we found $\beta_{\rm th} \approx 0.125, \, 0.265, \, 0.489$.

The discrepancies are striking---indeed, even the signs differ. These
results indicate that, for system sizes up to $L \approx 20$, the
post-QQ dynamics across FOQTs remains far from asymptotic
thermalization, despite the post-quench Hamiltonian already being
in the chaotic regime.  In this respect, the mere presence of spectral
chaos in the post-QQ Hamiltonian should not be regarded as a
sufficient condition for eventual thermalization.  This suggests
that QQ protocols across the FOQT line may not lead to thermalization
even in the thermodynamic limit, because the initial ground state
corresponding to opposite longitudinal fields is not effectively
connected to the typical states of the post-QQ Hamiltonian, when
they are separated by a FOQT. Alternatively, much larger sizes might
be required to observe thermal behavior, and the present results could
simply reflect preasymptotic effects.

\subsection{Comparison of results  for various values of $g$}
\label{sec:gvar}

%%%%%%%%%%%%%%%%%%%%%%%%%%%%%%%%%%%%%%%%%%%%%%%%%%%%%%%%%%%%%%%%%%%%%%%%%%%%%%%%%%
\begin{figure}[!t]
  \includegraphics[width=0.47\textwidth]{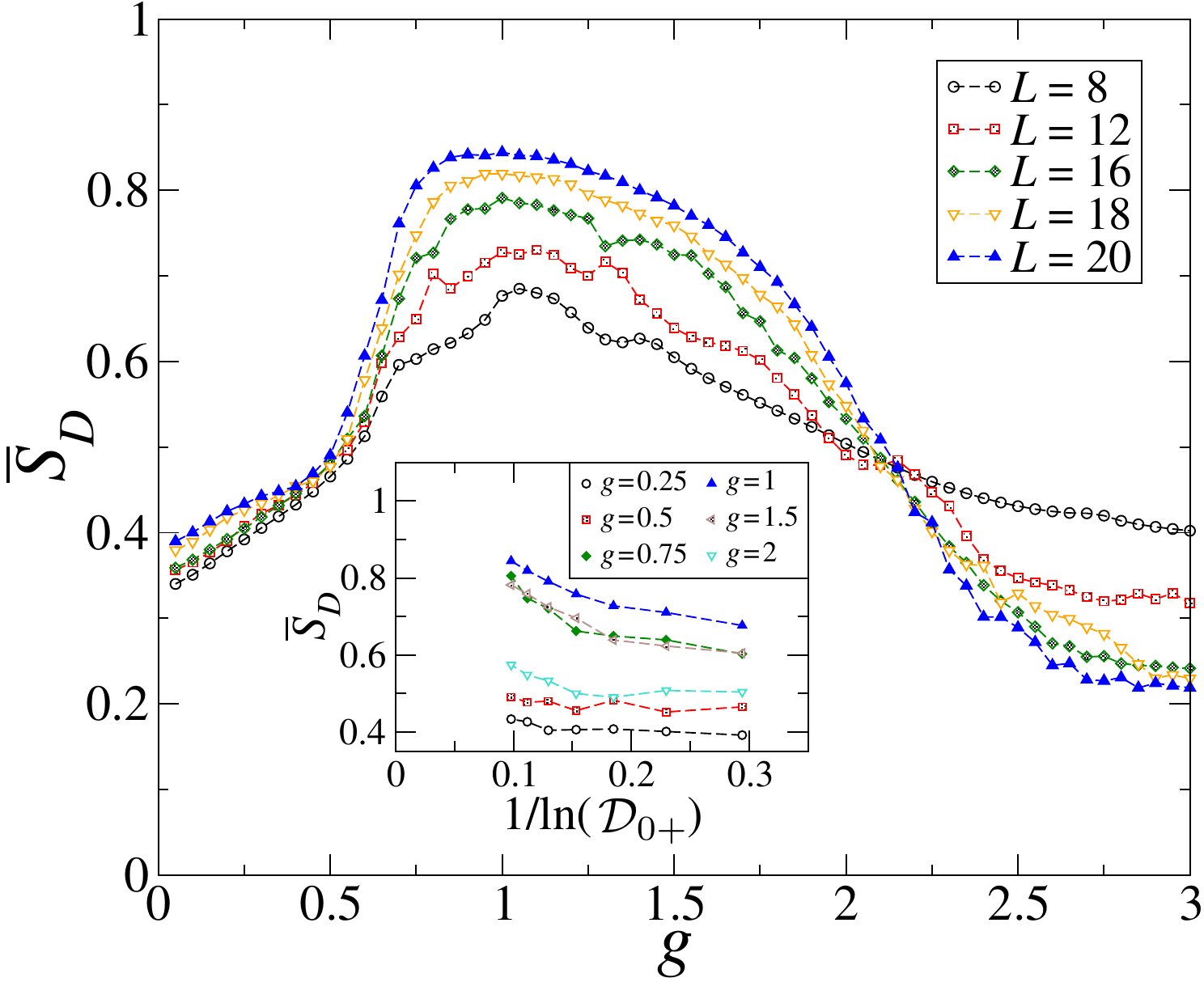}
  \caption{The ratio ${\bar S}_D\equiv S_D(L)/\ln {\cal D}_{0+}$
    between the diagonal entropy and its maximum value $\ln {\cal D}_{0+}$
    as a function of $g$, after QQs from $-h$ to $h$ with $h=1$.
    The inset shows the same data for some fixed value of $g$
    (see legend) versus $1/\ln {\cal D}_{0+}\sim 1/L$.}
    \label{dentroh1}
\end{figure}
%%%%%%%%%%%%%%%%%%%%%%%%%%%%%%%%%%%%%%%%%%%%%%%%%%%%%%%%%%%%%%%%%%%%%%%%%%%%%%%%%%

We conclude our numerical analysis by comparing results for different
values of the transverse field $g$, when quenching the longitudinal
field from $-h$ to $+h$, with $h=1$ in the chaotic region.
Figure~\ref{dentroh1} shows that the rescaled diagonal-ensemble entropy
${\bar S}_D$ exhibits a rather complex and nonmonotonic dependence on
$g$ at fixed $L$, reaching a maximum near the critical point
($g\approx 1$).  Most importantly, ${\bar S}_D$ is significantly
smaller for $g\lesssim 0.5$, i.e., deep in the FOQT region, at least
for lattice sizes up to $L=20$.  In particular, the data around $g=1$
apparently converge toward the maximal value ${\bar S}_D=1$ as $L$
increases, whereas in the FOQT region there is no indication of
convergence to this value (see the inset, displaying ${\bar S}_D$
as a function of $1/ \ln {\cal D}_{0+}$, for several values of $g$).
A similarly puzzling behavior emerges in the region $g \gtrsim 2$,
where the apparent nonmonotonic trend likely signals a departure from
the fully chaotic WD behavior at large transverse fields.

\begin{figure}[!t]
  \includegraphics[width=0.47\textwidth]{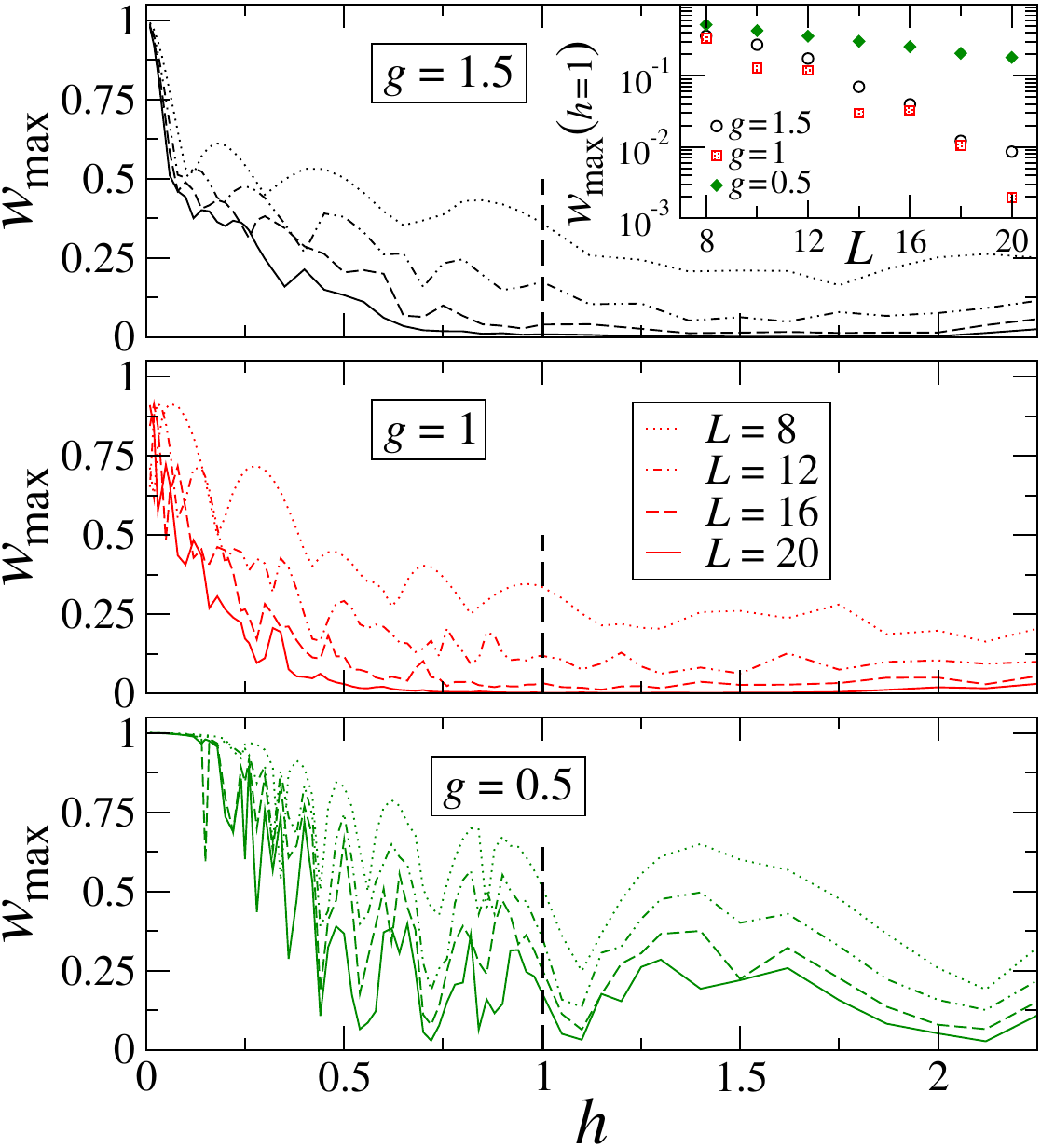} 
  \caption{The maximum overlap $w_{\rm max}$ as a function of
    $h$, for three values of $g$ : $1.5$ (top), $1$ (middle), and $0.5$
    (bottom).  Different curves are for various sizes (see legend).
    The inset in the top panel reports the dependence of
    $w_{\rm max}$ on $L$ in a log-linear scale, for fixed $h=1$ and the
    same values of $g$.}
  \label{maxoverh1}
\end{figure}

As already noted, the diagonal-ensemble entropy ${\bar S}_D$ also
provides information about the distribution $w_n$ of the overlaps
between the initial state and the eigenstates of the post-QQ
Hamiltonian. Larger values of ${\bar S}_D$ indicate a more homogeneous
distribution among the eigenstates (see, e.g.,
Fig.~\ref{overlapg1p5h12} for $g=1.5$ and Fig.~\ref{overlapg1p0h1} for
$g=1$), whereas smaller values typically correspond to more sharply
peaked distributions (see, e.g., Fig.~\ref{overlapg0p5} for $g=0.5$).
To quantify this different behavior, in Fig.~\ref{maxoverh1} we report
the maximum value $w_{\rm max}$ of the distribution as a function
of $h$, for $g=1.5$ (top), $g=1$ (middle), and $g=0.5$ (bottom).  On
average, $w_{\rm max}$ decreases with increasing $h$: for instance,
comparing the left and right panels in Fig.~\ref{overlapg1p5h12}
(corresponding to $h=1$ and $h=2$), one sees that, at fixed $L$, the
right-hand panels have smaller vertical scales and somewhat flatter
distributions.  Moreover, as expected, $w_{\rm max}$ decreases as $L$
increases; however, the decay rate depends markedly on $g$, being
significantly slower for $g=0.5$ than for $g \ge 1$, although it
remains at least exponential in $L$.  This behavior is clearly
illustrated in the inset of the top panel, which shows $w_{\rm max}$
versus $L$ at fixed $h=1$, for the three values of $g$. Similar
conclusions are obtained when considering the average of the first few
largest values of $w_n$.

%%%%%%%%%%%%%%%%%%%%%%%%%%%%%%%%%%%%%%%%%%%%%%%%%%%%%%%%%%%%%%%%%%%%%%%%%%%%%%%%%%
%%%%%%%%%%%%%%%%%%%%%%%%%%%%%%%%%%%%%%%%%%%%%%%%%%%%%%%%%%%%%%%%%%%%%%%%%%%%%%%%%%

\section{Conclusions}
\label{conclu}

We have investigated the out-of-equilibrium quantum dynamics induced
by QQs of the longitudinal field $h$ in quantum Ising chains at fixed
transverse field $g$, across the line $h=0$. For this purpose, we
focus on QQs across the CQT ($h=0$ and $g=g_c=1$) and the FOQT line
($h=0$ and $g<g_c=1$).  The dynamics starts from the ground state of
the Ising Hamiltonian $\hat H(g,h)$ with $h_i=-h<0$; then, at time
$t=0$, the longitudinal field is suddenly quenched to $h_f=h>0$, while
keeping $g$ fixed.  We analyze the subsequent unitary dynamics for $t
> 0$, focusing on the thermodynamic limit, which is identified by
examining the convergence of the post-QQ observables, such as the
magnetization~\eqref{mxm} and the excess bond energy~\eqref{boexdefF},
as the system size $L$ increases.  Our analysis primarily focuses on
values of $h$ for which the spectrum of the post-QQ Hamiltonian lies
in the chaotic regime, as signaled by a level-spacing distribution
following the WD statistics of the GOE.  This occurs for sufficiently
large (but not too large) values of $h$.  Under these conditions, the
system is generally expected to thermalize at long times, provided
that the initial state, namely the ground state at $h_i<0$, is
effectively connected with the typical eigenstates of $\hat H(g,h)$.

We consider systems with periodic boundary conditions, preserving
translational invariance and the ${\mathbb Z}_2$ symmetry when $h=0$.
The initial ground state and the entire post-QQ
evolution belong to the subspace ${\cal H}_{0+}$ of zero-momentum
states that are even under reflection parity, so only the spectrum of
the Ising Hamiltonian $\hat H(g,h)$ restricted to ${\cal H}_{0+}$ is
relevant for the post-QQ evolution. By using exact-diagonalization
methods, this allows us to determine the overlaps $w_n$ of the initial
state with the eigenstates of the post-QQ Hamiltonian, and the
diagonal-ensemble density matrix, which is expected to describe the
long-time behavior of the observables, for chains of length $L\le 20$.
The post-QQ evolution of the different observables can be computed on
significantly longer chains, up to $L=28$, as in that case one only
needs to determine the initial ground state (we use Lanczos
diagonalization).  The evolution is obtained by using a Runge-Kutta
integration of the Schr\"odinger equation.  This allows us to
effectively obtain thermodynamic-limit results up to quite long times,
possibly reaching $t = O(10^3)$, depending on the values of $g$ and
$h$.  We stress that the projection onto the zero-momentum Hilbert
subspace ${\cal H}_0$~\cite{PRV-25,Sandvik-10} has been crucial to
reach long chains.

We focus on two different types of quenches, namely, soft and (standard)
hard QQs. Soft QQs occur sufficiently close to the CQT and
FOQTs~\cite{PRV-18,PRV-20,RV-21,PV-24}, where the large-scale modes
driving the transition control the behavior of correlation functions,
in particular of the magnetization. In this case, the post-QQ dynamics
can be described in an OFSS framework, which requires the longitudinal
field $h$ to be appropriately tuned toward zero as $L$
increases: $h\sim L^{-y_h}$ with $y_h=15/8$ at the CQT and $h \sim
L^{-3/2} g^L$ at the FOQTs for $g<1$, see Sec.~\ref{softqu}.  In
contrast, OFSS regimes controlled by the universal features of the
transition are not expected to emerge in hard QQs, in which $h$ is
kept fixed as $L$ increases. The reason is that the energy $E_q$
injected by a hard QQ is extensive [$E_q=O(L)$], and thus it is much
larger than the energy scale $E_c \sim L^{-z}$ of the low-energy
critical modes at the CQT ($g=1$) or the scale $E_f \sim g^L$ at the
FOQTs ($g<1$).  Nevertheless, some peculiar singularities have been
observed in QQs of the transverse field at $g=g_c$ and
$h=0$~\cite{RV-20-qu,RV-21}, which may be associated with the CQT of
the quantum Ising chain.  However, these footprints associated with
the CQT are likely related to the integrability of the model at $h=0$.

We have extended the discussion to QQs involving sudden
changes of the longitudinal field $h$, such that the spectrum of the
post-QQ Hamiltonian $\hat H(g,h>0)$ lies in the chaotic regime.  We
investigate whether, under these more general conditions, qualitative
signatures of the universal low-energy features of the transitions can
still be observed in the post-QQ, energy-conserving, unitary dynamics.
In particular, we analyze the system behavior under QQs
across CQT ($g=1$) and FOQTs ($g<1$) for finite $h$ in the large-$L$ limit,
comparing it with the behavior observed for QQs in the disordered
phase ($g>1$), in order to identify qualitatively distinct behaviors
that may characterize QQs across CQTs and FOQTs.

We present a detailed analysis of the overlaps $w_n$ defined in
Eq.~\eqref{wndef}, comparing their behavior for QQs at fixed $g>g_c$,
$g=g_c$, and $g<g_c$.  A similar analysis has been performed for the
corresponding diagonal-ensemble density matrix $\hat \rho_D$ defined
in Eq.~\eqref{rhoddef}, which is expected to describe the long-time
post-QQ evolution of local observables, such as the magnetization and
the excess bond energy.  Our data confirm that the diagonal-ensemble
description becomes accurate in most cases after a relatively short
time, i.e., without requiring exponentially large times, as $L$
increases.  When an oscillatory behavior is present---particularly for
small chain sizes---the diagonal ensemble still provides an effective
description of time averages taken over sufficiently large time
windows. We also observe a general suppression of these oscillations
as the chain size increases.

To compare the main features of the distributions of $w_n$ in the
disordered region, across the CQT, and across the FOQTs, we determine
the diagonal-ensemble entropy $S_D$ [cf.~Eq.~\eqref{sddef}], the
largest values of $w_n$ [cf.~Eq.~\eqref{W1def}], and the effective
temperature, which can be defined assuming an eventual local
thermalization [cf.~Eq.~\eqref{tthdef}]. We do not observe significant
qualitative differences between the distributions of $w_n$ associated
with QQs in the disordered region and across the CQT. This confirms
that hard QQs, which involve an extended amount of energy exchange, do
not effectively probe the low-energy large-scale modes responsible for
the critical behavior unlike soft QQs.

On the other hand, the behavior at the FOQTs ($g<1$) exhibits features
that sharply distinguish it from the behavior for $g \gtrsim 1$. In
particular, the diagonal-ensemble entropy $S_D$ remains significantly
smaller, at least for the sizes for which we could analyze the full
Hamiltonian spectrum (see, for example, Fig.~\ref{dentroh1}), while
the decay of the largest values of $w_n$ is considerably slower
[cf.~Fig.~\ref{maxoverh1}].  Substantial differences also emerge when
comparing the observables obtained from the diagonal ensemble.  The
post-QQ long-time dynamics at FOQTs appears to be more intricate (at
least in the chaotic regime, which is observed for $0.3\lesssim h
\lesssim 1$), showing a pronounced sensitivity to small changes of
$h$. In addition, a stationary behavior is approached significantly
later than in the QQ protocols at $g\ge 1$ (see Figs.~\ref{quevg0p5}
and~\ref{diagquevg0p5}).  We also find significant discrepancies among
the effective temperatures obtained from different local observables
(such as the energy density, the excess bond energy, and the
magnetization) using the diagonal ensemble up to $L=20$, unlike for
$g\ge 1$.  This mismatch persists even in the region where the post-QQ
Hamiltonian exhibits clear signatures of spectral chaos. These findings
suggest that QQ protocols across the FOQT line may not lead to
thermalization even in the thermodynamic limit, as the initial ground
states on opposite sides of the FOQT may not be effectively
connected to the typical eigenstates of the post-QQ Hamiltonian.  This
may be related to the general loss of {\em ergodicity} at
first-order transitions in the thermodynamic limit, making
an analogous phenomenon at the FOQTs of the quantum Ising model at $h=0$
not entirely unexpected.  Nevertheless, since the post-QQ dynamics is
governed by a Hamiltonian with finite $h>0$, we cannot exclude that much
larger sizes are required to observe thermal behavior, or that
long-lived prethermalization regimes exist, with genuine
thermalization occurring only at sufficiently large sizes and at
very long times.

Therefore, the behavior of QQs across the FOQTs of the quantum Ising
chains cannot be considered as completely understood and we believe that
it would deserve further investigation, to strengthen our understanding
of the underlying dynamical mechanisms.  In this respect, numerical studies
of longer chains and time evolutions could be very useful. However, a
significant improvement in this respect represents a very hard task,
as the computations require either the calculation of the full
spectrum to construct the diagonal-ensemble matrix density, or the
time evolution over very long intervals.

We also mention that QQs across the FOQTs of quantum Ising models have
been discussed in terms of spinodal-like quantum nucleation phenomena,
assuming the emergence of some kind of metastabilities during the
out-of-equilibrium evolution, which may be interpreted as the decay of
a false vacuum (see, e.g.,
Refs.~\cite{Rutkevich-99, SCD-21, LSKC-21, MHV-25, JRCB-25, Moss-25}).
However, we
believe that such interpretations may be appropriate only in some
specific regimes, i.e., when the energy injected during the quench is
small, or when the protocols involve slow variations of the
Hamiltonian parameters. This is the case, for instance, in studies of
the Kibble-Zurek mechanism~\cite{Kibble-80,Zurek-96,RV-21}, where the
energy exchange remains much more limited along the quasi-adiabatic
variations of the parameters (see, e.g., Refs.~\cite{PRV-25,PRV-25b}).

Moreover, FOQTs in quantum Ising models are characterized by a
discontinuity in the magnetization [cf.~Eq.~\eqref{sigmasingexp}],
when varying the longitudinal field.  Different scenarios may
arise in other systems featuring FOQTs with a discontinuity in the
energy density, as in the quantum $q$-state Potts models for
sufficiently large $q$~\cite{SP-81,IS-83,CNPV-15,Wu-82} (in particular
for $q > 4$ in one dimension), which are driven by a homogeneous
transverse field.  Therefore, it would be tempting to extend the
study to QQs across such FOQTs driven by external fields that do not
break the global symmetry of the system.

We finally remark that all the numerical results presented
here have been obtained for relatively small system sizes
($L \leq 28$). Using exact-diagonalization methods, we retained
the full Hilbert-space subsector relevant to the quantum
dynamics, which prevented us from accessing larger sizes.
Nonetheless, this allowed us to extract several quantitative results
on the post-QQ behavior of the system in the thermodynamic limit,
irrespective of the entanglement growth, particularly when crossing
quantum transitions of different orders. The outcomes of this
work suggest that, in some circumstances, achieving high accuracy may
be more important than merely scaling to large sizes.
Moreover, this observation may be useful in guiding experimental tests
using, for instance, ultracold atoms in optical lattices~\cite{Bloch-08},
trapped ions~\cite{Debnath-etal-16}, Rydberg atoms in arrays of optical
microtraps~\cite{BL-20} or even quantum-computing architectures
based on superconducting qubits~\cite{Barends-etal-16}.
Recent experiments have begun to probe the dynamics, the excitation
spectrum, as well as nonequilibrium scaling properties of quantum Ising-type
chains~\cite{Gong-etal-16, CerveraLierta-18, LTDR-24, DMEFY-24,
  Ali-etal-24, Zhang-etal-25, WLL-25}, potentially opening new avenues
to observe and investigate the post-QQ dynamical behavior anticipated here.

\appendix

\section{Quantum quenches across the FOQT in the perturbative small-$g$ regime}
\label{Appendix}

In this appendix we analyze the quantum Ising system with a different
set of boundary conditions. We consider an open chain of length $L$
and fix the two boundary spins, restricting the Hilbert space
to eigenstates of $\hat \sigma_1^{(1)}$ and $\hat \sigma_L^{(1)}$
with eigenvalue $+1$ and $-1$, respectively. We consider small values
of $g$ and $h$, allowing us to study the dynamics within the
one-kink subspace, as was already done by the three of us for the Kibble-Zurek
dynamics~\cite{PRV-25,PRV-25b}.  More precisely, we focus on the
infinite-volume limit keeping $h L$ fixed, and restricting our analysis
to values satisfying $hL \lesssim 1$.  We then repeat the analyses
reported in Sec.~\ref{ququfoqt}, with the aim of understanding which
features are specific to the chaotic nature of the spectrum and which
instead stem solely from the fact that the QQ crosses a FOQT.

%%%%%%%%%%%%%%%%%%%%%%%%%%%%%%%%%%%%%%%%%%%%%%%%%%%%%%%%%%%%%%%%%%%%%%%%%%%%%%%%%%
\begin{figure}[!t]
  \includegraphics[width=0.47\textwidth]{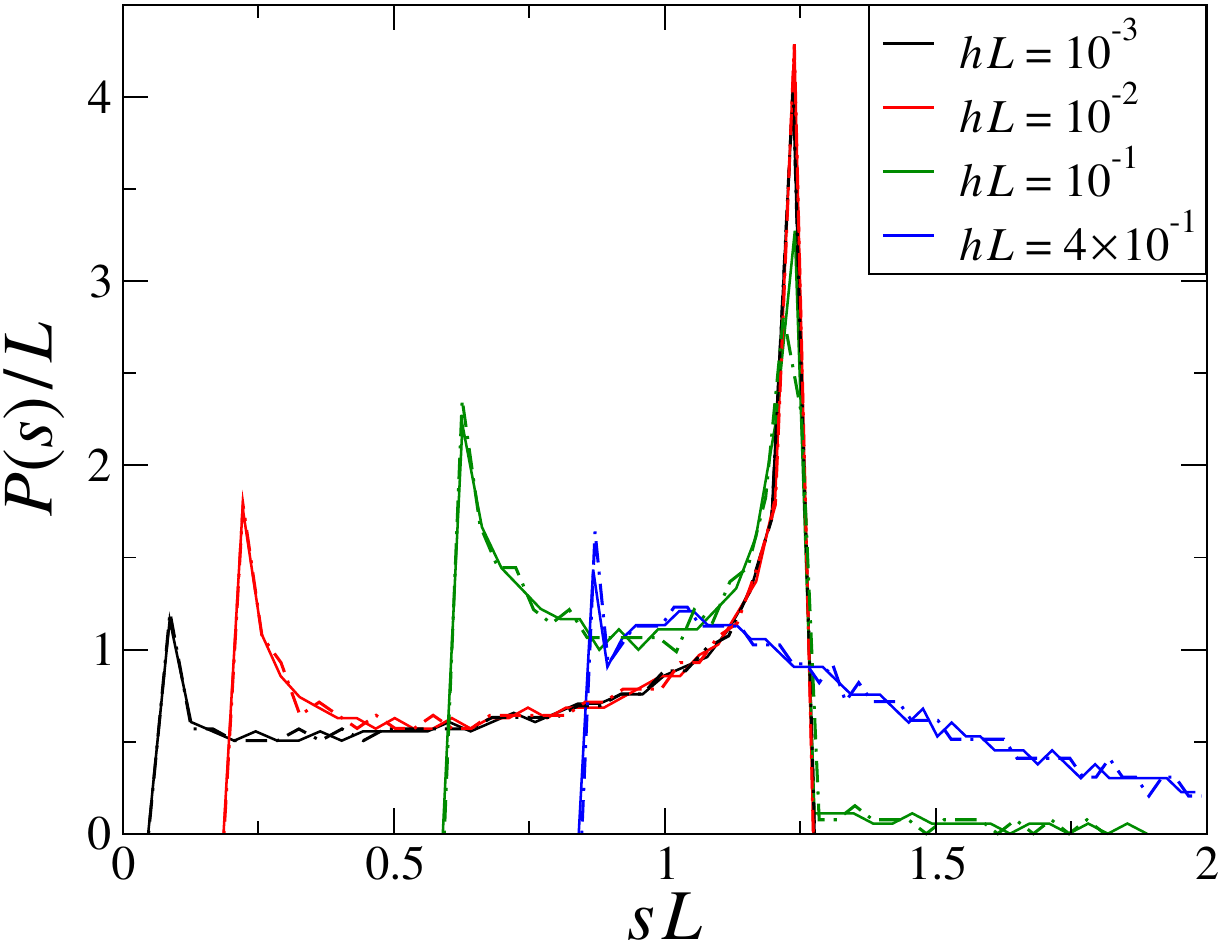} 
  \caption{The rescaled normalized distribution of the energy differences
    $s_n$ for different values of $hL$ and two system sizes,
    $L=1000$ (solid lines) and $L=800$ (dot-dashed lines),
    for a chain with oppositely fixed boundary conditions in
    the one-kink perturbative limit. We set $g=1/10$.}
  \label{diffEn-onekink}
\end{figure}
%%%%%%%%%%%%%%%%%%%%%%%%%%%%%%%%%%%%%%%%%%%%%%%%%%%%%%%%%%%%%%%%%%%%%%%%%%%%%%%%%%

In the limit considered here, the model is clearly integrable and thus
the spacings between consecutive energy levels should not follow the
WD statistics. This is confirmed by our data. In Fig.~\ref{diffEn-onekink}
we show the distribution of the energy differences $s_n$, which
exhibits the expected scaling behavior $P(s_n) = L^{-1} \pi(s_n L)$
as $L\to \infty$. Notably, the distribution displays two different shapes,
depending on the value of the scaling variable $hL$. For $hL \lesssim 0.2$,
it shows a two-peak shape, the higher one being insensitive of $hL$.
In contrast, for $hL \gtrsim 0.2$, it develops a single peak at small
values of $s_n$, followed by a long tail that decreases somewhat slowly.
In both regimes, we find no evidence of a Poissonian behavior.

%%%%%%%%%%%%%%%%%%%%%%%%%%%%%%%%%%%%%%%%%%%%%%%%%%%%%%%%%%%%%%%%%%%%%%%%%%%%%%%%%%
\begin{figure}[!t]
  \includegraphics[width=0.47\textwidth]{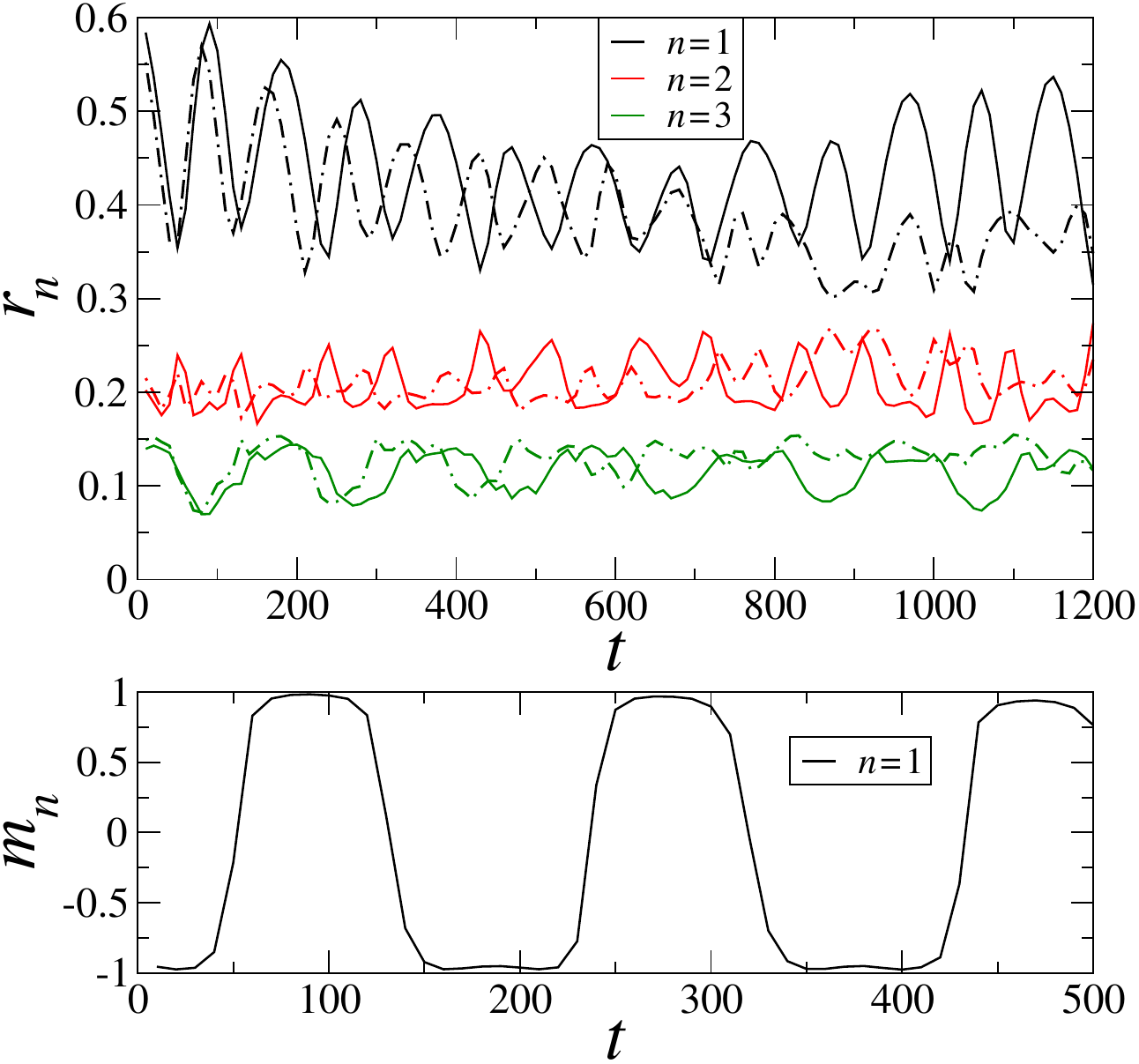} 
  \caption{(Top): The time evolution of the largest eigenvalues of the
    reduced density matrix for a subsystem of size $\ell=6$, for
    two different values of $L$ (continuous lines are for $L=22$,
    dot-dashed lines are for $L=20$).  (Bottom): The time evolution of
    the average magnetization $m_n = \langle r_n(t)|\hat M|r_n(t)\rangle$
    for a subsystem of size $\ell=6$; here $L=22$.
    All calculations refer to a chain with oppositely fixed boundary
    conditions in the one-kink perturbative limit, with $g=1/10$.}
  \label{canonico-onekink}
\end{figure}
%%%%%%%%%%%%%%%%%%%%%%%%%%%%%%%%%%%%%%%%%%%%%%%%%%%%%%%%%%%%%%%%%%%%%%%%%%%%%%%%%%

As already discussed in Sec.~\ref{secIII.A}, one of the main features
of the post-quench spectrum is a strongly peaked distribution of the
overlaps $w_n$ around a single energy value.  This feature is also
present in the one-kink limit.  In this case as well, the
normalized overlap distribution is sharply peaked, with a maximum
that increases with $L$, indicating that the state of the system
is a superposition of a few eigenstates of the post-quench Hamiltonian.
This conclusion is further supported by the behavior of the
diagonal-ensemble entropy. We find that $S_D(L)$ is essentially
independent of $L$, as $L$ increases (at least within the considered
range $50 \leq L \leq 1000$). For small values of $h L$,
$S_D(L)$ is close to 1; it reaches a maximum of $S_D(L)\approx 1.7$
for $hL \approx 0.45$, and then slowly decreases as $h L$ is
further increased. Assuming that the relevant eigenstates are equally
probable, this corresponds to an effective participation
of only about $3$-$5$ states.

Finally, we analyze the main features of the dynamics of small
subsystems.  For this purpose, we consider a subsystem of size
$\ell$, as described in Sec.~\ref{remarks}, and compute the
corresponding reduced density matrix $\hat{\rho}(t)$.
We then study its eigenvalues $r_1(t) > r_2(t) > r_3(t) > \ldots$
and eigenvectors $|r_i(t)\rangle$, as a function of time. We find a
remarkably simple structure, in which very few eigenvectors are relevant
(see Fig.~\ref{canonico-onekink}).  Interestingly, the
leading eigenvector behaves as if it were generated by a QQ in a single
chain with open boundary conditions. Indeed, it oscillates between the
two fully magnetized states, while its bond energy remains approximately
time-independent: $\langle r_1(t) | \hat{H}_s(0) | r_1(t) \rangle \approx -0.8$.
The second-largest eigenvalue corresponds to an
eigenvector that also has a large overlap with the fully magnetized
state; accordingly, its magnetization is typically large and of opposite
sign with respect to that of $| r_1(t) \rangle$. We have
also checked whether the reduced density matrix can be rewritten as
$e^{-\beta_{\rm th} \hat{H}_s}/Z$, for some effective temperature $\beta_{\rm th}$. 
Applying the procedure outlined in Sec.~\ref{remarks}, we obtain a
positive value for $\beta_{\rm th}$. However, when this value is inserted
back in the canonical-ensemble density matrix, the resulting eigenvalues
$r_i(t)$ are quite different from the observed values. This indicated that,
as expected, thermalization does not occur.


\begin{thebibliography}{99}

\bibitem{Fisher-74} M. E. Fisher, The renormalization group in the
  theory of critical behavior, Rev. Mod. Phys. {\bf 46}, 597 (1974);
  Erratum: Rev. Mod. Phys. {\bf 47}, 543 (1975).

\bibitem{Wilson-83} K. G. Wilson, The renormalization group and
  critical phenomena, Rev. Mod. Phys. {\bf 55}, 483 (1983).

\bibitem{Binder-87} K. Binder, Theory of first-order phase
  transitions, Rep. Prog. Phys. {\bf 50}, 783 (1987).

\bibitem{Ma-book} S.-k. Ma, {\em Modern theory of critical phenomena},
  (Routledge, New York, 2001).

\bibitem{PV-02} A. Pelissetto and E. Vicari, Critical phenomena and
   renormalization group theory, Phys. Rep. {\bf 368}, 549 (2002).

\bibitem{SGCS-97} S. L. Sondhi, S. M. Girvin, J. P. Carini, and
   D. Shahar, Continuous quantum phase transitions,
   Rev. Mod. Phys. {\bf 69}, 315 (1997).   

\bibitem{Sachdev-book} S. Sachdev, {\it Quantum Phase Transitions}, 2nd ed.
  (Cambridge University Press, Cambridge, 2011).

\bibitem{Pfleiderer-05} C. Pfleiderer, Why first order quantum phase
  transitions are interesting, J. Phys. Condens. Matter {\bf 17}, S987
  (2005).
  
\bibitem{RV-21} D. Rossini and E. Vicari, Coherent and dissipative
  dynamics at quantum phase transitions, Phys. Rep. {\bf 936}, 1
  (2021).  

\bibitem{PV-24} A. Pelissetto and E. Vicari, Scaling behaviors at quantum
  and classical first-order transitions, in {\em 50 years of the
    renormalization group}, chapter 27, dedicated to the memory of
  Michael E. Fisher, edited by A. Aharony, O. Entin-Wohlman, D. Huse,
  and L. Radzihovsky, World Scientific (2024) [arXiv:2302.08238] 

\bibitem{Kibble-80} T. W. B. Kibble, Some implications of a
  cosmological phase transition, Phys. Rep. {\bf 67}, 183 (1980).

\bibitem{Zurek-96} W. H. Zurek, Cosmological experiments in condensed
  matter systems, Phys. Rep. {\bf 276}, 177 (1996).

\bibitem{Bray-94} A. J. Bray, Theory of phase-ordering kinetics,
  Adv. Phys. {\bf 43}, 357 (1994).

\bibitem{CA-99} B. K. Chakrabarti and M. Acharyya, Dynamic transitions
  and hysteresis, Rev. Mod. Phys. {\bf 71}, 847 (1999).

\bibitem{CG-04} P. Calabrese and A. Gambassi, Ageing Properties of
  Critical Systems, J. Phys. A {\bf 38}, R133 (2005).

\bibitem{PSSV-11} A. Polkovnikov, K. Sengupta, A. Silva, and
  M. Vengalattore, Colloquium: Nonequilibrium dynamics of closed
  interacting quantum systems, Rev. Mod. Phys. {\bf 83}, 863 (2011).

\bibitem{Biroli-16} G. Biroli, Slow Relaxations and Non-Equilibrium
  Dynamics in Classical and Quantum Systems, in {\em Strongly
    interacting quantum systems out of equilibrium: Lecture notes of
    the Les Houches Summer School, Aug. 2012}, ed.  T. Giamarchi and
  A. J. Millis and O. Parcollet and H. Saleur and L. F. Cugliandolo,
  vol. 99, chapter. 3, page 208 (Oxford University Press, 2016);
  arXiv:1507.05858.

\bibitem{CPV-14} M. Campostrini, A. Pelissetto, and E. Vicari,
  Finite-size scaling at quantum transitions, Phys. Rev. B {\bf 89},
  094516 (2014).

\bibitem{GMGW-98} T. Guhr, A. M\"uller, Groeling, and
  H. A. Weidenm\"uller, Random Matrix Theories in Quantum Physics:
  Common Concepts, Phys. Rep. {\bf 299}, 189 (1998).

\bibitem{Alhassid-00}
  Y. Alhassid, The statistical theory of quantum dots,
  Rev. Mod. Phys. {\bf 72}, 895 (2000).

\bibitem{Mehta-book} M. L. Mehta, {\em Random Matrices} (Amsterdam:
  Elsevier/Academic Press, 2004).
  
\bibitem{DKPR-16} L. D'Alessio, Y. Kafri, A. Polkovnikov, and
  M. Rigol, From Quantum Chaos and Eigenstate Thermalization to
  Statistical Mechanics and Thermodynamics, Adv. Phys. {\bf 65}, 239
  (2016).

\bibitem{Izrailev-90} F. M. Izrailev, Simple models of quantum chaos:
  Spectrum and eigenfunctions, Phys. Rep. {\bf 196}, 299 (1990).

\bibitem{ZBFH-96} V. Zelevinsky, B. A. Brown, N. Frazier, and
  M. Horoi, Phys.  Rep. {\bf 276}, 85 (1996).
  
\bibitem{BT-77} M. V. Berry and M. Tabor, Level clustering in
  the regular spectrum,  Proc. Roy. Soc. A {\bf 356}, 375 (1977).

\bibitem{DH-90} B. Dietz and F. Haake, Taylor and Padé analysis of the
  level spacing distributions of random-matrix ensembles, Z. Phys. B
  {\bf 80}, 153 (1990).

\bibitem{OH-07} V. Oganesyan and D. A. Huse, Localization of
  interacting fermions at high temperature, Phys. Rev. B {\bf 75},
  155111 (2007).

\bibitem{RDO-08} M. Rigol, V. Dunjko, and M. Olshanii, Thermalization
  and its mechanism for generic isolated quantum systems, Nature {\bf
    452}, 854 (2008).

\bibitem{SR-10} L. F. Santos and M. Rigol, Onset of quantum chaos in
  one-dimensional bosonic and fermionic systems and its relation to
  thermalization, Phys. Rev. E {\bf 81}, 036206 (2010).

\bibitem{BCH-11} M. C. Ba\~nuls, J. I. Cirac, and M. B. Hastings,
  Strong and Weak Thermalization of Infinite Nonintegrable Quantum
  Systems, Phys. Rev. Lett. {\bf 106}, 050405 (2011).

\bibitem{SPR-11}  
  L. F. Santos, A. Polkovnikov, and M. Rigol,
  Entropy of Isolated Quantum Systems after a Quench,
  Phys. Rev. Lett. {\bf 107}, 040601 (2011).

\bibitem{RS-12} M. Rigol and M. Srednicki, Alternatives to Eigenstate
  Thermalization, Phys. Rev. Lett. {\bf 108}, 110601 (2012).
  
\bibitem{SBI-12} L. F. Santos, F. Borgonovi, and F. M. Izrailev, Chaos
  and Statistical Relaxation in Quantum Systems of Interacting
  Particles, Phys. Rev. Lett.  {\bf 108}, 094102 (2012).

\bibitem{SBI-12b} L. F. Santos, F. Borgonovi, and F. M. Izrailev,
  Onset of chaos and relaxation of isolated systems of interacting
  spins: Energy shell approach, Phys. Rev. E {\bf 85}, 036209 (2012).
  
\bibitem{ABGR-13} Y. Y. Atas, E. Bogomolny, O. Giraud, and G. Roux,
  Distribution of the Ratio of Consecutive Level Spacings in Random
  Matrix Ensembles, Phys. Rev. Lett. {\bf 110}, 084101 (2013).

\bibitem{KH-13} H. Kim and D. A. Huse, Ballistic Spreading of
  Entanglement in a Diffusive Nonintegrable System,
  Phys. Rev. Lett. {\bf 111}, 127205 (2013).

\bibitem{KIH-14} H. Kim, T. N. Ikeda, and D. A. Huse, Testing whether
  all eigenstates obey the eigenstate thermalization hypothesis,
  Phys. Rev. E {\bf 90}, 052105 (2014).

\bibitem{MIKU-17} T. Mori, T. N. Ikeda, E. Kaminishi, and N. Ueda,
  Thermalization and prethermalization in isolated quantum systems: a
  theoretical overview, J. Phys. B {\bf 51}, 112001 (2018).
  
\bibitem{LM-17} C.-J. Liu and O. I. Motrunich, Quasiparticle
  explanation of the weak-thermalization regime under quench in a
  nonintegrable quantum spin chain, Phys. Rev. A {\bf 95}, 023621
  (2017).

\bibitem{PC-22} C. Peng and X. Cui, Bridging quantum many-body scars
  and quantum integrability in Ising chains with transverse and
  longitudinal fields, Phys. Rev. B {\bf 106}, 214311 (2022).

\bibitem{Greiner-02} M. Greiner, O. Mandel, T. Esslinger,
  T. W. H\"ansch, and I. Bloch, Quantum phase transition from a
  superfluid to a Mott insulator in a gas of ultracold atoms, Nature
  {\bf 415}, 39 (2002).

\bibitem{Weiss-06} T. Kinoshita, T. Wenger, and D. S. Weiss, A quantum
  Newton's cradle, Nature {\bf 440}, 900 (2006).

\bibitem{Schmiedmayer-07} S. Hofferberth, I. Lesanovsky, B. Fischer,
  T. Schumm, and J. Schmiedmayer, Non-equilibrium coherence dynamics
  in one-dimensional Bose gases, Nature {\bf 449}, 324 (2007).

\bibitem{Trotzky-12} S. Trotzky, Y.-A. Chen, A. Flesch,
  I. P. McCulloch, U. Schollw\"ock, J. Eisert, and I. Bloch, Probing
  the relaxation towards equilibrium in an isolated strongly
  correlated one-dimensional Bose gas, Nat. Phys. {\bf 8}, 325 (2012).

\bibitem{Cheneau-12} M. Cheneau, P. Barmettler, D. Poletti, M. Endres,
  P. Scbau\ss, T. Fukuhara, C. Gross, I. Bloch, C. Kollath, and
  S. Kuhr, Light-cone-like spreading of correlations in a quantum
  many-body system, Nature {\bf 481}, 484 (2012).

\bibitem{Schmiedmayer-12} M. Gring, M. Kuhnert, T. Langen,
  T. Kitagawa, B. Rauer, M. Schreitl, I. Mazets, D. Adu Smith,
  E. Demler, and J. Schmiedmayer, Relaxation and Prethermalization in
  an Isolated Quantum System, Science {\bf 337}, 1318 (2012).

\bibitem{Guardado-etal-18} E. Guardado-Sanchez, P. T. Brown, D. Mitra,
  T. Devakul, D. A. Huse, P. Schauss, and W. S. Bakr, Probing the
  quench dynamics of antiferromagnetic correlations in a 2D quantum
  Ising spin system, Phys. Rev. X {\bf 8}, 021069 (2018).
  
\bibitem{Ali-etal-24}
  A. Ali, H. Xu, W. Bernoudy, A. Nocera, A. D. King, and A. Banerjee,
  Quantum quench dynamics of geometrically frustrated Ising models,
  Nat Commun {\bf 15}, 10756 (2024). 

\bibitem{Niemeijer-67} T. Niemeijer, Some exact calculations on a
  chain of spins $1/2$, Physica {\bf 36}, 377 (1967); Some exact
  calculations on a chain of spins $1/2$ II, Physica {\bf 39}, 313
  (1968).

\bibitem{BMD-70} E. Barouch, B. McCoy, and M. Dresden, Statistical
  mechanics of the XY model. I, Phys. Rev. A {\bf 2}, 1075 (1970).

\bibitem{DMCF-06} G. De Chiara, S. Montangero, P. Calabrese, and
  R. Fazio, Entanglement entropy dynamics of Heisenberg chains,
  J. Stat. Mech. P03001 (2006).

\bibitem{RDYO-07} M. Rigol, V. Dunjko, V. Yurovsky, and M. Olshanii,
  Relaxation in a Completely Integrable Many-Body Quantum System: An
  Ab Initio Study of the Dynamics of the Highly Excited States of 1D
  Lattice Hard-Core Bosons, Phys. Rev. Lett. {\bf 98}, 050405 (2007).

\bibitem{KLA-07} C. Kollath, A. M. L\"auchli, and E. Altman, Quench
  Dynamics and Nonequilibrium Phase Diagram of the Bose-Hubbard Model,
  Phys. Rev. Lett. {\bf 98}, 180601 (2007).

\bibitem{MWNM-07} S. R. Manmana, S. Wessel, R. M. Noack, and
  A. Muramatsu, Strongly Correlated Fermions after a Quantum Quench,
  Phys. Rev. Lett. {\bf 98}, 210405 (2007).

\bibitem{ZPP-08} M. \v{Z}nidari\v{c}, T. Prosen, and P. Prelov\v{s}ek,
  Many-body localization in the Heisenberg XXZ magnet in a random
  field, Phys. Rev. B {\bf 77}, 064426 (2008).

\bibitem{Bloch-08}
  I. Bloch, Quantum coherence and entanglement with
  ultracold atoms in optical lattices, Nature {\bf 453}, 1016 (2008).
  
\bibitem{PZ-09}
  T. Prosen and M. \v{Z}nidari\v{c},
  Matrix product simulations of non-equilibrium steady states
  of quantum spin chains, J. Stat. Mech. (2009) P02035.

\bibitem{Roux-10} G. Roux, Finite-size effects in global quantum
  quenches: Examples from free bosons in an harmonic trap and the
  one-dimensional Bose-Hubbard model, Phys. Rev. A {\bf 81}, 053604
  (2010).
  
\bibitem{IR-11} F. Igl\'oi and H. Rieger, Quantum Relaxation after a
  Quench in Systems with Boundaries, Phys. Rev. Lett. {\bf 106},
  035701 (2011); H. Rieger and F. Igl\'oi, Semiclassical theory for
  quantum quenches in finite transverse Ising chains, Phys. Rev. B
  {\bf 84}, 165117 (2011).

\bibitem{CEF-12-1} P. Calabrese, F. H. Essler, and M. Fagotti, Quantum
  quench in the transverse field Ising chain: I. Time evolution of
  order parameter correlators, J. Stat. Mech. (2012) P07016.

\bibitem{CEF-12-2} P. Calabrese, F. H. Essler, and M. Fagotti, Quantum
  quenches in the transverse field Ising chain: II. Stationary state
  properties, J. Stat. Mech. (2012) P07022.

\bibitem{CE-13} J.-S. Caux and F. H. L. Essler, Time evolution of
  local observables after quenching to an integrable model,
  Phys. Rev. Lett. {\bf 110}, 257203 (2013).
  
\bibitem{FCEC-14}
  M. Fagotti, M. Collura, F. H. L. Essler, and P. Calabrese,
  Relaxation after quantum quenches in the spin-1/2 Heisenberg XXZ chain,
  Phys. Rev. B {\bf 89}, 125101 (2014).
  
\bibitem{NH-15} R. Nandkishore and D. A. Huse, Many body localization
  and thermalization in quantum statistical mechanics,
  Annu. Rev. Condens. Matter Phys. {\bf 6}, 15 (2015).

\bibitem{CTGM-15} A. Chiocchetta, M. Tavora, A. Gambassi, and
  A. Mitra, Short-time universal scaling and light-cone dynamics after
  a quench in an isolated quantum system in $d$ spatial dimensions,
  Phys. Rev. B {\bf 94}, 134311 (2016).

\bibitem{BDD-15} S. Bhattacharyya, S. Dasgupta, and A. Das, Signature
  of a continuous quantum phase transition in non-equilibrium energy
  absorption: Footprints of criticality on higher excited states,
  Sci. Rep. {\bf 5}, 16490 (2015).

\bibitem{CC-16} P. Calabrese and J. Cardy, Quantum quenches in $1+1$
  dimensional conformal field theories, J. Stat. Mech. (2016) 064003.

\bibitem{BD-16} D. Bernard and B. Doyon, Conformal field theory out of
  equilibrium: a review, J. Stat. Mech. (2016) 064005.

\bibitem{IMPZ-16} E. Ilievski, M. Medenjak, T. Prosen, and L. Zadnik,
  Quasilocal charges in integrable lattice systems,
  J. Stat. Mech. (2016) 064008.

\bibitem{LGS-16} T. Langen, T. Gasenzer, and J. Schmiedmayer,
  Prethermalization and universal dynamics in near-integrable quantum
  systems, J. Stat. Mech. (2016) 064009.

\bibitem{VM-16} R. Vasseur and J. E. Moore, Nonequilibrium quantum
  dynamics and transport: from integrability to many-body
  localization, J. Stat. Mech. (2016) 064010.

\bibitem{EF-16} F. H. L. Essler and M. Fagotti, Quench dynamics and
  relaxation in isolated integrable quantum spin chains,
  J. Stat. Mech. (2016) 064002.

\bibitem{CKT-16}
  M. Collura, M. Kormos, and G. Tak{\'a}cs,
  Dynamical manifestation of the Gibbs paradox after a quantum quench,
  Phys. Rev. A {\bf 98}, 053610 (2018).
  
\bibitem{NRVH-17} A. Nahum, J. Ruhman, S. Vijay, and J. Haah, Quantum
  Entanglement Growth under Random Unitary Dynamics, Phys. Rev. X {\bf
    7}, 031016 (2017).
  
\bibitem{Heyl-17} M. Heyl, Dynamical quantum phase transitions: a
  review, Rep. Prog. Phys. {\bf 81}, 054001 (2018).

\bibitem{RMD-17} S. Roy, R. Moessner, and A. Das, Locating topological
  phase transitions using nonequilibrium signatures in local bulk
  observables, Phys. Rev. B {\bf 95}, 041105(R) (2017).

\bibitem{KCTC-17}
  M. Kormos, M. Collura, G. Tak{\'a}cs, and P. Calabrese,  Real-time
  confinement following a quantum quench to a non-integrable model,
  Nat. Phys {\bf 13}, 246 (2017).
  
\bibitem{PRV-18} A. Pelissetto, D. Rossini, and E. Vicari, Dynamic
  finite-size scaling after a quench at quantum transitions,
  Phys. Rev. E {\bf 97}, 052148 (2018).
  
 \bibitem{HPD-18} M. Heyl, F. Pollmann, and B. Dora, Detecting
   equilibrium and dynamical quantum phase transitions in Ising chains
   via out-of-time-ordered correlators, Phys. Rev. Lett. {\bf 121},
   016801 (2018).

\bibitem{TIGGG-19} P. Titum, J. T. Iosue, J. R. Garrison,
  A. V. Gorshkov, and Z.-X. Gong, Probing ground-state phase
  transitions through quench dynamics, Phys. Rev. Lett. {\bf 123},
  115701 2019.

\bibitem{Silva-08} A. Silva, Statistics of the Work Done on a Quantum
  Critical System by Quenching a Control Parameter, Phys.
  Rev. Lett. {\bf 101}, 120603 (2008).

\bibitem{GS-12} A. Gambassi and A. Silva, Large Deviations and
  Universality in Quantum Quenches, Phys. Rev. Lett. {\bf 109}, 250602
  (2012).

\bibitem{MS-14} J. Marino and A. Silva, Non-Equilibrium Dynamics of a
  Noisy Quantum Ising Chain: statistics of the work and
  prethermalization after a sudden quench of the transverse field,
  Phys. Rev. B {\bf 89}, 024303 (2014).

\bibitem{MR-18} K. Mallayya and M. Rigol, Quantum Quenches and
  Relaxation Dynamics in the Thermodynamic Limit,
  Phys. Rev. Lett. {\bf 120}, 070603 (2018).

\bibitem{MRD-19}  
  K. Mallayya, M. Rigol, and W. De Roeck, Prethermalization
  and thermalization in isolated quantum systems, Phys. Rev. X
  {\bf 9}, 021027 (2019).

\bibitem{NRV-19-w} D. Nigro, D. Rossini, and E. Vicari, Dynamic
  scaling of work fluctuations after quenches near quantum
  transitions, J. Stat. Mech. (2019) 023104.

\bibitem{NRV-19} D. Nigro, D. Rossini, and E. Vicari, Competing
  coherent and dissipative dynamics close to quantum criticality,
  Phys. Rev. A {\bf 100}, 052108 (2019); D. Rossini and E. Vicari,
  Scaling behavior of stationary states arising from dissipation at
  continuous quantum transitions, Phys. Rev. B {\bf 100}, 174303
  (2019).

\bibitem{STT-19} J. Surace, L. Tagliacozzo, and E. Tonni, Operator
  content of entanglement spectra after global quenches in the
  transverse field Ising chain, Phys. Rev. B {\bf 101}, 241107(R)
  (2020).

\bibitem{RV-20-qu} D. Rossini and E. Vicari, Dynamics after quenches
  in one-dimensional quantum Ising-like systems, Phys. Rev. B {\bf
    102}, 054444 (2020).

\bibitem{MR-21} K. Mallayya and M. Rigol, Prethermalization,
  thermalization, and Fermi’s golden rule in quantum many-body
  systems, Phys. Rev. B {\bf 104}, 184302 (2021).

\bibitem{HMHPRD-21}
  A. Haldar, K. Mallayya, M. Heyl, F. Pollmann, M. Rigol, and A. Das,
  Signatures of quantum phase transitions after quenches
  in quantum chaotic one-dimensional systems, Phys. Rev. X {\bf 11},
  031062 (2021).

\bibitem{MW-78} B. M. McCoy and T. T. Wu, Two-dimensional Ising field
  theory in a magnetic field: breakup of the cut in the two-point
  function, Phys. Rev. D {\bf 4}, 1259 (1978).

\bibitem{Rutkevich-99} S. B. Rutkevich, Decay of the metastable phase
  in $d=1$ and $d=2$ Ising models, Phys. Rev. B {\bf 60}, 14525
  (1999).

\bibitem{PRV-18-def} A. Pelissetto, D. Rossini, and E. Vicari,
  Out-of-equilibrium dynamics driven by localized time-dependent
  perturbations at quantum phase transitions, Phys. Rev. B {\bf 97},
  094414 (2018).

\bibitem{LZW-19} Q. Luo, J. Zhao, and X. Wang, Intrinsic jump
  character of first-order quantum phase transitions, Phys. Rev. B
  {\bf 100}, 121111(R) (2019).

\bibitem{PRV-20} A. Pelissetto, D. Rossini, and E. Vicari, Scaling
  properties of the dynamics at first-order quantum transitions when
  boundary conditions favor one of the two phases, Phys. Rev. E {\bf
    102}, 012143 (2020). 

\bibitem{DRV-20} G. Di Meglio, D. Rossini, and E. Vicari, Dissipative
  dynamics at first-order quantum transitions, Phys. Rev. B {\bf 102},
  224302 (2020).

\bibitem{RV-20-qm}  
  D. Rossini and E. Vicari, Measurement-induced dynamics of
  many-body systems at quantum transitions, Phys. Rev. B {\bf 102}
  035119 (2020).

\bibitem{SCD-21} A. Sinha, T. Chanda, and J. Dziarmaga, Nonadiabatic
  dynamics across a first-order quantum phase transition: Quantized
  bubble nucleation, Phys. Rev. B {\bf 103}, L220302 (2021).

\bibitem{LSKC-21}
  G. Lagnese, F. M. Surace, M. Kormos, and P. Calabrese,
  False vacuum decay in quantum spin chains,
  Phys. Rev. B {\bf 104}, L201106 (2021).

\bibitem{TV-22} F. Tarantelli and E. Vicari, Out-of-equilibrium
  dynamics arising from slow round-trip variations of Hamiltonian
  parameters across quantum and classical critical points,
  Phys. Rev. B {\bf 105}, 235124 (2022).
  
\bibitem{TS-23} F. Tarantelli and S. Scopa, Out-of-equilibrium scaling
  behavior arising during round-trip protocols across a quantum
  first-order transition, Phys. Rev. B {\bf 108}, 104316 (2023).

\bibitem{Surace-etal-24} F. M. Surace, A. Lerose, O. Katz,
  E. R. Bennewitz, A. Schuckert, De Luo, A. De, B. Ware, W. Morong,
  K. Collins, C. Monroe, Z. Davoudi, and A. V. Gorshkov,
  String-breaking dynamics in quantum adiabatic and diabatic
  processes, arXiv:2411.10652.  
  
\bibitem{PRV-25} A. Pelissetto, D. Rossini, and E. Vicari,
  Out-of-equilibrium dynamics across the first-order quantum
  transitions of one-dimensional quantum Ising models, 
  Phys. Rev. B {\bf 111}, 224306 (2025).

\bibitem{PRV-25b} A. Pelissetto, D. Rossini, and E. Vicari,
  Kibble-Zurek dynamics across the first-order quantum transitions of
  one-dimensional quantum Ising models in the thermodynamic limit,
  Phys. Rev. B {\bf 112}, 104309 (2025).

\bibitem{MHV-25} D. Maertens, J. Haegeman, and K. Van Acoleyen,
  Real-time bubble nucleation and growth for false vacuum decay on the
  lattice, arXiv:2508.13645.

\bibitem{JRCB-25}
  C. Johansen, A. Recati, I. Carusotto, and A. Biella,
  Many-body theory of false vacuum decay in quantum spin chains,
  arXiv:2508.13780.

\bibitem{Moss-25} I. G. Moss, Notes on false vacuum decay in quantum
  Ising models, arXiv:2510.12592.

\bibitem{ZDZ-05} W. H. Zurek, U. Dorner, and P. Zoller, Dynamics of a
  quantum phase transition, Phys. Rev. Lett. {\bf 95}, 105701 (2005).

\bibitem{PG-08} A. Polkovnikov and V. Gritsev, Breakdown of the
  adiabatic limit in low-dimensional gapless systems, Nature
  Phys. {\bf 4}, 477 (2008).

\bibitem{SUXF-06}
  R. Sch{\" u}tzhold, M. Uhlmann, Y. Xu, and U. R. Fischer,
  Sweeping from the superfluid to the Mott phase in the Bose-Hubbard model,
  Phys. Rev. Lett. {\bf 97}, 200601 (2006).

\bibitem{USF-10}
  M. Uhlmann, R. Sch{\" u}tzhold, and U. R. Fischer,
  System size scaling of topological defect creation in a second-order
  dynamical quantum phase transition,
  New J. Phys. {\bf 12}, 095020 (2010).
  
\bibitem{CEGS-12} A. Chandran, A. Erez, S. S. Gubser, and
  S. L. Sondhi, Kibble-Zurek problem: Universality and the scaling
  limit, Phys. Rev. B {\bf 86}, 064304 (2012).

\bibitem{RV-20} D. Rossini and E. Vicari, Dynamic Kibble-Zurek scaling
  framework for open dissipative many-body systems crossing quantum
  transitions, Phys. Rev. Research {\bf 2}, 023611 (2020).
  
\bibitem{DV-23} F. De Franco and E. Vicari, Out-of-equilibrium
  finite-size scaling in generalized Kibble-Zurek protocols crossing
  quantum phase transitions in the presence of symmetry-breaking
  perturbations, Phys. Rev. B {\bf 107}, 115175 (2023).
  
\bibitem{RV-24} D. Rossini and E. Vicari, Interplay between
  short-range and critical long-range fluctuations in the
  out-of-equilibrium behavior of the particle density at quantum
  transitions, Phys. Rev. B {\bf 110} 035126 (2024).

\bibitem{Sandvik-10}
  A. W. Sandvik, Computational studies of quantum spin systems,
  AIP Conf. Proc. {\bf 1297}, 135 (2010). 

\bibitem{Pfeuty-70} P. Pfeuty, The one-dimensional Ising model with a
  transverse field, Ann. Phys. {\bf 57}, 79 (1970).

\bibitem{BILV-21}
  M. Baldovin, S. Iubini, R. Livi, and A. Vulpiani,
  Statistical Mechanics of Systems with Negative Temperature,
  Phys. Rep. {\bf 923}, 1 (2021).

\bibitem{Necklaces-ref1} G. Polya, R. C.  Read, and D. Aeppli, {\em
  Combinatorial enumeration of groups, graphs, and chemical compounds}
  (Springer-Verlag, 1989).
  
\bibitem{Necklaces-ref2}
  F. Ruskey, C. Savage, and T. M. Y.  Wang,
  Generating necklaces, Jornal of Algorithms {\bf 13}, 414 (1992)

\bibitem{Bohigas-91} O. Bohigas, in Les Houches Lecture Series, edited
  by M.J.  Giannoni, A. Voros, and J. Zinn-Justin (North-Holland,
  Amsterdam, 1991), Vol. 52.
  
\bibitem{BCMS-03} G. Benenti, G. Casati, S. Montangero, and
  D. L. Shepelyansky, Statistical properties of eigenvalues for an
  operating quantum computer with static imperfections,
  Eur. Phys. J. D {\bf 22}, 285 (2003).

\bibitem{Rutkevich-08} S. B. Rutkevich, Energy spectrum of
  bound-spinons in the quantum Ising spin-chain ferromagnet,
  J. Stat. Phys. {\bf 131}, 917 (2008).

\bibitem{AC-09} M. H. S. Amin and V. Choi, First-order quantum phase
  transition in adiabatic quantum computation, Phys. Rev. A {\bf 80},
  062326 (2009).

\bibitem{YKS-10} A. P. Young, S. Knysh, and V. N. Smelyanskiy, First
  order phase transition in the quantum adiabatic algorithm,
  Phys. Rev. Lett. {\bf 104}, 020502 (2010).

\bibitem{JLSZ-10} T. J\"org, F. Krzakala, G. Semerjian, and
  F. Zamponi, First-order transitions and the performance of quantum
  algorithms in random optimization problems, Phys. Rev. Lett. {\bf
    104}, 207206 (2010).

\bibitem{Coldea-etal-10}
  R. Coldea, D. A. Tennant, E. M. Wheeler, E. Wawrzynska, D. Prabhakaran,
  M. Telling, K. Habicht, P. Smeibidl, and K. Kiefer,
  Quantum criticality in an Ising chain: experimental evidence of the
  emergent $E_8$ symmetry, Science {\bf 327}, 177 (2010).

\bibitem{Rutkevich-10} S. B. Rutkevich, On the weak confinement of
  kinks in the one-dimensional quantum ferromagnet CoNb$_2$O$_6$,
  J. Stat. Mech. (2010) P07015.

\bibitem{LMSS-12} C. R. Laumann, R. Moessner, A. Scardicchio, and
  S. L. Sondhi, Quantum adiabatic algorithm and scaling of gaps at
  first-order quantum phase transitions, Phys. Rev. Lett. {\bf 109},
  030502 (2012).

\bibitem{CNPV-14} M. Campostrini, J. Nespolo, A. Pelissetto, and
  E. Vicari, Finite-size scaling at first-order quantum transitions,
  Phys. Rev. Lett. {\bf 113}, 070402 (2014).

\bibitem{CPV-15} M. Campostrini, A. Pelissetto, and E. Vicari, Quantum
  transitions driven by one-bond defects in quantum Ising rings,
  Phys. Rev. E {\bf 91}, 042123 (2015). 

\bibitem{CPV-15b} M. Campostrini, A. Pelissetto, and E. Vicari,
  Quantum Ising chains with boundary fields, J. Stat. Mech. 
  P11015 (2015). 

\bibitem{RV-18} D. Rossini and E. Vicari, Ground-state fidelity at
  first-order quantum transitions, Phys. Rev. E {\bf 98}, 062137
  (2018).
  
\bibitem{PRV-18-fowb} A. Pelissetto, D. Rossini, and E. Vicari,
  Finite-size scaling at first-order quantum transitions when boundary
  conditions favor one of the two phases, Phys. Rev. E {\bf 98},
  032124 (2018). 

\bibitem{FB-82} M. E. Fisher and A. N. Berker, Scaling for first-order
  phase transitions in thermodynamic and finite systems, Phys. Rev. B
  {\bf 26}, 2507 (1982).

\bibitem{PF-83} V. Privman and M. E. Fisher, Finite-size effects at
  first-order transitions, J. Stat. Phys. {\bf 33}, 385 (1983).

\bibitem{FP-85} M. E. Fisher and V. Privman, First-order transitions
  breaking O$(n)$ symmetry: Finite-size scaling, Phys. Rev. B {\bf
    32}, 447 (1985).

\bibitem{CLB-86} M. S. S. Challa, D. P. Landau, and K. Binder,
  Finite-size effects at temperature-driven first-order transitions,
  Phys. Rev. B {\bf 34}, 1841 (1986).

\bibitem{BK-90} C. Borgs and R. Koteck{\'y}, A rigorous theory of
  finite-size scaling at first-order phase transitions,
  J. Stat. Phys. {\bf 61}, 79 (1990).

\bibitem{LK-91} J. Lee and J. M. Kosterlitz, Finite-size scaling and
  Monte Carlo simulations of first-order phase
  transitions. Phys. Rev. B {\bf 43}, 3265 (1991).

\bibitem{BK-92} C. Borgs and R. Koteck{\'y}, Finite-Size Effects at
  Asymmetric First-Order Phase Transitions. Phys. Rev. Lett. {\bf 68},
  1734 (1992).
  
\bibitem{VRSB-93} K. Vollmayr, J. D. Reger, M. Scheucher, and
  K. Binder, Finite size effects at thermally-driven first order phase
  transitions: A phenomenological theory of the order parameter
  distribution, Z. Phys. B {\bf 91}, 113 (1993).

\bibitem{RTMS-94} P. A. Rikvold, H. Tomita, S. Miyashita, and
  S. W. Sides, Metastable lifetimes in a kinetic Ising model:
  Dependence on field and system size, Phys. Rev. E {\bf 49}, 5080
  (1994).
  
\bibitem{NIW-11} T. Nogawa, N. Ito, and H. Watanabe, Static and
  dynamical aspects of the metastable states of first order transition
  systems, Physics Procedia {\bf 15}, 76 (2011).
 
\bibitem{TB-12} A. Tr\"oster and K. Binder, Microcanonical
  determination of the interface tension of flat and curved interfaces
  from Monte Carlo simulations, J. Phys.: Condens. Matter {\bf 24},
  284107 (2012).

\bibitem{PV-17} A. Pelissetto and E. Vicari, Dynamic off-equilibrium
  transition in systems slowly driven across thermal first-order
  phase transitions, Phys. Rev. Lett. {\bf 118}, 030602 (2017).

\bibitem{PV-17-dyn} A. Pelissetto and E. Vicari, Dynamic finite-size
  scaling at first-order transitions, Phys. Rev. E {\bf 96}, 012125
  (2017).

\bibitem{SW-18} S. Scopa and S. Wald, Dynamical off-equilibrium
  scaling across magnetic first-order phase transitions,
  J. Stat. Mech. 113205 (2018).

\bibitem{Fontana-19} P. Fontana, Scaling behavior of Ising systems at
  first-order transitions, J. Stat. Mech. 063206 (2019).
  
\bibitem{PV-25} A. Pelissetto and E. Vicari,
  Out-of-equilibrium spinodal-like scaling behaviors across the
  magnetic first-order transitions of two-dimensional and
  three-dimensional Ising systems, Phys. Rev. E {\bf 113}, 014107 (2026).

\bibitem{PRV-25c} A. Pelissetto, D. Rossini, and E. Vicari,
  Out-of-equilibrium spinodal-like scaling behaviors at the thermal
  first-order transitions of three-dimensional $q$-state Potts models,
  arXiv:2511.16510.

\bibitem{SP-81} J. S\'olyom and P. Pfeuty, Renormalization-group study
  of the Hamiltonian version of the Potts model, Phys. Rev. B {\bf
    24}, 218 (1981).

\bibitem{IS-83} F. Igl\'oi and J. S\'olyom, First-order transition for
  the (1+1)-dimensional $q>4$ Potts model from finite lattice
  extrapolation, J. Phys. C: Solid State Phys.  {\bf 16}, 2833 (1983).

\bibitem{CNPV-15} M. Campostrini, J. Nespolo, A. Pelissetto, and
  E. Vicari, Finite-size scaling at the first-order quantum transitions
  of quantum Potts chains, Phys. Rev. E {\bf 91}, 052103 (2015).

\bibitem{Wu-82}
  F. Y. Wu, The Potts model, Rev. Mod. Phys. {\bf 54}, 235 (1982).

\bibitem{Debnath-etal-16} S. Debnath, N. M. Linke, C. Figgatt,
  K. A. Landsman, K. Wright, and C. Monroe, Demonstration of a small
  programmable quantum computer with atomic qubits, Nature {\bf 536},
  63 (2016).

\bibitem{BL-20}
  A. Browaeys and T. Lahaye, Many-body physics with individually
  controlled Rydberg atoms, Nat. Phys. {\bf 16}, 132 (2020).

\bibitem{Barends-etal-16}
  R. Barends, A. Shabani, L. Lamata, J. Kelly, A. Mezzacapo,
  U. L. Heras, R. Babbush, A. G. Fowler, B. Campbell, Y. Chen,
  {\it et al.}, Digitized adiabatic quantum computing with
  a superconducting circuit, Nature {\bf 534}, 222 (2016).

\bibitem{Gong-etal-16} M. Gong, X. Wen, G. Sun, D.-W. Zhang, D. Lan,
  Y. Zhou, Y. Fan, Y. Liu, X. Tan, H. Yu, Y. Yu, S.-L. Zhu, S. Han,
  and P. Wu, Simulating the Kibble-Zurek mechanism of the Ising model
  with a superconducting qubit system, Sci. Rep. {\bf 6}, 22667
  (2016).

\bibitem{CerveraLierta-18} A. Cervera-Lierta, Exact Ising model
  simulation on a quantum computer, Quantum {\bf 2}, 114 (2018).

\bibitem{LTDR-24}
  C. Lamb, Y. Tang, R. Davis, and A. Roy, Ising meson spectroscopy
  on a noisy digital quantum simulator, Nat Commun {\bf 15}, 5901 (2024). 

\bibitem{DMEFY-24}
  C. B. Dag, H. Ma, P. Myles Eugenio, F. Fang, and S. F. Yelin,
  Emergent disorder and sub-ballistic dynamics in quantum simulations
  of the Ising model using Rydberg atom arrays,
  Phys Rev. Lett. {\bf 135}, 250403 (2025).

\bibitem{Zhang-etal-25}
  T. Zhang, H. Wang, W. Zhang, Y. Wang, A. Du, Z. Li, Y. Wu, C. Li,
  J. Hu, H. Zhai, W. Chen,  Observation of near-critical Kibble-Zurek
  scaling in Rydberg atom arrays, Phys. Rev. Lett. {\bf 135}, 093403 (2025).

\bibitem{WLL-25}
  H. Wang, X. Li, and C. Li, Tricritical Kibble-Zurek scaling
  in Rydberg atom ladders,  Nat. Commun. {\bf 16}, 10584 (2025).
  
\end{thebibliography}
\end{document}